\documentclass{aastex631}
\usepackage[utf8]{inputenc}

\usepackage{scrextend}
\usepackage{rotating}
\usepackage{amsmath}
\usepackage{hyperref}

\title{Analysis of CN Coma Morphology Features of Comet 21P/Giacobini-Zinner}

\shorttitle{CN features of 21P/Giacobini Zinner}
\shortauthors{Goldberg and Lejoly et al.}

\begin{document}

\title{Analysis of CN Coma Morphology Features of Comet 21P/Giacobini-Zinner}
\author[0000-0002-4901-7693]{Charles Goldberg}
\affil{Lunar and Planetary Laboratory, University of Arizona \\
1629 E University Blvd \\
Tucson AZ 85721-0092, USA }

\author[0000-0003-0165-7701]{Cassandra Lejoly}
\affil{Lunar and Planetary Laboratory, University of Arizona \\
1629 E University Blvd \\
Tucson AZ 85721-0092, USA }

\author[0000-0001-8925-7010]{Nalin Samarasinha}
\affil{Planetary Science Institute \\
1700 East Fort Lowell Rd \\
Tucson AZ 85719-2395, USA }

\begin{abstract}
We analyze jet features found in the coma of Comet 21P/Giacobini-Zinner (21P/GZ) during its 2018 perihelion passage using narrowband CN photometric imaging in order to determine the comet's rotational period, and constrain the CN gas outflow velocity and rotational state through the analysis of azimuthally enhanced morphological features. We find that 21P/GZ has a period of either 7.39 $\pm$ 0.01 or 10.66 $\pm$ 0.01 hours. We measure a lower limit to the outflow velocity for the Northern Jet of 730 $\pm$ 30 m s$^{-1}$ and for the Southern Jet of 740 $\pm$ 30 m s$^{-1}$. We analyze the morphologies of the jet features and determine that the Northern jet possesses a corkscrew pattern, and utilize that knowledge to determine a rotational pole position at a Right Ascension of ${169^{+28}_{-23}}^{\circ}$ and a Declination of ${73^{+5}_{-11}}^{\circ}$, with undetermined sense of rotation.
\end{abstract}
\keywords{Comets --- CN --- 21P/GZ }

\section{Introduction}\label{introduction}
\label{Periodicity} \label{outflow_velo} \label{spin_axis} \label{x_param}
 Since its discovery in 1900 by Michel Giacobini, and rediscovery by Ernst Zinner in 1913, Jupiter Family Comet 21P/Giacobini-Zinner (21P/GZ) has been monitored over its many apparitions, including throughout its 2018 perihelion passage. During the 2018 apparition, we obtained four nights of photometric imaging data spanning from 18 August 2018 to 12 September 2018. Utilizing narrowband cyanogen (CN) photometric imaging to highlight 21P/GZ's jet features, we are able to constrain the comet's gas outflow velocity and spin properties, then compare to previously obtained rotational periods to confirm standard torque effects.  

The rotation period of a cometary nucleus can provide clues to the understanding of the comet's structure, evolution, and formation. There are multiple methods for determining the rotation period of cometary nuclei, including obtaining a lightcurve of the bare nucleus when it is far away from the Sun and minimal outgassing is present, or by obtaining a lightcurve of the coma itself when it is active and assume that variations in the coma are representative of the nucleus' rotational state. These methods have been used for many comets in the past including comet 1P/Halley (e.g.: \citet{Millis}, \citet{Leibowitz_Halley}) or even more recently for a variety of comets (e.g.: \citet{Kokotanekova}, \citet{Bodewits_41P}). Other methods of determining the period of cometary nuclei include the cross-correlation of jet features (e.g.: \citet{Ivanova_2012}), the image cross-correlation of CN shells (e.g.: \citet{Waniak}), repeatability of jet features (e.g.: \citet{Knight_2011}, \citet{Bodewits_41P}), and even radar observations (e.g.: \citep{Ellen_DPS_2017}).

For the 1985 apparition of comet 21P/GZ, the lightcurve method was used on the 1.0 m reflector of the Wise Observatory to obtain a rotation period of 9.5 $\pm$ 0.2 hours from 21 July 1985 to 11 August, 1985 \citep{Leibowitz_1986}. In contrast, we utilize a novel method, which measures the radial location of the peak flux of a jet feature as a function of time. This method is particularly effective for corkscrew-like features or morphological features that fluctuate with time. The measured peak flux locations versus time are transformed by a Lomb-Scargle Periodogram to find possible periodicities within the data. This method relies upon the assumption that there is a repeatable brightness fluctuation present. Evidence that this assumption is justified, as well as a more in depth explanation of the method, can be found in \S \ref{periodicity_measurement}.

The velocity at which gas is ejected from the nucleus is essential to predict molecular production rates and torques, among other cometary parameters. \citet{Cochran_1993} showed that the outflow velocity of gases can be estimated by the relationship $v \approx 0.85 r_h^{-0.5}$ km sec$^{-1}$, which for typical distances at which outflow velocities are measured, produces outflow velocities from a few hundred meters per second to greater than a kilometer per second. Such estimates have been proven accurate over the years through different outflow velocity measuring techniques, such as modeling sub-millimeter cometary emission lines (\citet{Cordiner_2022}), or from the OH 18-cm emission line (e.g. \citet{Tseng_2007}). To measure the CN outflow velocity we assume that the CN features are moving radially outward.

The pole vector of cometary nuclei can help interpret the rotational features seen, and help better understand activity patterns on the surface. There are several methods to constrain the pole vector of a comet. One such method is analyzing a polar jet over time (e.g.: \citet{samarasinha_2002}, \citet{Farnham_2002}). We use a similar method, constraining the pole position with regards to the jet features after a deep understanding of the jet features has been acquired (see \S \ref{Pole_vector_section}).

Sublimating gases, whether in hemispherical outgassing or jet-like features, create torques on nuclei that slowly change comets' rotational periods over time. \citet{x_param} introduced a parameter, $X$, described in \S \ref{x_parameter}, that relates the change in a comet's rotational period per orbit to the rotational period itself, its nucleus size, and its water production per orbit per unit area of the nucleus. \citet{x_param} claim that such a parameter should be fairly constant for all comets within a factor of a few unity. \citet{Jewitt_2021} empirically shows that the $X$ parameter's approximate constancy is due to the approximately opposing trends in the dependence on the active fraction of features and the moment arms of torques on the size of the nucleus. We use the rotational period we have found, in correlation with the previously found period from \citet{Leibowitz_1986} to determine the $X$ parameter of 21P/GZ.

\section{Data}\label{data}
We observed 21P/GZ over four nights spanning from 28 August to 12 September 2018. Comet 21P/GZ's closest approach to Earth occurred on 10 September 2018 at 14:16 UT at a geocentric distance of 0.39 au, and its perihelion occurred on the same day at 6:43 UT at a heliocentric distance of 1.01 au \citep{Horizon}. Table \ref{Dates_observed} includes the date and time of each image frame utilized, the peak brightness distances measured both for the Northern and Southern Jets, the heliocentric ($r_h$) and geocentric ($\Delta$) distances, the solar position angles \textemdash the skyplane-projected position angle of the sun with respect to the comet (position angle is measured from North through East), the solar phase angles, and the rotational phases based upon our two derived periods. Our images were obtained using the 1.54 m Kuiper Telescope, located on Mount Bigelow, just north of Tucson, Arizona. The Telescope has a focal length of 9.6 meters, and the imaging system has a pixel scale of 0.42 arcsec pixel$^{-1}$ (in our 3 $\times$ 3 binning setup) with a field of view of 9.7 $\times$ 9.7 arcmin$^2$ \citep{Kuiper}.

\begin{table}

\caption{Measurements of peak projected radial brightness for the Northern and Southern Jets for each image. Certain measurements are unavailable due to the insufficient signal-to-noise ratio (SNR). The rotational cycles are based off of periods of 7.39 $\pm$ 0.01 and 10.66 $\pm$ 0.01 hours and the integer portion refers to the number of rotations since the reference time and the decimal portion refers to the rotational phase. The reference time for both periods is 28 August 2018 at 05:12 UTC and chosen such that the minimum measured jet distance corresponds to a phase of zero for the 10.66 hour period. $\Delta$ is the geocentric distance and $r_h$ is the heliocentric distance for each image. \label{Dates_observed}}
\hspace{-2.3cm}
\begin{tabular}{ccccccccc}

\textbf{} &
\textbf{Northern Jet} &
\textbf{Southern Jet} &
\textbf{} &
\textbf{} &
\textbf{Solar} &
\textbf{Solar} &
\textbf{7.39 Hour} &
\textbf{10.66 Hour}\\

\textbf{} &
\textbf{Distance} &
\textbf{Distance} &
\textbf{} &
\textbf{} &
\textbf{Position} &
\textbf{Phase} &
\textbf{Rotational} &
\textbf{Rotational}\\

\textbf{UTC Date and Time} &
\textbf{($\times 10^3$km)} &
\textbf{($\times 10^3$km)} &
\textbf{$\Delta$ (au)} &
\textbf{$r_h$ (au)} &
\textbf{Angle} &
\textbf{Angle} &
\textbf{Cycle} &
\textbf{Cycle}

\\ \hline
28 Aug. 2018, 08:58 & 14.6 $\pm$ 0.4 & 13.0 $\pm$ 0.3   & 0.42                        & 1.03                        & 86.22 & 75.52 & 0.51 & 0.35 \\
28 Aug. 2018, 09:03 & 16.1 $\pm$ 0.3 & 8.6 $\pm$ 0.3  & 0.42                        & 1.03                        & 86.22 & 75.53 & 0.52 & 0.36 \\
28 Aug. 2018, 09:08 & 15.5 $\pm$ 0.4 & 11.3 $\pm$ 0.3 & 0.42                        & 1.03                        & 86.22 & 75.53 & 0.53 & 0.37 \\
28 Aug. 2018, 09:14 & 15.7 $\pm$ 0.3 & 12.1 $\pm$ 0.3 & 0.42                        & 1.03                        & 86.23 & 75.53 & 0.55 & 0.38 \\
28 Aug. 2018, 09:19 & 16.5 $\pm$ 0.3 & 12.7 $\pm$ 0.3 & 0.42                        & 1.03                        & 86.23 & 75.53 & 0.56 & 0.39 \\
28 Aug. 2018, 10:08 & 18.5 $\pm$ 0.4 & 14.6 $\pm$ 0.3 & 0.42                        & 1.03                        & 86.25 & 75.54 & 0.67 & 0.46 \\
28 Aug. 2018, 10:14 & 18.3 $\pm$ 0.2 & 14.0 $\pm$ 0.4   & 0.42                        & 1.03                        & 86.26 & 75.54 & 0.68 & 0.47 \\
28 Aug. 2018, 10:19 & 19.2 $\pm$ 0.3 & 12.1 $\pm$ 0.3 & 0.42                        & 1.03                        & 86.26 & 75.55 & 0.69 & 0.48 \\
28 Aug. 2018, 10:24 & 19.8 $\pm$ 0.3 & 14.7 $\pm$ 0.3 & 0.42                        & 1.03                        & 86.27 & 75.55 & 0.71 & 0.49 \\
28 Aug. 2018, 10:30 & 19.4 $\pm$ 0.3 & 13.3 $\pm$ 0.3 & 0.42                        & 1.03                        & 86.27 & 75.55 & 0.72 & 0.50 \\
28 Aug. 2018, 10:54 & 20.2 $\pm$ 0.3 & 13.8 $\pm$ 0.3 & 0.42                        & 1.03                        & 86.28 & 75.56 & 0.77 & 0.54 \\
28 Aug. 2018, 10:59 & 20.4 $\pm$ 0.3 & 16.7 $\pm$ 0.3 & 0.42                        & 1.03                        & 86.28 & 75.56 & 0.78 & 0.54 \\
28 Aug. 2018, 11:05 & 21.2 $\pm$ 0.3 & 14.8 $\pm$ 0.3 & 0.42                        & 1.03                        & 86.29 & 75.56 & 0.80 & 0.55 \\
28 Aug. 2018, 11:10 & 20.5 $\pm$ 0.2 & 17.6 $\pm$ 0.4 & 0.42                        & 1.03                        & 86.29 & 75.56 & 0.81 & 0.56 \\
28 Aug. 2018, 11:15 & 21.2 $\pm$ 0.3 & 15.3 $\pm$ 0.3 & 0.42                        & 1.03                        & 86.29 & 75.56 & 0.82 & 0.57\\
4 Sep. 2018, 09:48  & 13.1 $\pm$ 0.4 & 10.5 $\pm$ 0.2 & 0.40                        & 1.02                        & 89.79 & 77.56 & 23.36 & 16.19\\
4 Sep. 2018, 09:53  & 13.0 $\pm$ 0.2   & 10.5 $\pm$ 0.1 & 0.40                        & 1.02                        & 89.79 & 77.56 & 23.38 & 16.20\\
4 Sep. 2018, 09:58  & 12.7 $\pm$ 0.2 & 10.9 $\pm$ 0.2 & 0.40                        & 1.02                        & 89.79 & 77.56 & 23.39 & 16.21 \\
4 Sep. 2018, 10:03  & 13.3 $\pm$ 0.2 & 11.8 $\pm$ 0.2 & 0.40                        & 1.02                        & 89.79 & 77.56 & 23.40 & 16.22\\
4 Sep. 2018, 10:09  & 16.2 $\pm$ 0.3 & 11.0 $\pm$ 0.2   & 0.40                        & 1.02                        & 89.79 & 77.56 & 23.41 & 16.23\\
4 Sep. 2018, 10:58  & 15.4 $\pm$ 0.4 & 11.7 $\pm$ 0.2 & 0.40                        & 1.02                        & 89.80 & 77.57 & 23.52 & 16.30\\
4 Sep. 2018, 11:03  & 16.4 $\pm$ 0.3 & 12.0 $\pm$ 0.2   & 0.40                        & 1.02                        & 89.80 & 77.57 & 23.53 & 16.31\\
4 Sep. 2018, 11:08  & 16.0 $\pm$ 0.2   & 13.1 $\pm$ 0.2 & 0.40                        & 1.02                        & 89.80 & 77.57 & 23.55 & 16.32\\
4 Sep. 2018, 11:13  & 16.4 $\pm$ 0.2 & 12.9 $\pm$ 0.2 & 0.40                        & 1.02                        & 89.81 & 77.57 & 23.56 & 16.33\\
4 Sep. 2018, 11:19  & 16.5 $\pm$ 0.2 & 13.8 $\pm$ 0.3 & 0.40                        & 1.02                        & 89.81 & 77.57 & 23.57 & 16.33\\
4 Sep. 2018, 11:41  & 17.1 $\pm$ 0.2 & 15.5 $\pm$ 0.2 & 0.40                        & 1.02                        & 89.81 & 77.57 & 23.62 & 16.37 \\
4 Sep. 2018, 11:46  & 17.1 $\pm$ 0.2 & 14.7 $\pm$ 0.2 & 0.40                        & 1.02                        & 89.81 & 77.57 & 23.63 & 16.38\\
4 Sep. 2018, 11:51  & 17.9 $\pm$ 0.3 & 14.6 $\pm$ 0.2 & 0.40                        & 1.02                        & 89.81 & 77.57 & 23.64 & 16.39\\
7 Sep. 2018, 10:12  & -              & 5.4 $\pm$ 0.2  & 0.39                        & 1.01                        & 90.50     & 77.95  & 33.16 & 22.98\\
7 Sep. 2018, 10:18  & -              & 5.9 $\pm$ 0.1  & 0.39                        & 1.01                        & 90.50     & 77.95  & 33.17 & 22.99\\
7 Sep. 2018, 10:22  & 9.0 $\pm$ 0.2    & 7.2 $\pm$ 0.1  & 0.39                        & 1.01                        & 90.50 & 77.95 & 33.19 & 23.00\\
7 Sep. 2018, 10:28  & 10.1 $\pm$ 0.3 & 6.4 $\pm$ 0.2  & 0.39                        & 1.01                        & 90.50 & 77.95 & 33.20 & 23.01\\
7 Sep. 2018, 10:33  & 9.3 $\pm$ 0.3  & 6.7 $\pm$ 0.2  & 0.39                        & 1.01                        & 90.50     & 77.95  & 33.21 & 23.02\\
7 Sep. 2018, 11:21  & 10.7 $\pm$ 0.2 & 7.1 $\pm$ 0.1  & 0.39                        & 1.01                        & 90.50 & 77.97 & 33.32 & 23.09\\
7 Sep. 2018, 11:30  & 10.7 $\pm$ 0.2 & 8.0 $\pm$ 0.1    & 0.39                        & 1.01                        & 90.50 & 77.97 & 33.34 & 23.11 \\
7 Sep. 2018, 11:36  & 11.6 $\pm$ 0.3 & 7.8 $\pm$ 0.1  & 0.39                        & 1.01                        & 90.51 & 77.95 & 33.35 & 23.12\\
7 Sep. 2018, 11:41  & 10.6 $\pm$ 0.2 & 8.5 $\pm$ 0.1  & 0.39                        & 1.01                        & 90.51 & 77.95 & 33.36 &23.12\\
7 Sep. 2018, 11:46  & 11.2 $\pm$ 0.2 & 8.3 $\pm$ 0.2  & 0.39                        & 1.01                        & 90.51 & 77.95 & 33.38 & 23.13\\
12 Sep. 2018, 09:55 & 11.4 $\pm$ 0.2 & -              & 0.39                        & 1.01                        & 91.05 & 77.84 & 49.37 & 34.22 \\
12 Sep. 2018, 10:00 & 11.7 $\pm$ 0.4 & -              & 0.39                        & 1.01                        & 91.06 & 77.84 & 49.38 & 34.22\\
12 Sep. 2018, 10:06 & 11.7 $\pm$ 0.2 & -              & 0.39                        & 1.01                        & 91.06     & 77.84     & 49.39 & 34.23 \\
12 Sep. 2018, 10:11 & 11.9 $\pm$ 0.2 & 8.2 $\pm$ 0.1  & 0.39                        & 1.01                        & 91.06 & 77.84 & 49.40 & 34.24\\
12 Sep. 2018, 10:16 & 12.1 $\pm$ 0.5 & 8.6 $\pm$ 0.1  & 0.39                        & 1.01                        & 91.06 & 77.83 & 49.42 & 34.25\\
12 Sep. 2018, 11:26 & 14.1 $\pm$ 0.2 & 10.5 $\pm$ 0.2 & 0.39                        & 1.01                        & 91.06 & 77.83 & 49.57 & 34.36\\
12 Sep. 2018, 11:31 & 15.9 $\pm$ 0.3 & 12.9 $\pm$ 0.2 & 0.39                        & 1.01                        & 91.06 & 77.83 & 49.59 & 34.37\\
12 Sep. 2018, 11:37 & 14.7 $\pm$ 0.3 & 12.2 $\pm$ 0.2 & 0.39                        & 1.01                        & 91.06 & 77.83 & 49.60 &34.37
\end{tabular}
\end{table}

\subsection{Image Enhancement}

\begin{figure}
    \centering
    \includegraphics[width=0.4\textwidth]{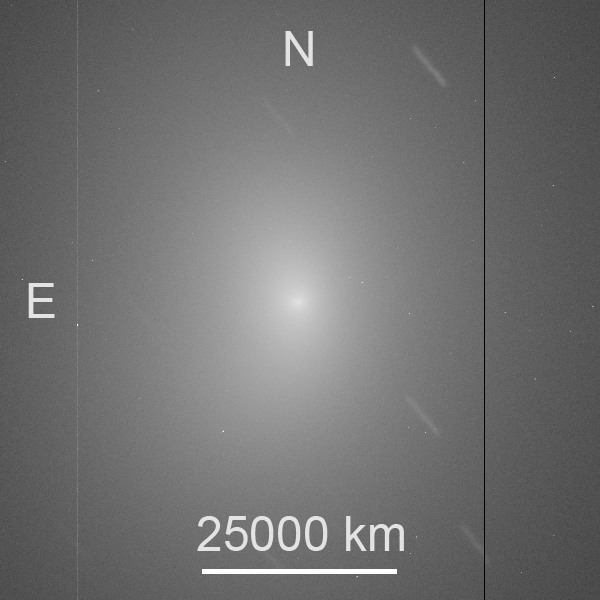} 
    \hspace{0.5cm}
    \includegraphics[width=0.4\textwidth]{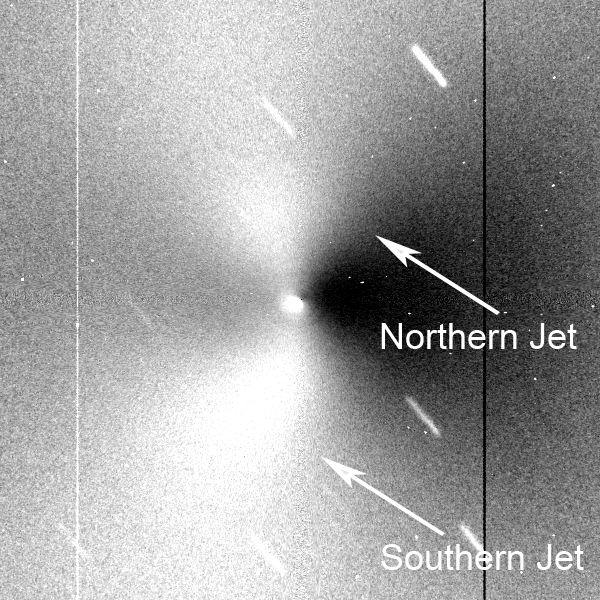}
    \caption{\textbf{Left:} Unenhanced image of 21P/GZ from 28 August 2018 that has undergone the basic data reduction. \textbf{Right:} Same image enhanced by an azimuthal median division, where white denotes regions of higher flux. The white streaks are star trails.}
    \label{enhanced_jets}
\end{figure}

\begin{figure}[ht]
\centering
\begin{minipage}[t]{.45\textwidth}
\centering
\includegraphics[width=\textwidth]{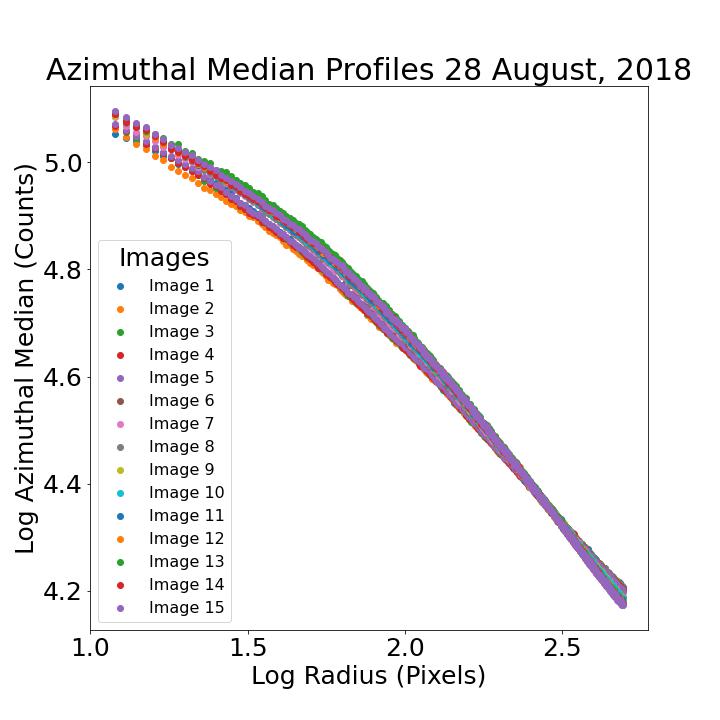} 
\includegraphics[width=\textwidth]{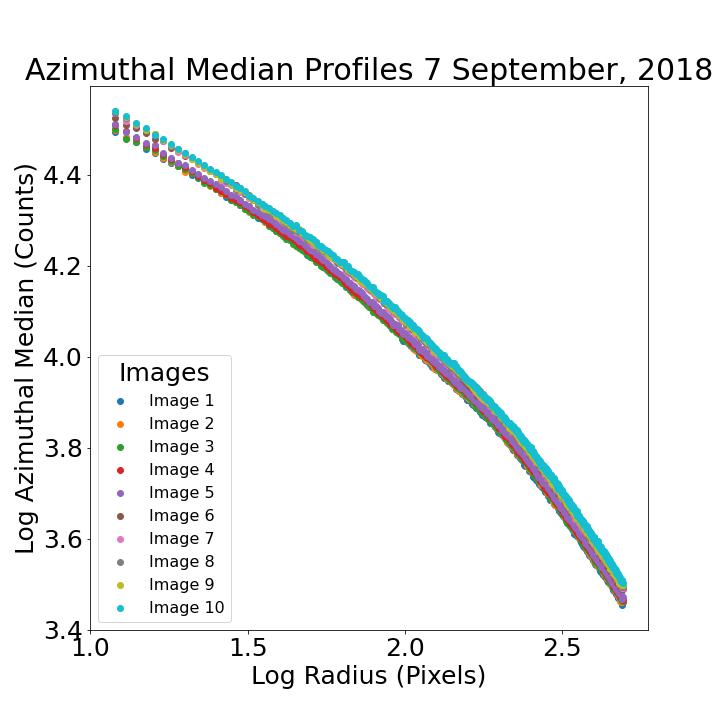}
\end{minipage}
\hspace{0.05\textwidth}
\begin{minipage}[t]{.45\textwidth}
\centering
\includegraphics[width=\textwidth]{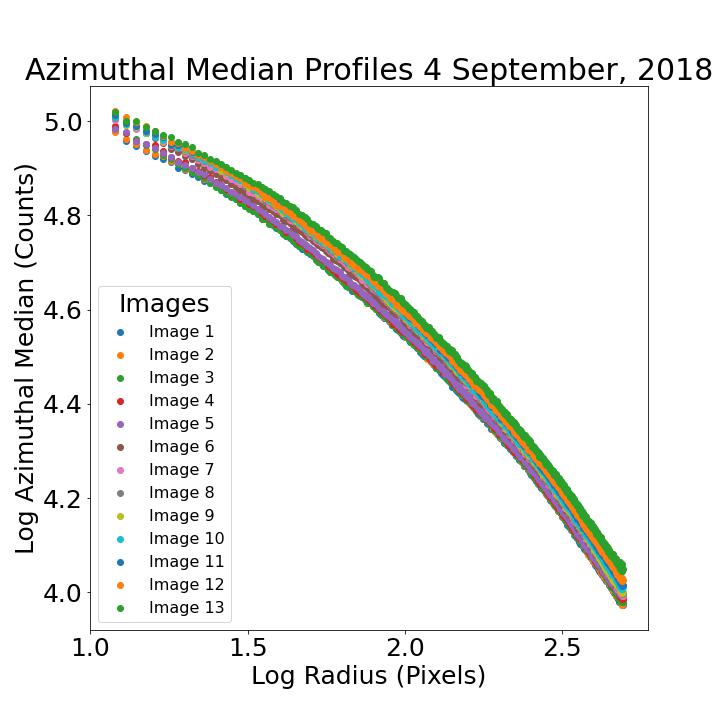} 
\includegraphics[width=\textwidth]{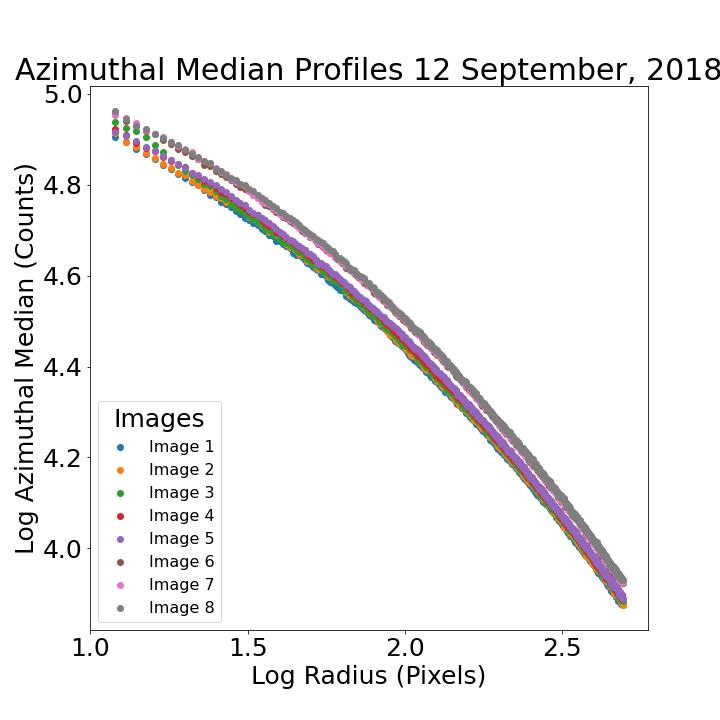}
\end{minipage}
\centering
\caption{Median azimuthal brightness versus the radius in log-log form for each image from each night showing a consistent and smooth relationship between projected radial distance and median azimuthal brightness. The brightness variations are caused by variable sky conditions.\label{azimuthal_median}}
\label{azimuthal_enhanced}

\end{figure}

All images underwent a basic data reduction process which included a subtraction of the bias level and a division of the flat field. Dark noise is negligible as the camera is liquid nitrogen cooled. In our analysis, we utilize enhancement techniques of the coma to highlight coma morphology features and determine their repeatabilities. \citet{Nalin_Larson} explain the benefits of enhancement through several different methods, including:  dividing the images by an azimuthal median, and dividing or subtracting two temporally close images by one another to reveal temporal variations. We use a combination of these techniques to observe the behavior of CN jet features in the gas coma. Specifically, to first highlight the cometary jet features, we use a division by the azimuthal median profile as explained in \citet{psi_enhancement} as seen in Figure \ref{azimuthal_enhanced}. We create these enhanced images using the online tool provided in \citet{Martin} (available at \url{https://www.psi.edu/research/cometimen}). Enhancement by azimuthal median division allows us to distinguish two distinct CN jet features from the overarching coma: a Northern and a Southern jet, as visible in Figure \ref{enhanced_jets}.

To confirm that an azimuthal median does not affect the shape of the features but simply removes the overarching coma, we analyze the azimuthal median removed from each image. As is visible in Figure \ref{azimuthal_median}, there does not appear to be any significant radial structure that could affect the shape of the features, specifically the distance at which the feature peaks in brightness. The azimuthal median division is a proxy for the overlaying coma removal without an actual measurement of it. The nature of Figure \ref{azimuthal_median} shows a good correlation between a radially expanding coma and the azimuthal median measurements. For this reason, and because the coma obscures faint jet features, the measurements described in \S \ref{peak_distance} are taken from azimuthally enhanced images.

\subsection{Jet feature Morphology \label{peak_distance}}
The morphology of jet features, made clearer through enhancement, can be studied to constrain the activity of the coma and the nucleus. We characterize each jet's behavior both radially and azimuthally. 

\begin{figure}[ht]
\centering
\includegraphics[width=0.6\textwidth]{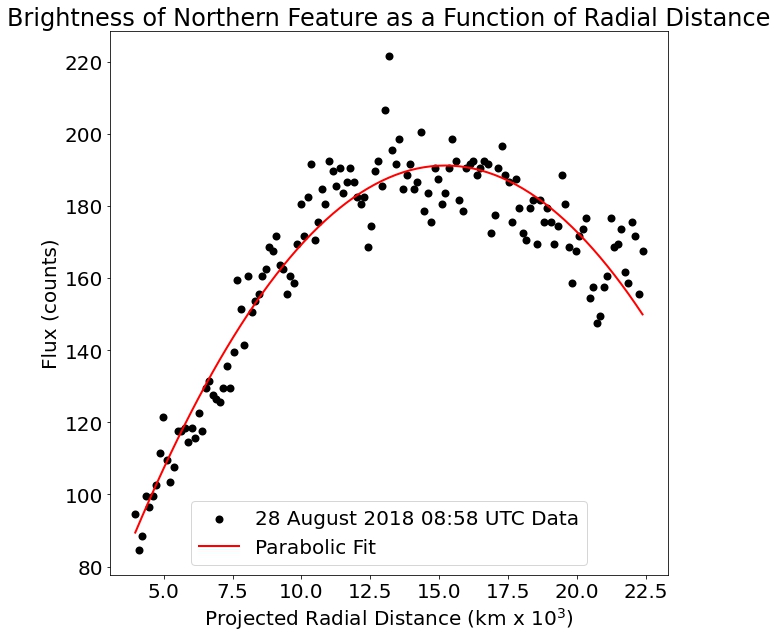}
\caption{Demonstrative fitting algorithm of Northern Jet feature showing a parabolic trend. \label{brightness_VS_distance} }
\end{figure}

To characterize the projected radial behavior of each jet, we measure the projected radial distance at which the flux of each jet is at a maximum. This is achieved using a wedge centered on each jet where the total flux at each radius is measured, then fitting a parabola to the data as seen in Figure \ref{brightness_VS_distance} near the brightness peak to determine the location of the maximum flux. A parabola is chosen to fit the data because it closely matches the shape of the measured brightness profile, however, it is unclear why this is. The one-sigma error bars quoted in Table \ref{Dates_observed} result from a combination of the goodness of the fit and the SNR of the brightness of the feature. Measurements where stars can not be properly removed are omitted.

To characterize the azimuthal behavior of the jet features, we choose a set of projected radial distances from the nucleus and measure the azimuthal locations of the jets at each distance for each image. This allows us to track the azimuthal motion of the jet features with time. From this, we detect patterns and behaviors within each jet feature. The analysis of these measurements are explained in \S \ref{jet_morphology_results}.

\subsection{Obtaining the CN Gas Outflow Velocity \label{ouflow_velocity_method}}

\begin{figure}
    \centering
    \includegraphics[width=.7\textwidth]{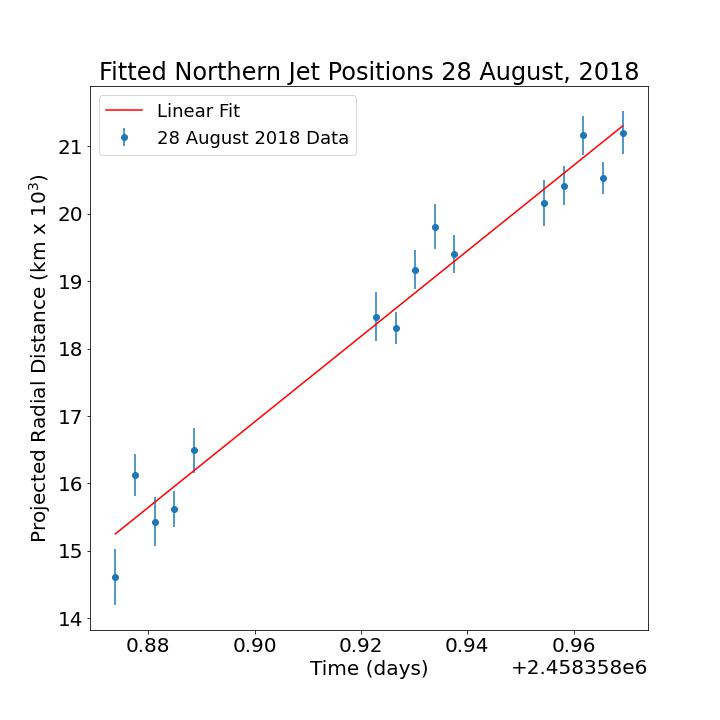}
    \caption{Projected radial distance of the peak brightness versus time (in Julian days) for the Northern Jet taken on 28 August 2018 showing a feature moving radially away from the nucleus. A linear fit is shown over the data with the slope representing the measured CN outflow velocity. }
    \label{fitted_jet1}
\end{figure}

During the course of our investigation, we measured the projected radial distance of the peak brightness of each jet moving away from the nucleus, suggesting that some amount of material was released in that projected radial direction and was moving away from the comet. This behavior allows us to track the material's movement over time, and thus its projected radial outflow velocity. If we assume the outflow velocity of the material to be constant for each jet over the course of a given night, we can fit a straight line to the projected radial distance of the peak brightness versus time, where the slope is the projected outflow velocity of the CN gas for that feature. Figure \ref{fitted_jet1} shows the linear fit for 28 August 2018 for the Northern Jet. The errors of the outflow velocities are determined using a Gaussian distribution of the errors on the measurements and a Markov Chain Monte Carlo approach. In this method, the data points are randomly placed along their error bars according to a Gaussian distribution centered on the measurement point. For each iteration, the outflow velocity is measured and recorded. After 100,000 iterations, the standard deviation of the recorded velocities is taken as the error of the velocity measurement. Because of the projection effects, our measured outflow velocities are all lower limits on what the actual CN outflow velocities may be provided the jet widths are narrow \citep[cf.][]{Samarasinha_2000}.

\subsection{Determination of the Periodicity \label{periodicity_measurement}}

To determine the periodicity, we rely on the assumption that there exist repeatable brightness fluctuations consistent with radial expansion. This could consist, for example, of continuous jet features with rotationally driven variations in outflow, or even line-of-sight enhancements due to the geometry that specific directions of outflow make with the observer. In any case, we observe radially expanding features that are periodic in nature with brightness fluctuations, where the brightest location of the features are shown in Figure \ref{fitted_jet1}.

To determine the periodicity, we perform a Lomb-Scargle Periodogram (which specializes in discrete and uneven intervals of time) with the peak brightness location versus time data for each jet (see \S \ref{peak_distance}) to generate a power density spectrum (PDS) of the Fourier transform of our data for each jet individually. From each PDS, we extract several most probable periodicities \citep{LS_periodogram}. Once we determine the periodicities, we phase our data by them to find the most probable period. The phasing equation as described in \citet{Stellingwerf} is given by:

\begin{equation}
    Phase = \frac{\text{mod }(t, P) }{P},
\end{equation}
where $t=JD-JD_{\circ}$, the Julian date ($JD$) minus some referenced or starting point date ($JD_{\circ}$), and $P$ is the period in days. In this equation, we divide the time - which is an adjusted day - by the period, and take the remainder. This remainder is then divided by the period again to give us a number between zero and one. This is a common method for obtaining the rotational phase in other period search methods (e.g.: \citet{Schleicher_2}, \citet{Eisner}). Each phase is between zero and one, and each cycle increases as our time baseline increases, with $JD_0 =$ 28 August 2018 at 05:12 UTC. As an example, the first image was taken at 28 August 2018 at 08:58 UTC, which was 3.77 hours after $JD_{\circ}$. Taking the modulus of this with respect to a 10.66 hour period yields a remainder of 3.77 hours. Dividing this by the 10.66 hour period, the phase is found to be 0.35. Thus, the nucleus has a phase of 0.35, or similarly, was 35\% through a 10.66 hour rotation at the time the image was taken.

\section{Results}\label{results}
Comet 21P/GZ has two persistent CN jet features: a Northern feature and a Southern feature (see Figure \ref{enhanced_jets}). For both features, we examine the projected radial peak brightness distance as a function of time, which is used to determine the periodicity and the projected CN outflow velocity (see \S \ref{data} for methodology).

\subsection{Projected Radial Peak Brightness distance of CN Features}
\begin{figure}[h]
\centering
\begin{minipage}[t]{.45\textwidth}
\centering
\includegraphics[width=\textwidth]{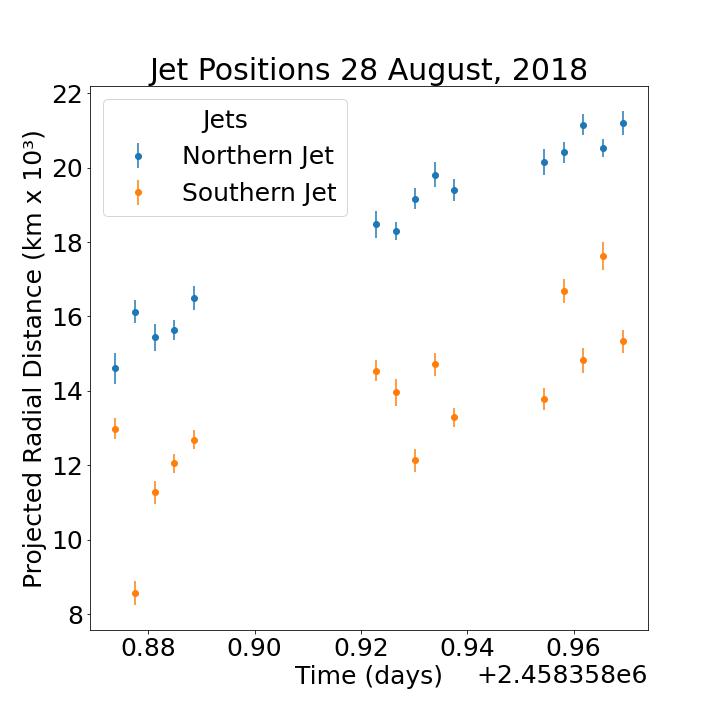} 
\includegraphics[width=\textwidth]{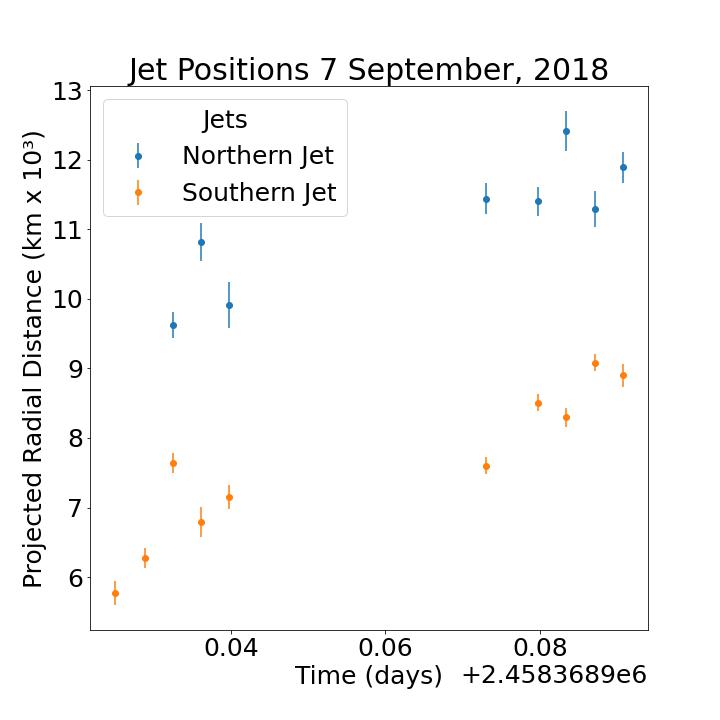}
\end{minipage}
\hspace{0.05\textwidth}
\begin{minipage}[t]{.45\textwidth}
\centering
\includegraphics[width=\textwidth]{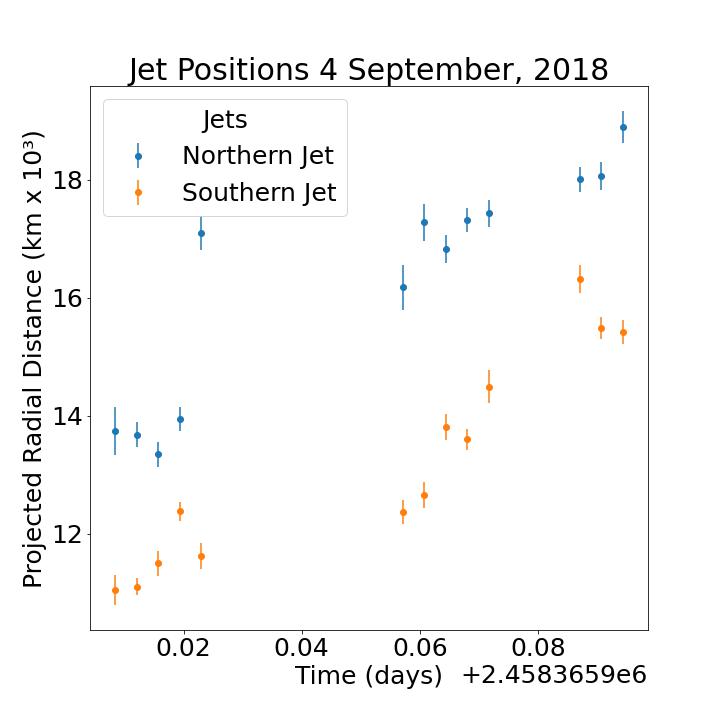} 
\includegraphics[width=\textwidth]{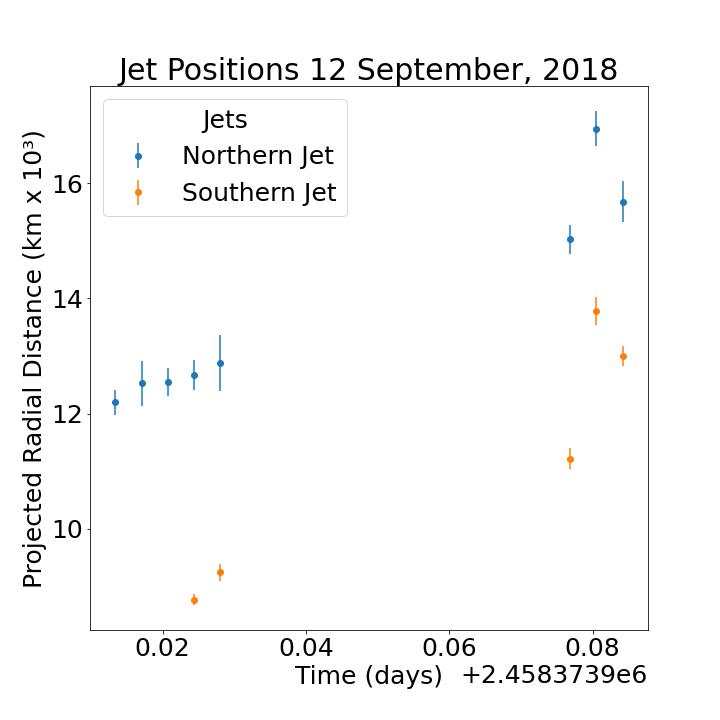}
\end{minipage}
\caption{Projected radial distances of peak brightnesses for the Northern and Southern Jets for the four nights of measurements. The blue circles represent the Northern Jet and the orange circles represent the Southern Jet.
\label{All_distance_v_time_data}}
\end{figure}

Figure \ref{All_distance_v_time_data} shows the projected radial peak brightness distance measured for each CN jet feature for the four days of data that we have obtained, with the blue circles representing the Northern jet and the orange circles representing the Southern Jet. Apart from the scatter in measurements, the brightest regions of both jets appear to be moving radially outward. 

\subsection{Periodicity}
\begin{figure}[ht]
\centering
\begin{minipage}{.49\textwidth}
  \centering
   \includegraphics[width=1.0\textwidth]{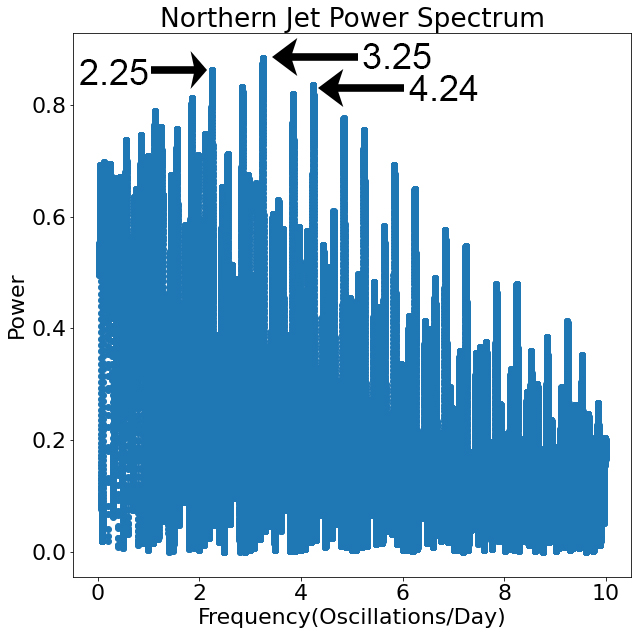}
\end{minipage}%
\begin{minipage}{.49\textwidth}
  \centering
    \includegraphics[width=1.0\textwidth]{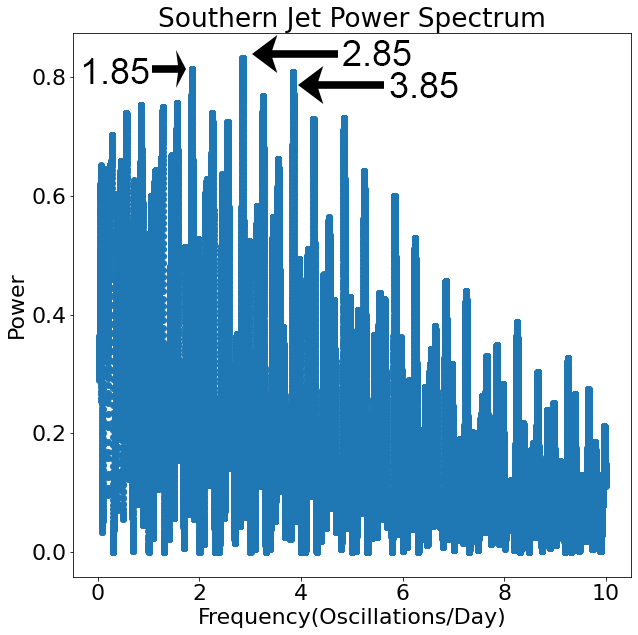}
\end{minipage}
\\
\hspace{.01cm}
\begin{minipage}{.49\textwidth}
  \centering
    \includegraphics[width=1.0\textwidth]{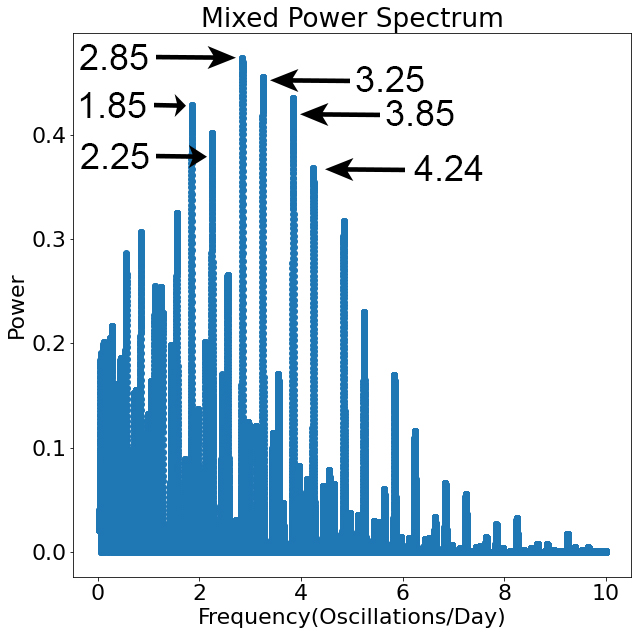}
\end{minipage}
\begin{minipage}{.49\textwidth}
  \centering
    \caption{\textbf{Upper Left:} Power density spectrum for the Northern jet. Prominent peaks located at 2.25, 3.25, and 4.24 oscillations per day are labeled, corresponding to periods of 10.66, 7.39, and 5.7 hours. \textbf{Upper Right:} Power density spectrum for the Southern Jet. Prominent peaks located at 1.85, 2.85, and 3.85 oscillations per day are labeled, corresponding to periods of 13.0, 8.41, and 6.24 hours. Potentially significant peaks for both jets are labeled, although the peaks at 4.24 and 1.85 oscillations per day for the Northern and Southern Jets, respectively, result in less compelling phase diagrams and image divisions than the others. Note that in both power spectra, the most prominent peaks are spaced by 1 oscillation per day. This is a common result of periodograms due to daily aliasing, and performing a window clean did not eliminate the primary frequencies observed due to the limited number of data points. In addition, the most prominent peaks in either power spectrum appear as weaker frequencies in the other. \textbf{Lower Left:} Mixed power spectrum created by multiplying the power spectra of the Northern and Southern Jets. The six most powerful frequencies in the mixed spectrum correspond to the three most powerful frequencies in each of the Northern and Southern Jet power spectra. \label{power_spectrum}}
\end{minipage}
\end{figure}

As explained in \S \ref{periodicity_measurement}, we obtain the periodicity of the Northern and Southern Jets by creating a Lomb-Scargle Periodogram of their peak brightness location data versus time. As shown in Figure \ref{power_spectrum}, the power spectrum for the Northern Jet yields a maximum power at a frequency of 3.25 oscillations per day, corresponding to a period of 7.39 $\pm$ 0.01 hours, with secondary peaks at 10.66 $\pm$ 0.01 hours and 5.7 $\pm$ 0.1 hours corresponding to frequencies of 2.25 and 4.24 oscillations per day, respectively. The power spectrum for the Southern Jet yields a maximum power at a frequency of 2.85 oscillations per day, corresponding to a period of 8.41 $\pm$ 0.01 hours, with secondary peaks at 6.24 $\pm$ 0.01 hours and 13.0 $\pm$ 0.4 hours corresponding to frequencies of 3.85 and 1.85 oscillations per day, respectively. The peaks seen in the power spectrum for the Northern Jet can be seen as tertiary peaks in the power spectrum for the Southern Jet, and vice versa. This indicates that the peak frequencies for each power spectrum are significant and are not artifacts of the periodogram. In addition, mixing the power spectra for the Northern and Southern Jets by multiplying them by each other results in the same six most powerful frequencies, although it is unclear why the most prominent peaks have different strengths for the two jets.

\begin{figure}[h]
\vspace{1cm}
\hspace{-2cm}
\centering
\begin{minipage}[t]{.45\textwidth}
\centering
\includegraphics[width=.73\textwidth]{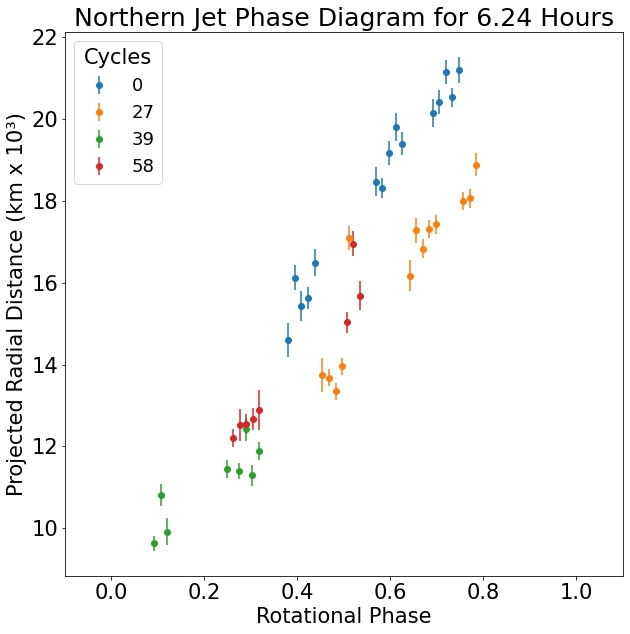}
\includegraphics[width=.73\textwidth]{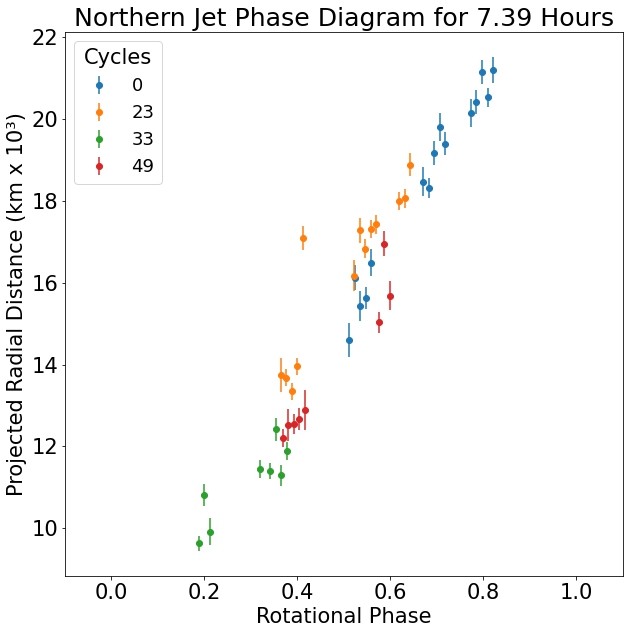}
\includegraphics[width=.73\textwidth]{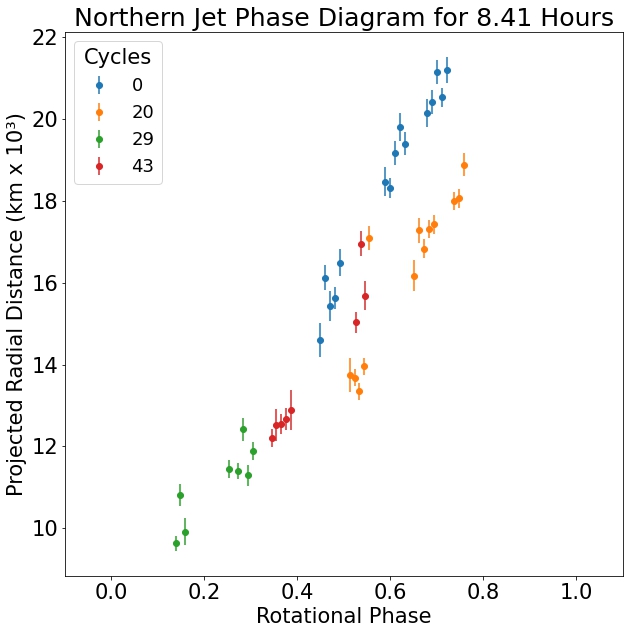}
\includegraphics[width=.73\textwidth]{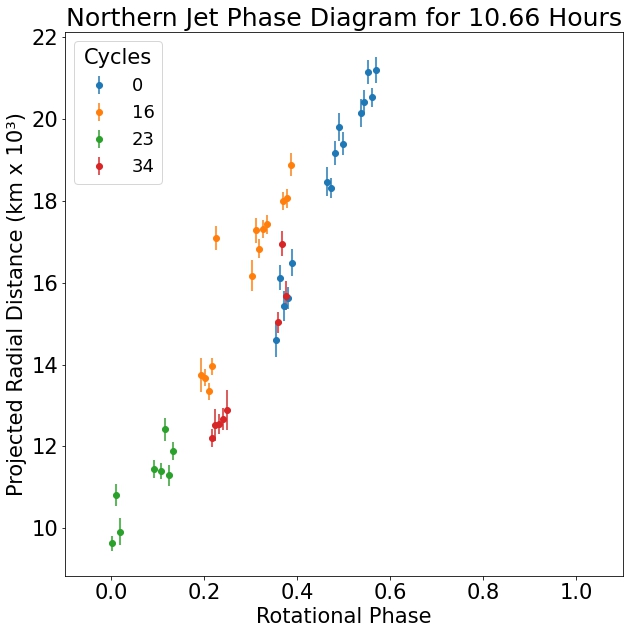}
\end{minipage}
\hspace{-1.5cm}
\begin{minipage}[t]{.45\textwidth}
\centering
\includegraphics[width=.73\textwidth]{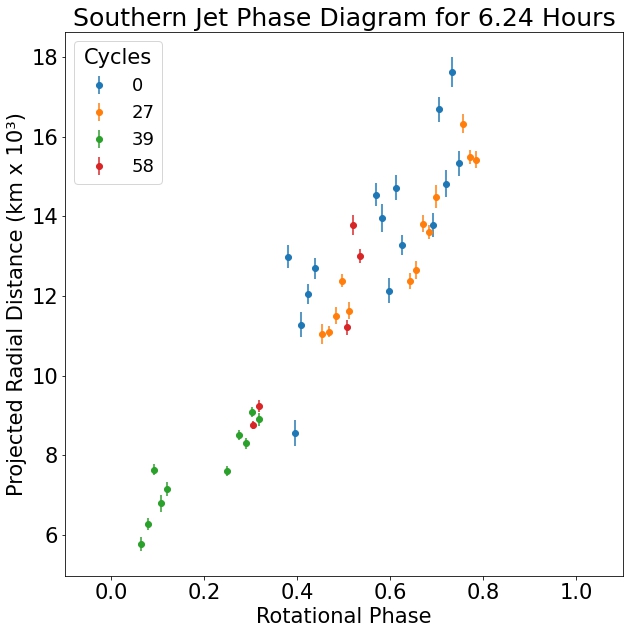} 
\includegraphics[width=.73\textwidth]{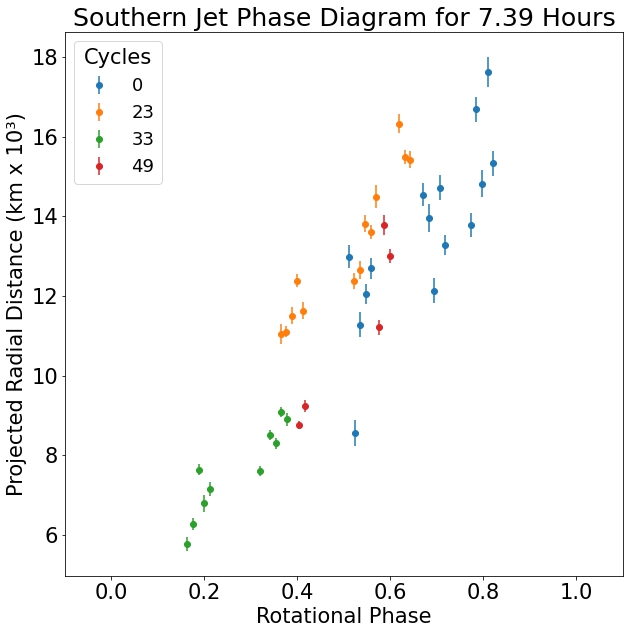}
\includegraphics[width=.73\textwidth]{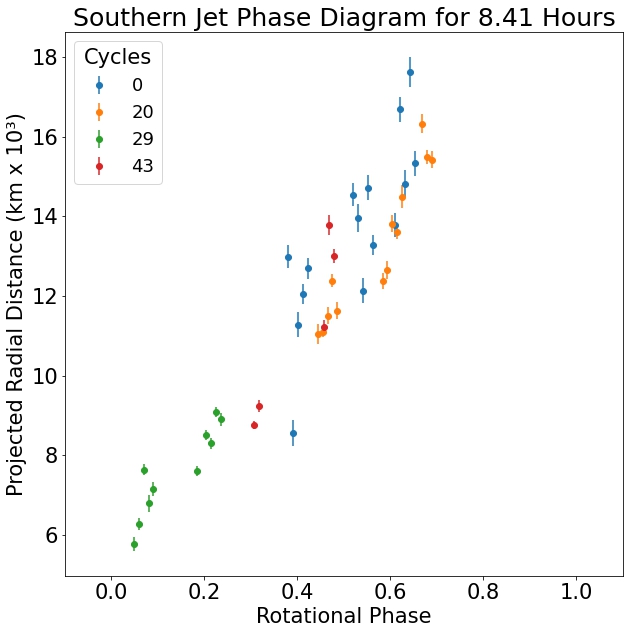}
\includegraphics[width=.73\textwidth]{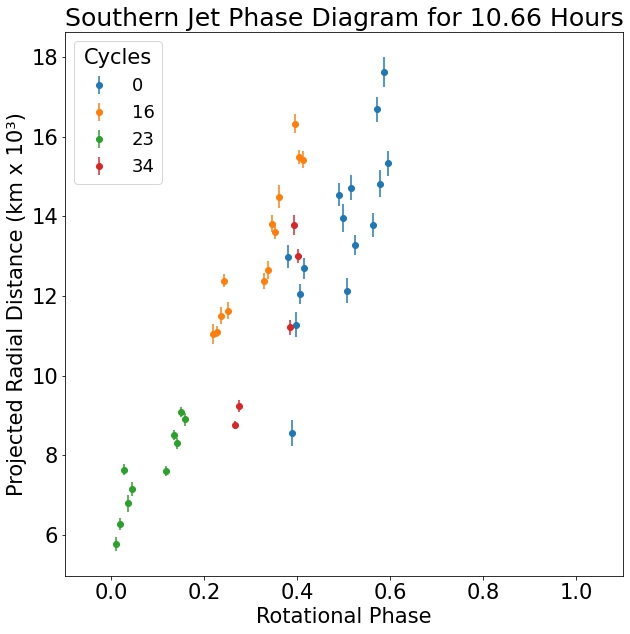}
\vspace{2.5cm}
\end{minipage}
\begin{minipage}[t]{.25\textwidth}
\caption{Phase diagrams of the Northern and Southern Jets plotted for 6.24 $\pm$ 0.01 hours, 7.39 $\pm$ 0.01 hours, 8.41 $\pm$ 0.01 hours, and 10.66 $\pm$ 0.01 hours, which are the more prominent peaks of the power density spectra in Figure \ref{power_spectrum}. From these phase diagrams, the 7.39 hour and 10.66 hour figures have the strongest correlation, as seen by a clear overlap of data from all cycles. \label{phase_diagram}}
\end{minipage}
\end{figure}

These peak frequencies are then used to phase the peak radial brightness of each feature as explained in \S \ref{periodicity_measurement} to begin eliminating weaker periodicities from consideration for 21P/GZ's rotational period. Phase diagrams using the four frequencies resulting in the best phase-location correlations are shown in Figure \ref{phase_diagram}, with the greatest correlations corresponding to periodicities of 7.39 $\pm$ 0.01 and 10.66 $\pm$ 0.01 hours for the Northern Jet. The 5.7 $\pm$ 0.1 and 13.0 $\pm$ 0.4 hour periods had the weakest phase-location correlations are were eliminated from consideration.

\begin{figure}[ht]
\centering
\begin{minipage}[t]{1\textwidth}
\centering
\includegraphics[width=.24\textwidth]{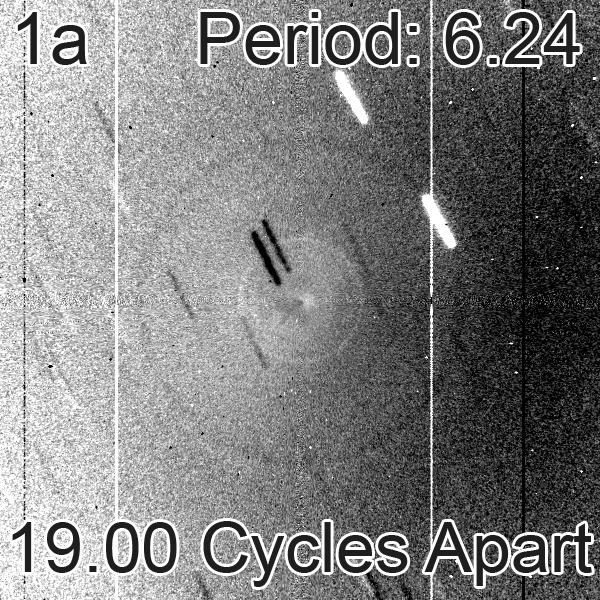}
\includegraphics[width=.24\textwidth]{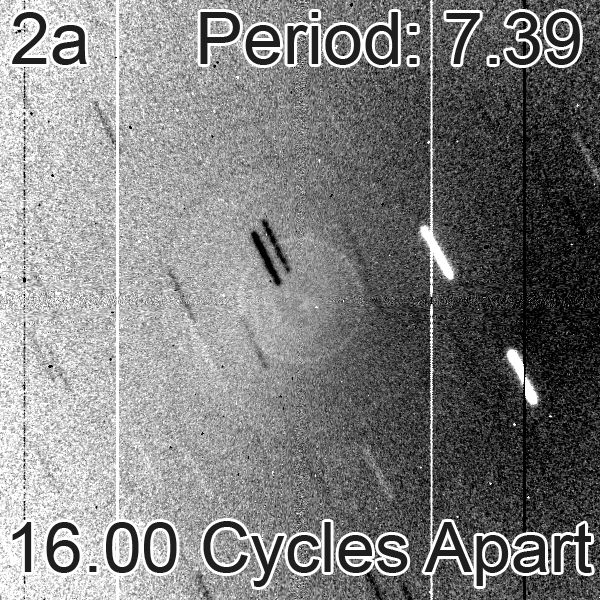}
\includegraphics[width=.24\textwidth]{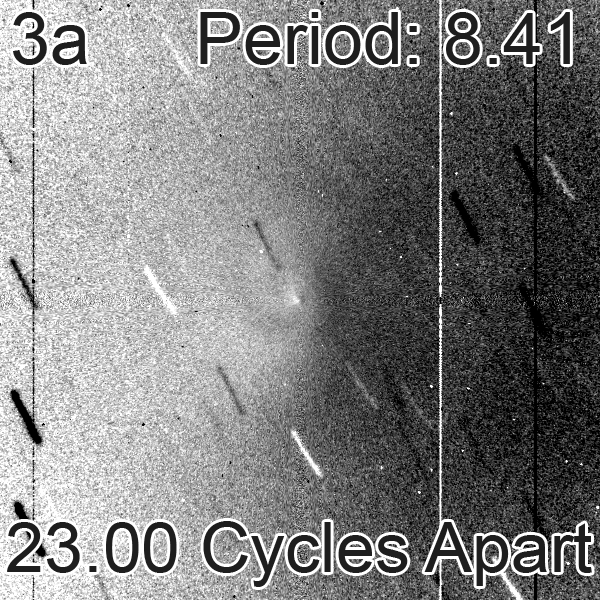}
\includegraphics[width=.24\textwidth]{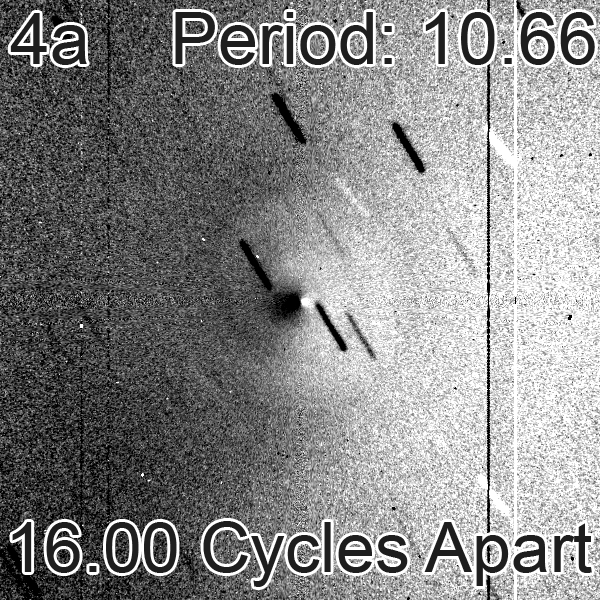}
\end{minipage}
\vspace{-.5cm}
\begin{minipage}[t]{1\textwidth}
\centering
\includegraphics[width=.24\textwidth]{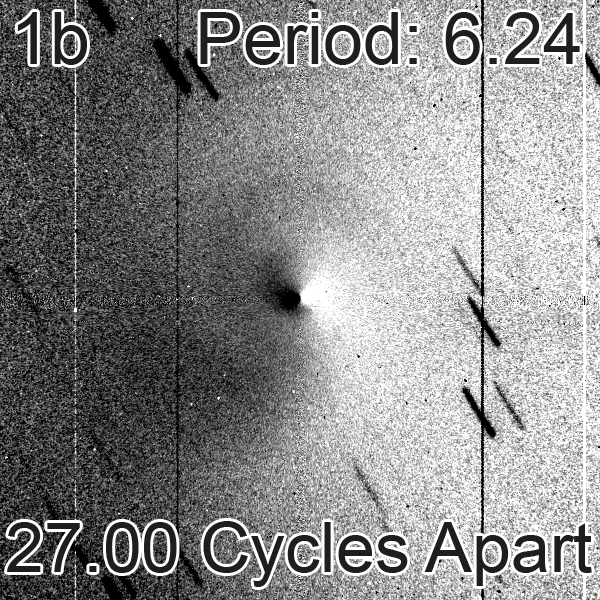}
\includegraphics[width=.24\textwidth]{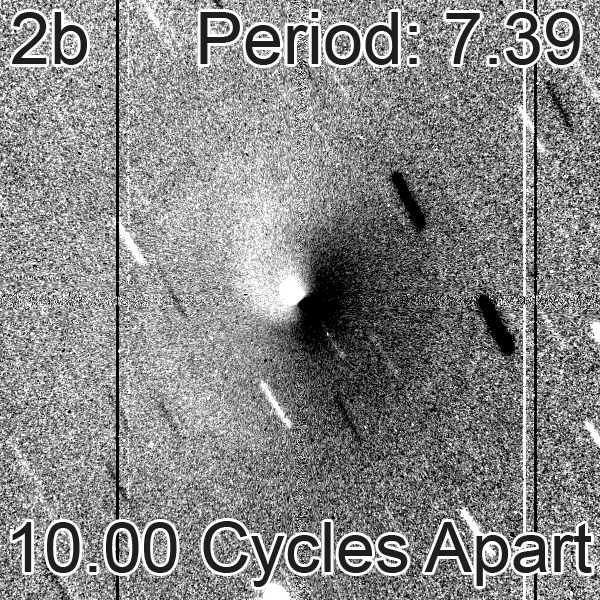}
\includegraphics[width=.24\textwidth]{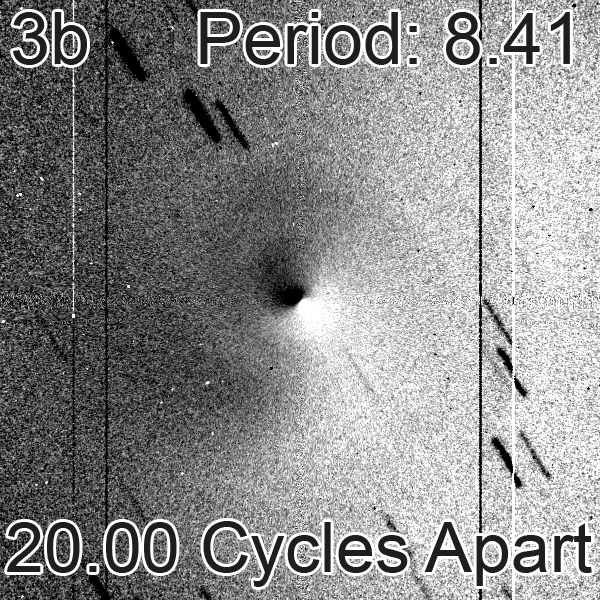}
\includegraphics[width=.24\textwidth]{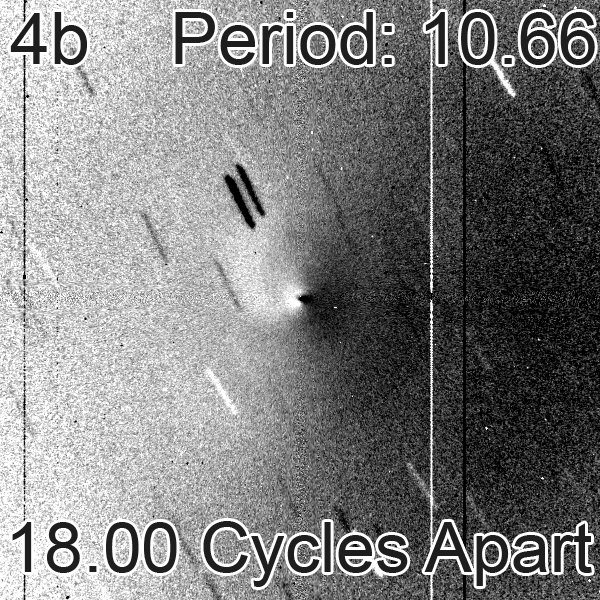}
\end{minipage}

\caption{Same phase divisions for the four periods, with the divisions with the least residual features in the top row, and divisions with the most residual features in the bottom row. Columns a, b, c, and d show divisions for the 6.24 $\pm$ 0.01, 7.39 $\pm$ 0.01, 8.41 $\pm$ 0.01, and 10.66 $\pm$ 0.01 hour periods, respectively.}
\label{same_subtracted}
\end{figure}

\begin{figure}[ht]
\centering
\includegraphics[width=.24\textwidth]{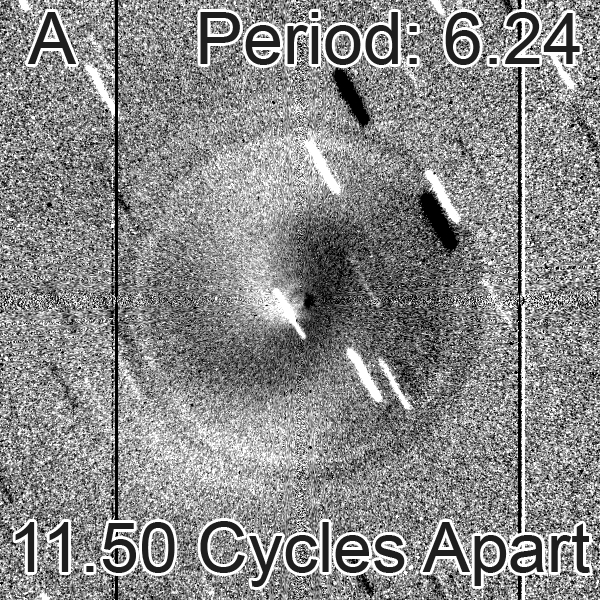}\hfill
\includegraphics[width=.24\textwidth]{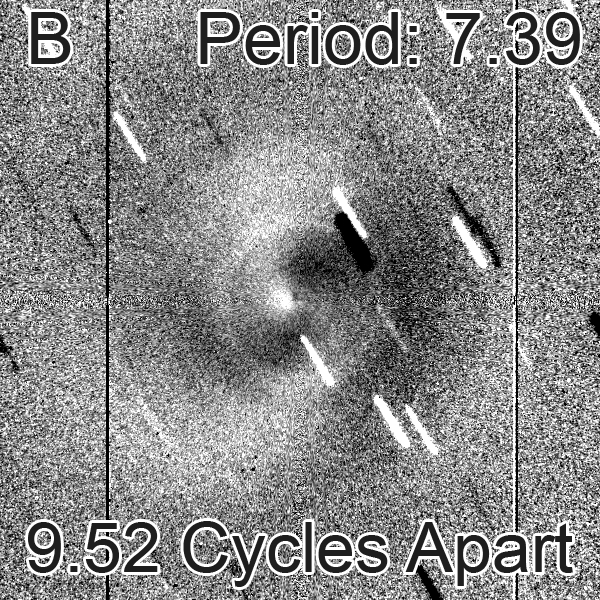}\hfill
\includegraphics[width=.24\textwidth]{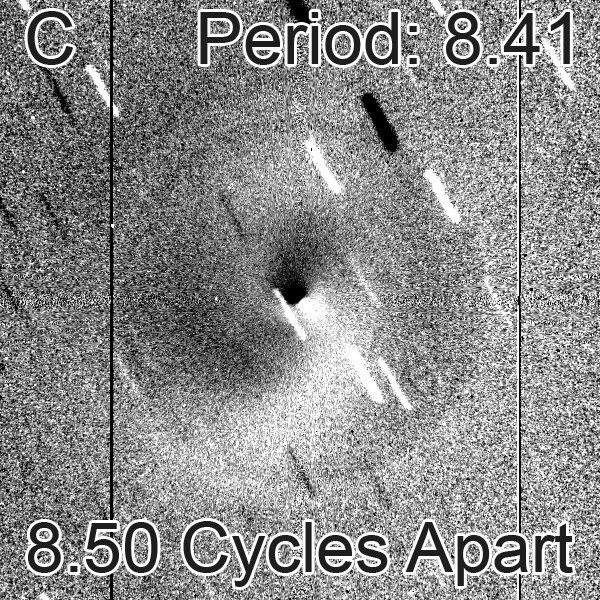}\hfill
\includegraphics[width=.24\textwidth]{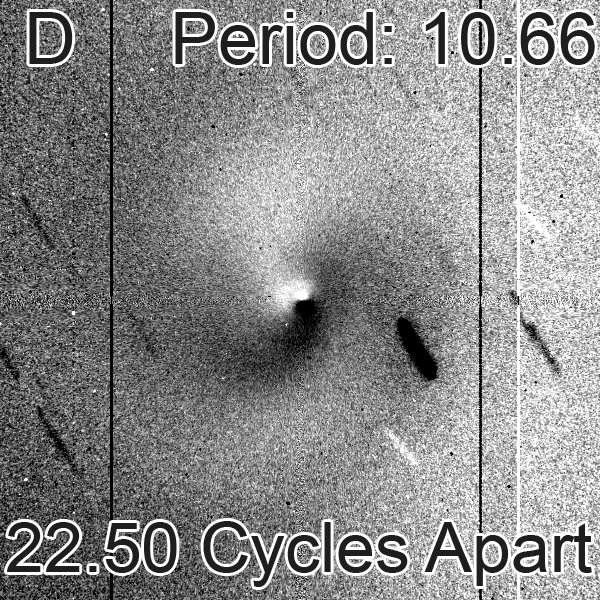}

\caption{Divisions of image pairs offset by half rotations for each period. Columns A, B, C, and D show divisions for the 6.24 $\pm$ 0.01, 7.39 $\pm$ 0.01, 8.41 $\pm$ 0.01, and 10.66 $\pm$ 0.01 hour periods, respectively.}
\label{different_subtracted}
\end{figure}

To continue testing the different periodicities, we divide different sets of images, one set with images of the same phase and one set with images half a phase apart, and compare the results as described in \citet{Nalin_Larson}. Dividing two images of the same phase ---a perfect period match--- with no significant viewing geometry change nor other abnormality, would result in a flat image with no residual. For a half phase division, the results would be a maximum residual. In our study, we test 6.24 $\pm$ 0.01 hours, 7.39 $\pm$ 0.01 hours, 8.41 $\pm$ 0.01 hours and 10.66 $\pm$ 0.01 hours as shown in Figure \ref{same_subtracted} for both the full phase division and half phase division, noting that the viewing geometry changed less than approximately 15$^\circ$ for each division. Figure \ref{same_subtracted} shows the best and worst same phase divisions for each period. All same phase divisions yielded few residual features, indicating each period is plausible. Furthermore, Figure \ref{different_subtracted} shows the half phase divisions for the four periods, all showing significant residual features. Because the same phase and half phase divisions for all four periods show the expected residual features, none can be ruled out by Figures \ref{same_subtracted} and \ref{different_subtracted}. As a result, the 7.39 $\pm$ 0.01 and 10.66 $\pm$ 0.01 hour periods are chosen as the most likely periods for Comet 21P/GZ because they result in the strongest phase plots for both jets in Figure \ref{phase_diagram}, although the 6.24 $\pm$ 0.01 and 8.41 $\pm$ 0.01 hour periods remain plausible.

Full phase divisions also allow for the reduction of the errors for the four periods tested. This is done using a reference image (called image A) to divide by another image of the same phase (called image B). Images of increasingly different phase (image C) are then divided from the reference image until there is a noticeable increase in residual features. Next, the time elapsed between images B and C is calculated. This time represents the magnitude of the error accrued over the integer number of rotations, N, elapsed between images A and B. To find the error in the period per rotation, the time elapsed between B and C is divided by N. This was rounded to the nearest hundredth for the four periods used, and had a rounded value of 0.01 hours per rotation for each.

\subsection{Projected Outflow Velocity of CN Gas}

\begin{table}[ht]
\centering
\caption{CN projected outflow velocity measurements. \label{outflow_velocities}}
\begin{tabular}{ccccc}
\vtop{\hbox{\strut \textbf{ }}\hbox{\strut \textbf{Date}}\hbox{\strut  }} & \vtop{\hbox{\strut \textbf{Northern Jet}}\hbox{\strut \textbf{Velocity}}\hbox{\strut (m s$^{-1}$)}}  & \vtop{\hbox{\strut \textbf{Southern Jet}}\hbox{\strut \textbf{Velocity}}\hbox{\strut (m s$^{-1}$)}}  \\ \hline
2018-08-28    & 730 $\pm$ 30   & 590 $\pm$ 30 \\
2018-09-04    & 610 $\pm$ 30   & 580 $\pm$ 20 \\
2018-09-07    & 350 $\pm$ 40   & 400 $\pm$ 20 \\
2018-09-12    & 590 $\pm$ 40   & 740 $\pm$ 30 \\
            
\end{tabular}
\end{table}

Figure \ref{All_distance_v_time_data} shows that both the Northern and Southern Jets are moving radially away from the nucleus as projected onto the plane of the sky for each night. As explained in \S \ref{ouflow_velocity_method}, we measure the CN projected outflow velocity, summarized in Table \ref{outflow_velocities}, for both the Northern and Southern Jets. The velocities measured represent the projected outflow velocities onto the plane of the sky and are thus lower limits on the outflow velocities for each CN jet. However, as explained in \citet{Samarasinha_2000}, if the CN jets are broad, these may be representing direct outflow measurements. 

\section{Discussion}\label{discussion}
As stated in \S \ref{results}, we find that comet 21P/GZ has a rotation period of either 7.39 $\pm$ 0.01 or 10.66 $\pm$ 0.01 hours in the August-September 2018 time-frame. Additionally, we find that the maximum projected outflow velocity of the CN gas is 730 $\pm$ 30 m s$^{-1}$ for the Northern Jet, and 740 $\pm$ 30 m s$^{-1}$ for the Southern Jet. We can further analyze the morphology of the CN features to constrain the rotational state of the nucleus and gain valuable information about the nature of the jet features themselves.

\subsection{CN Jet Feature Morphology \label{jet_morphology_results}}
\begin{figure}[h]
\centering
\begin{minipage}[t]{.45\textwidth}
\centering
    \includegraphics[width=0.9\textwidth]{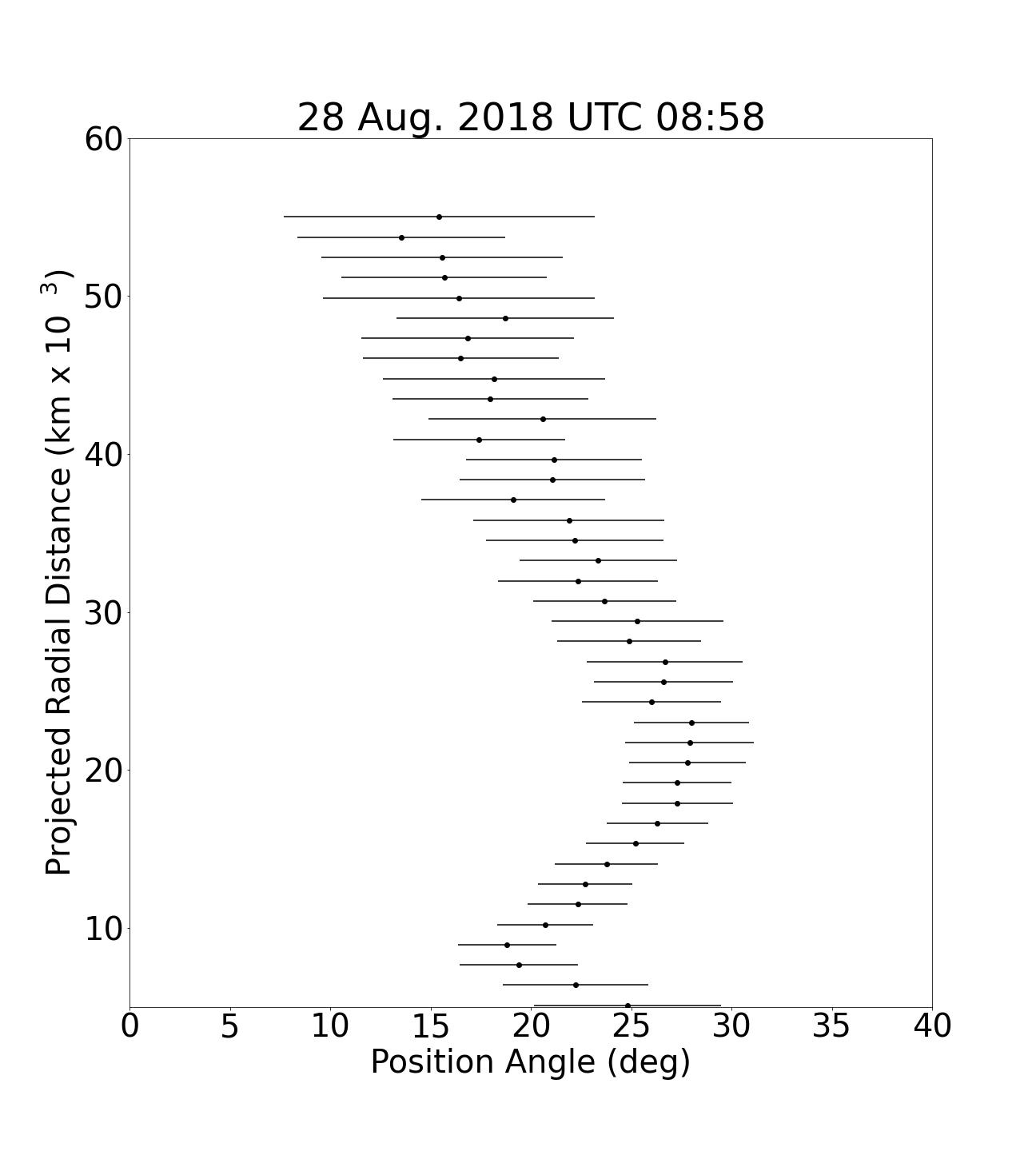}
    \includegraphics[width=0.9\textwidth]{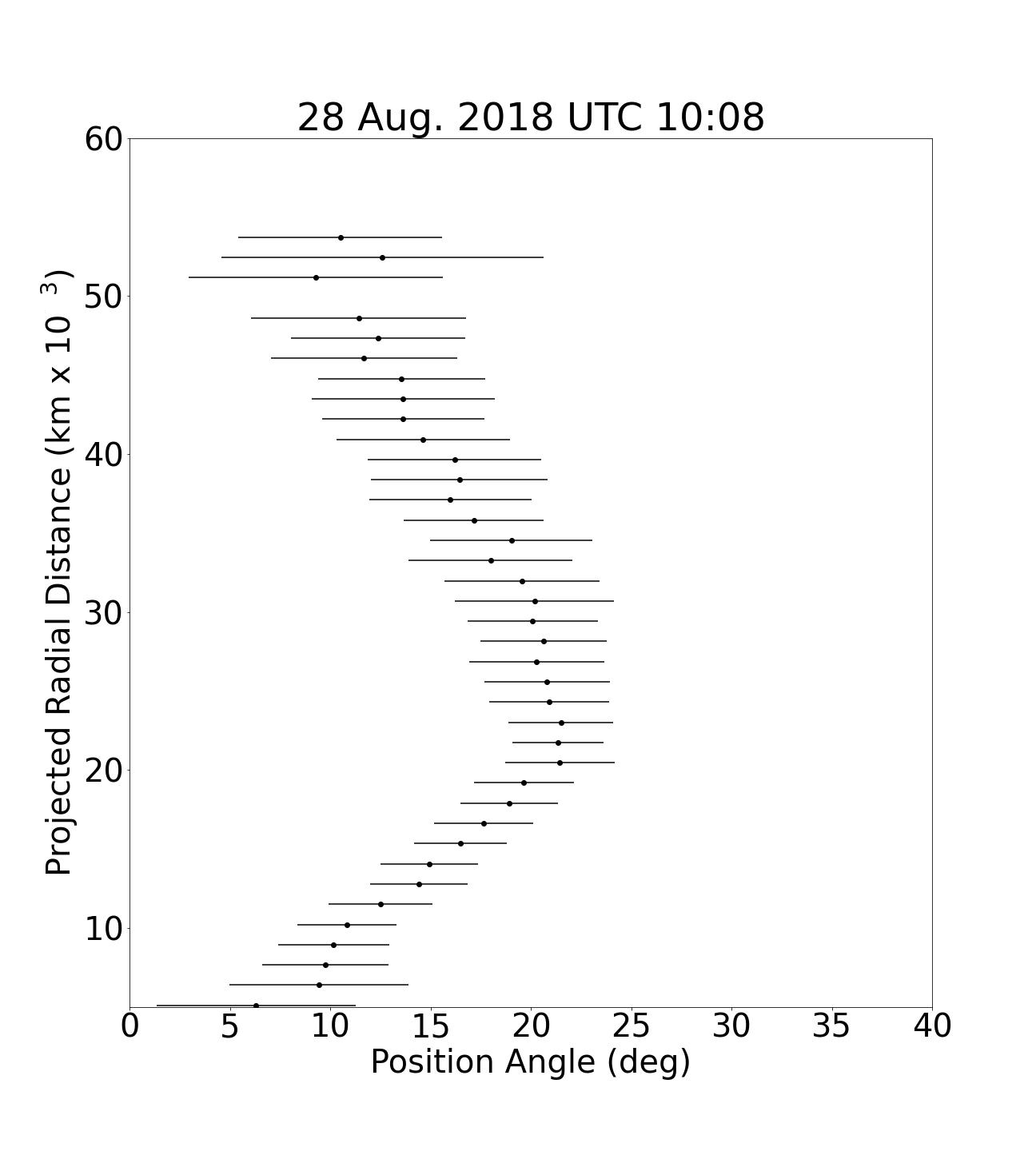}
    \includegraphics[width=0.9\textwidth]{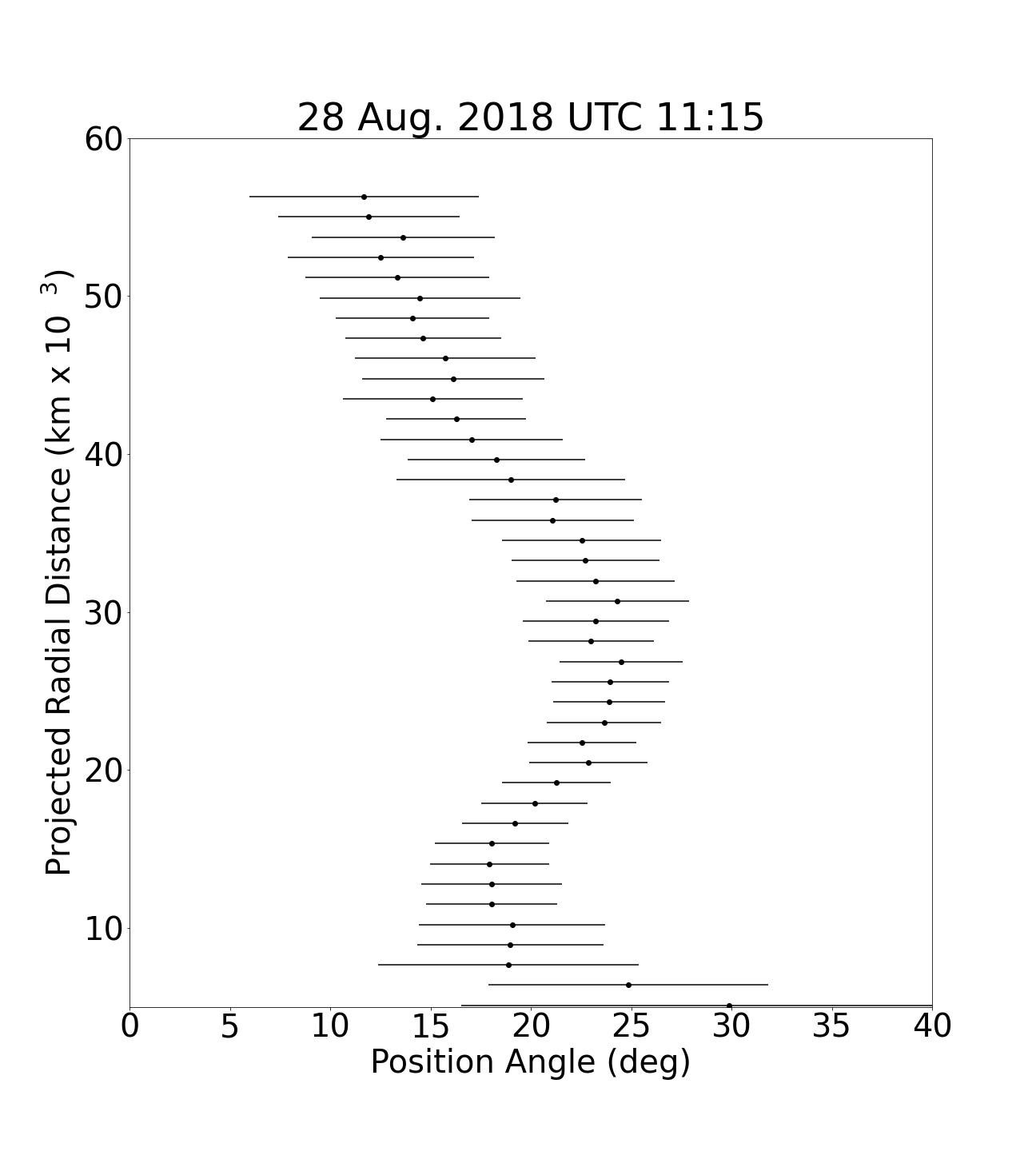}
\end{minipage}
\hspace{0.05\textwidth}
\begin{minipage}[t]{.45\textwidth}
\centering
    \includegraphics[width=0.9\textwidth]{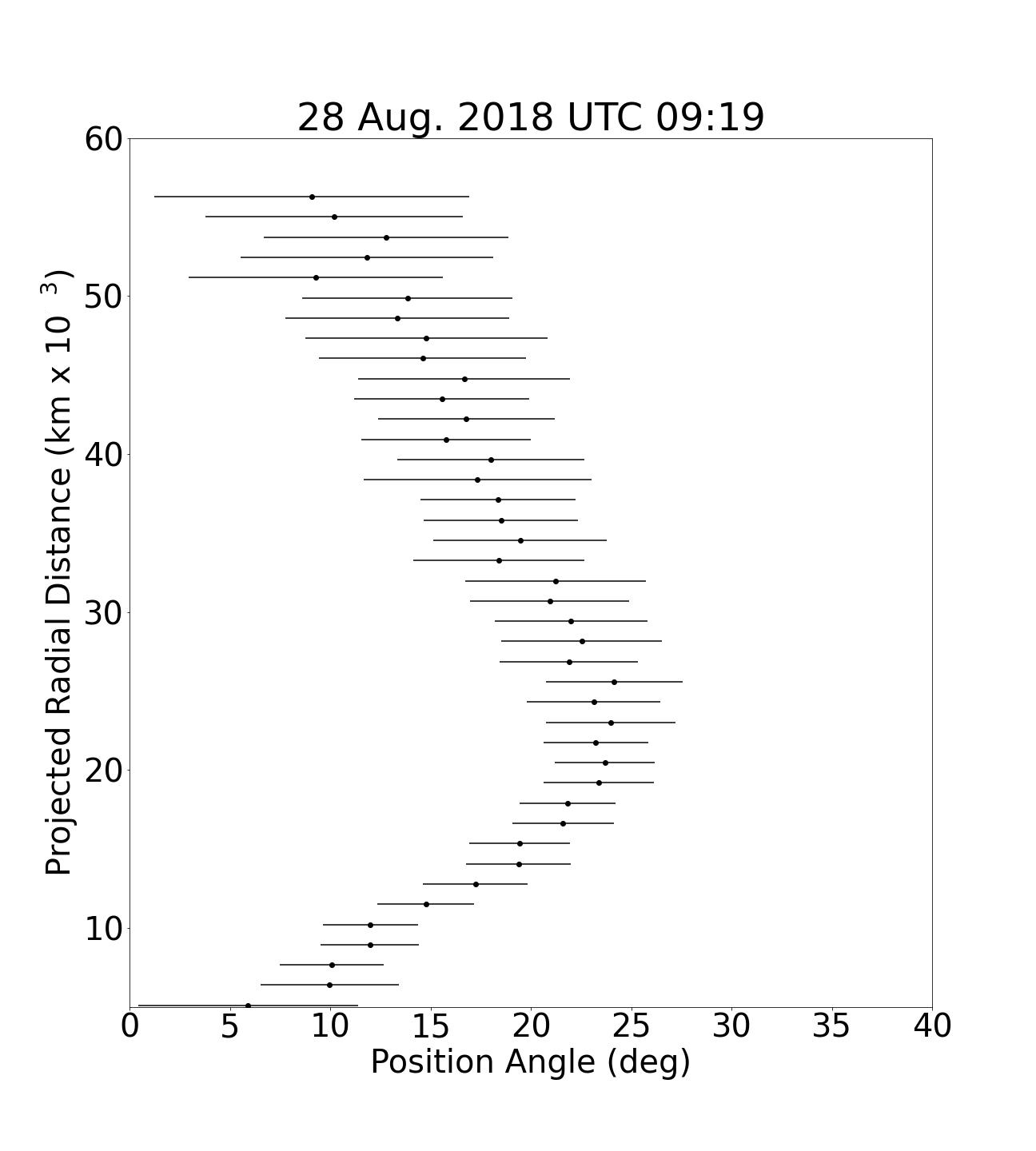}
    \includegraphics[width=0.9\textwidth]{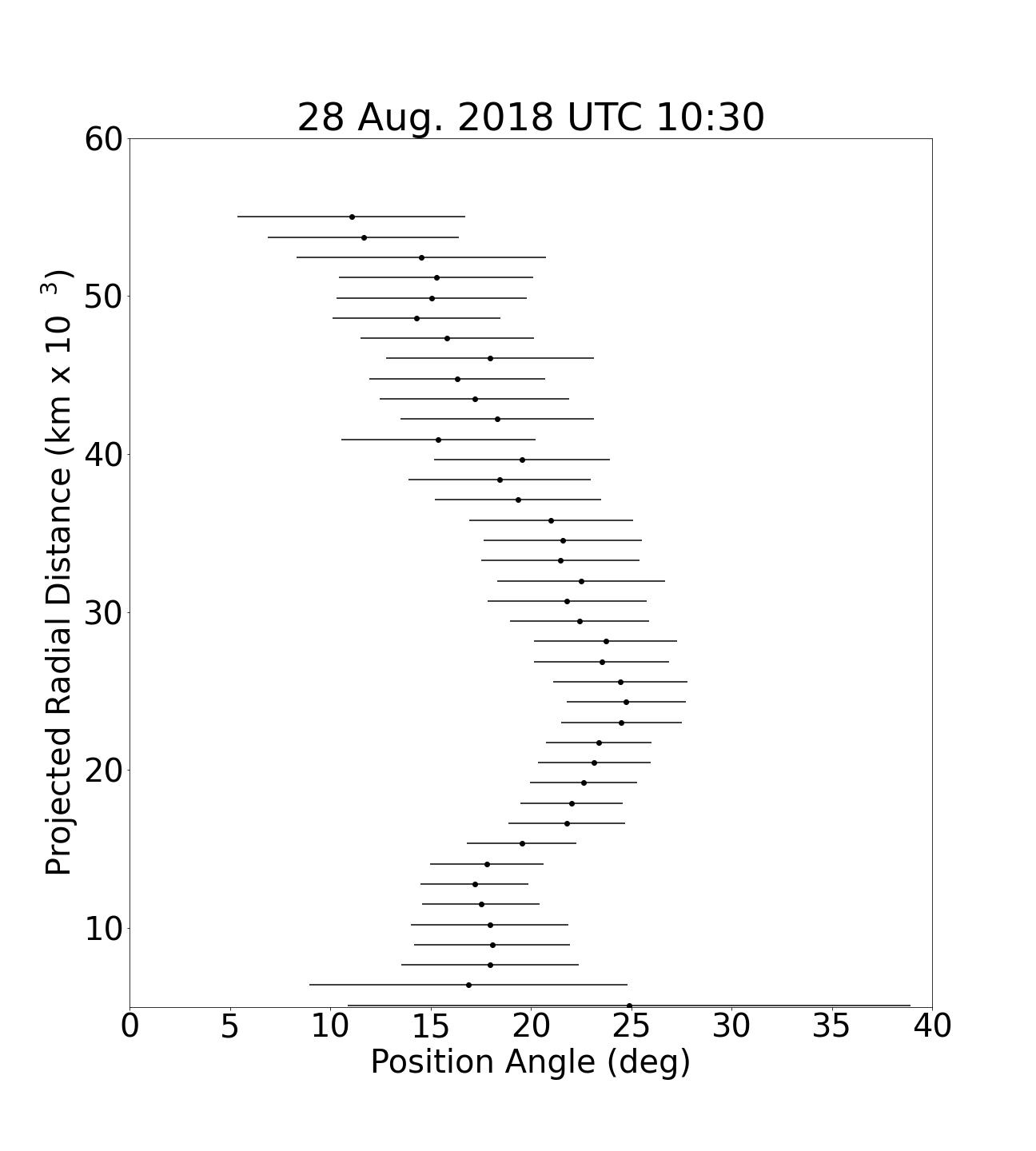}
    \vspace{2.5cm}
    \caption{Northern Jet morphology for 28 August 2018 at five different times spread out across the night, showing outward movement and a clear corkscrew pattern.\label{morphology_night1}}
    \vspace{1cm}
\end{minipage}
\end{figure}

\begin{figure}[h]
\centering
\begin{minipage}[t]{.45\textwidth}
\centering
    \includegraphics[width=0.9\textwidth]{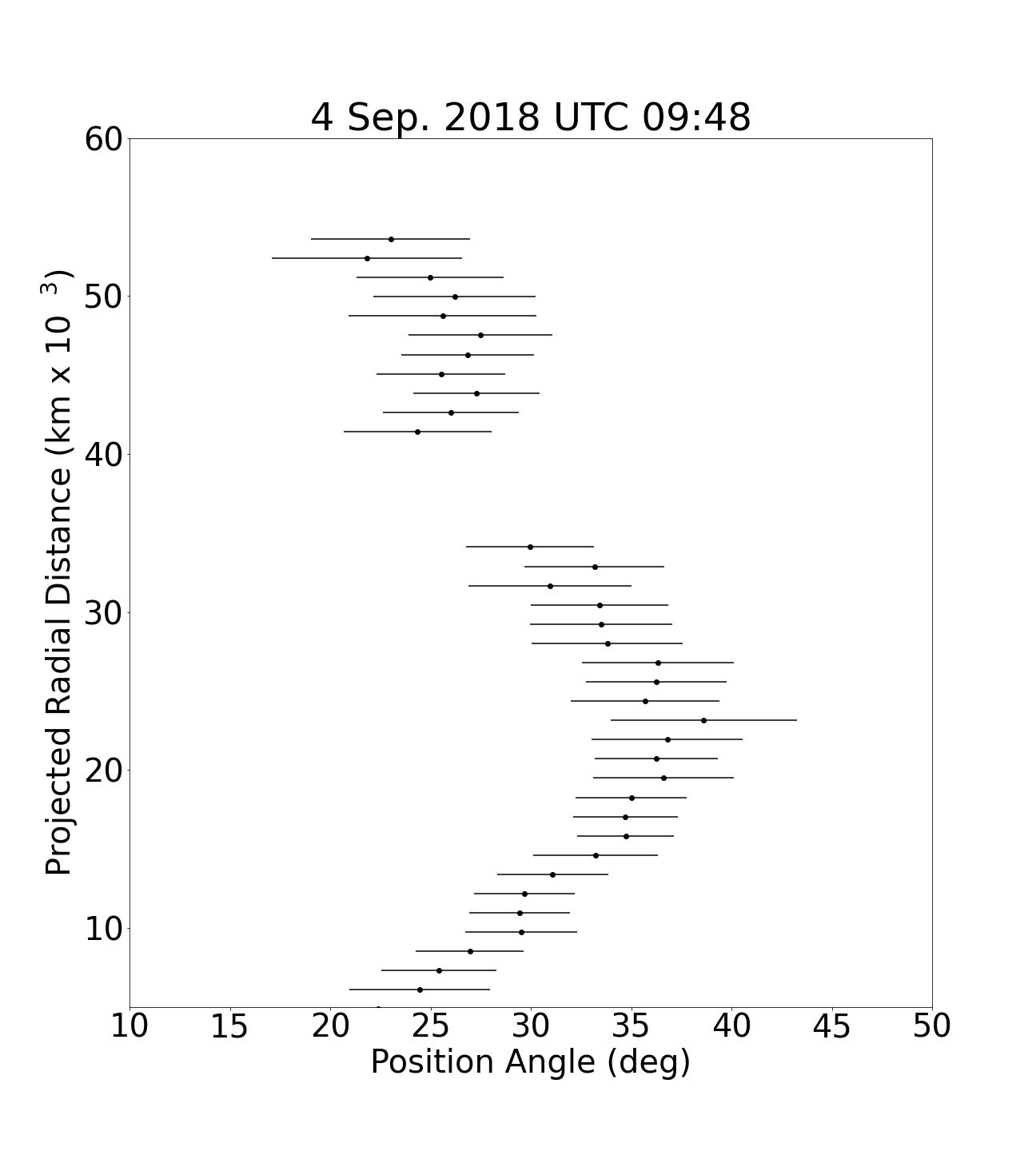}
    \includegraphics[width=0.9\textwidth]{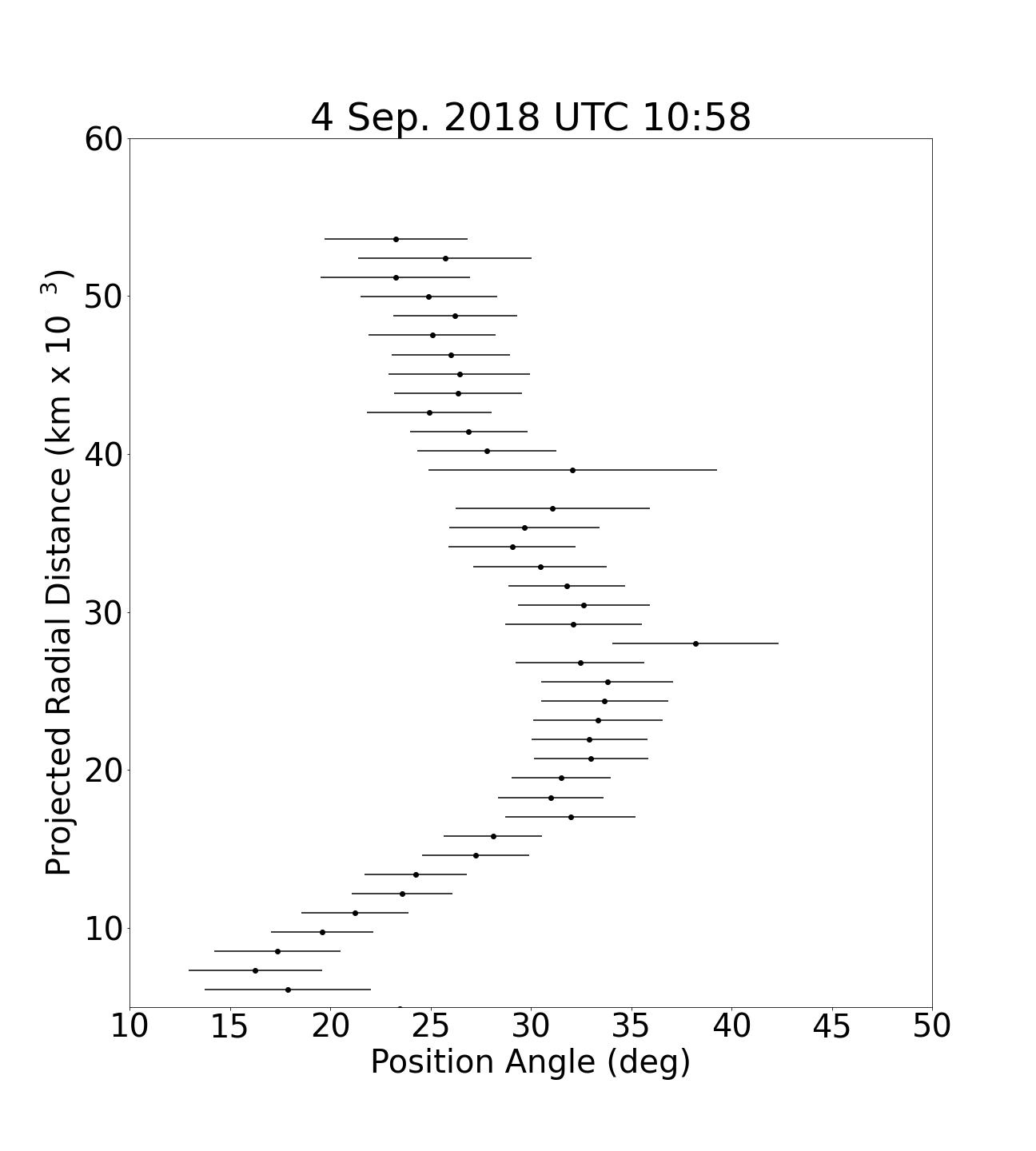}
    \includegraphics[width=0.9\textwidth]{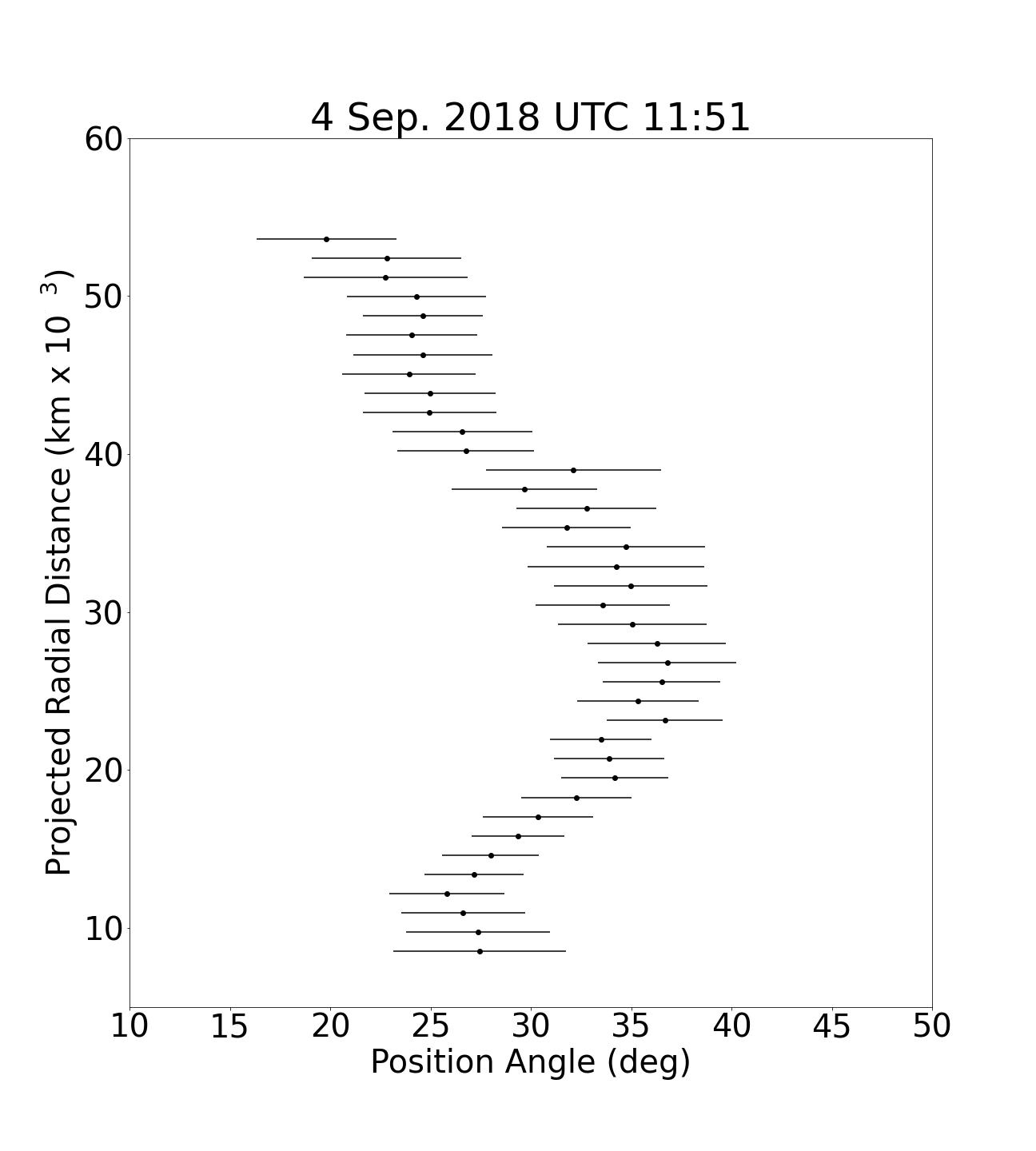}
\end{minipage}
\hspace{0.05\textwidth}
\begin{minipage}[t]{.45\textwidth}
\centering
    \includegraphics[width=0.9\textwidth]{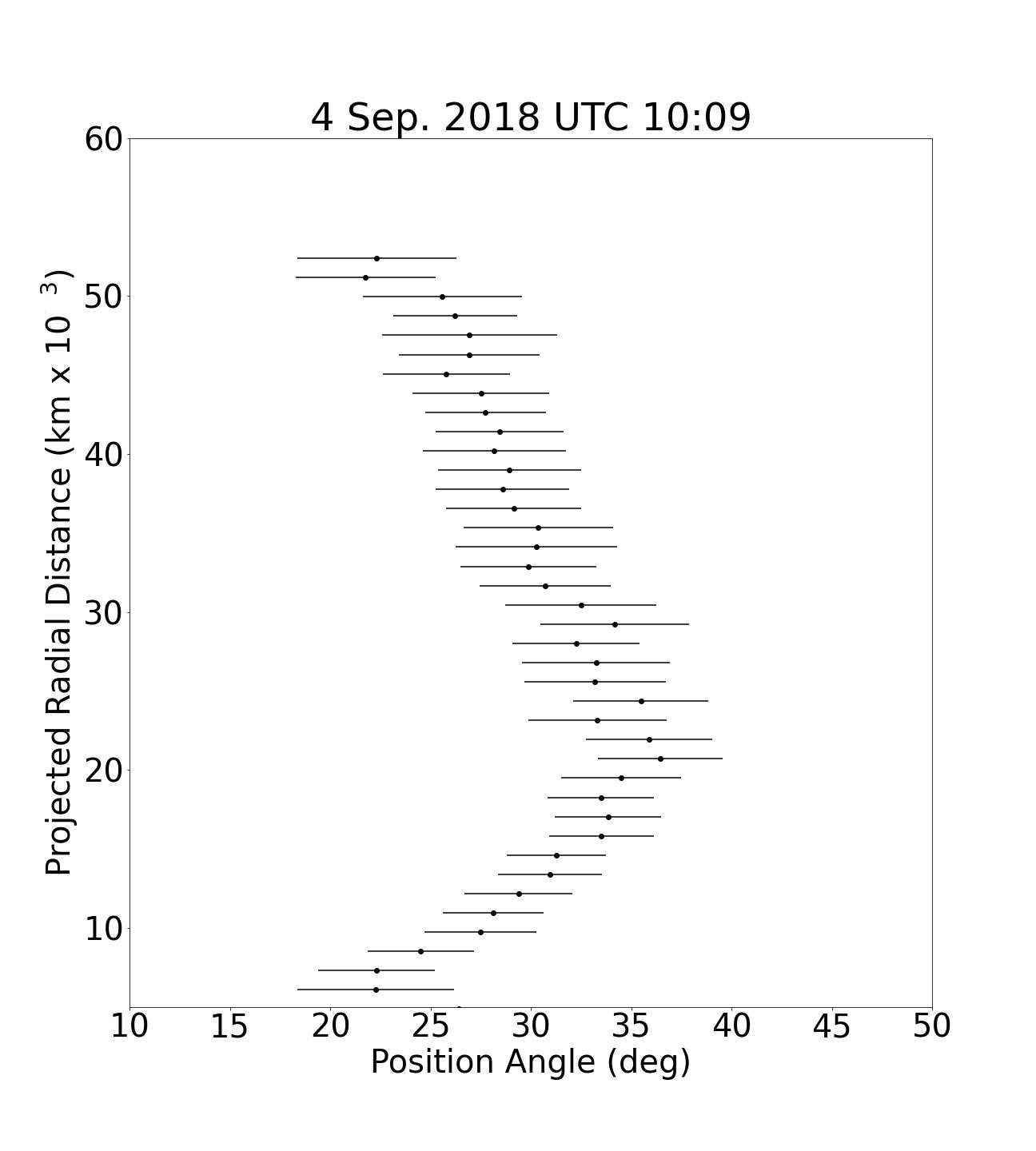}
    \includegraphics[width=0.9\textwidth]{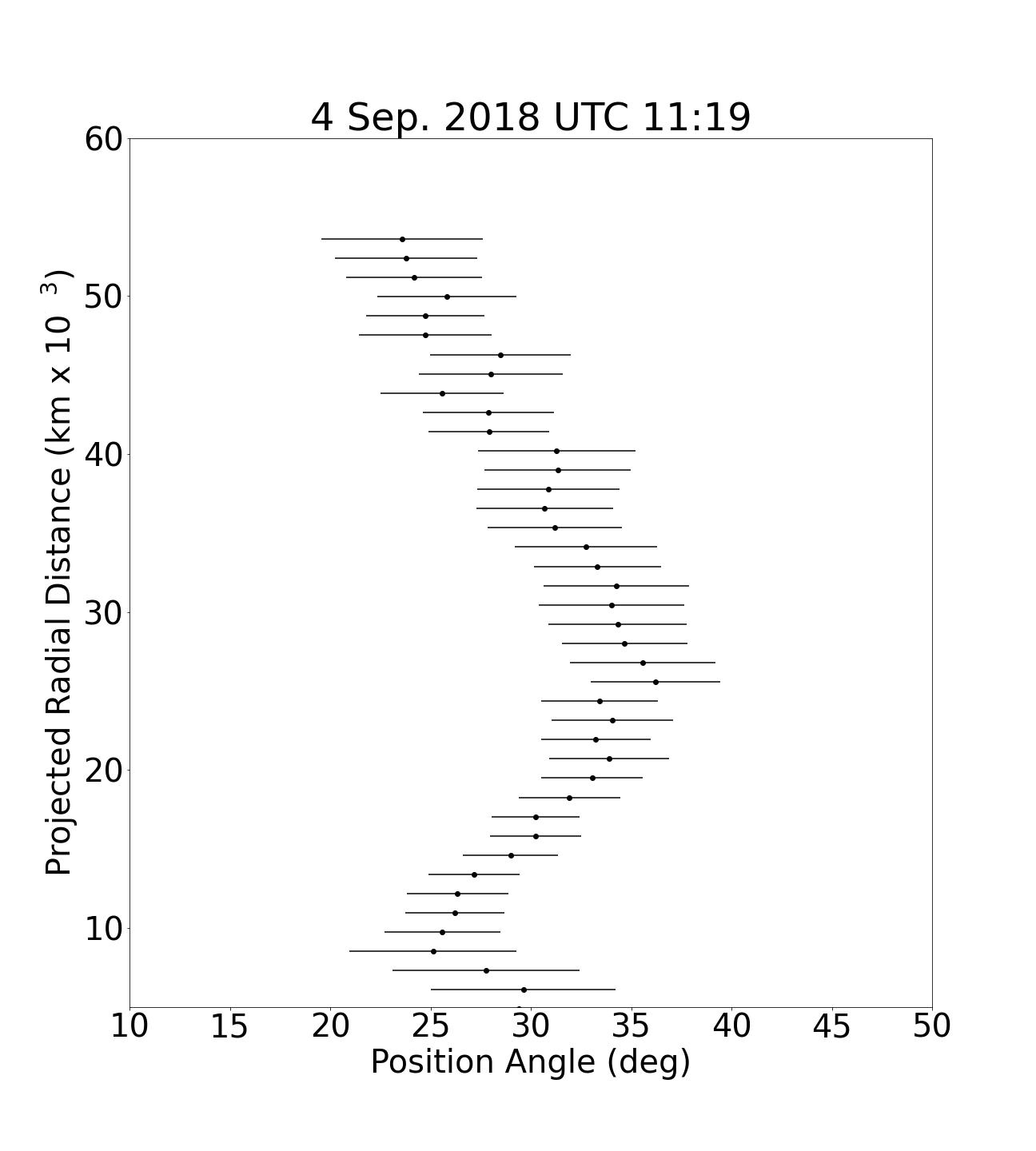}
    \vspace{2.5cm}
    \caption{Northern Jet morphology for 4 September 2018 at five different times spread out across the night, showing outward movement and a clear corkscrew pattern.\label{morphology_night2}}
    \vspace{1cm}
\end{minipage}
\end{figure}

\begin{figure}[h]
\centering
\begin{minipage}[t]{.45\textwidth}
\centering
    \includegraphics[width=0.9\textwidth]{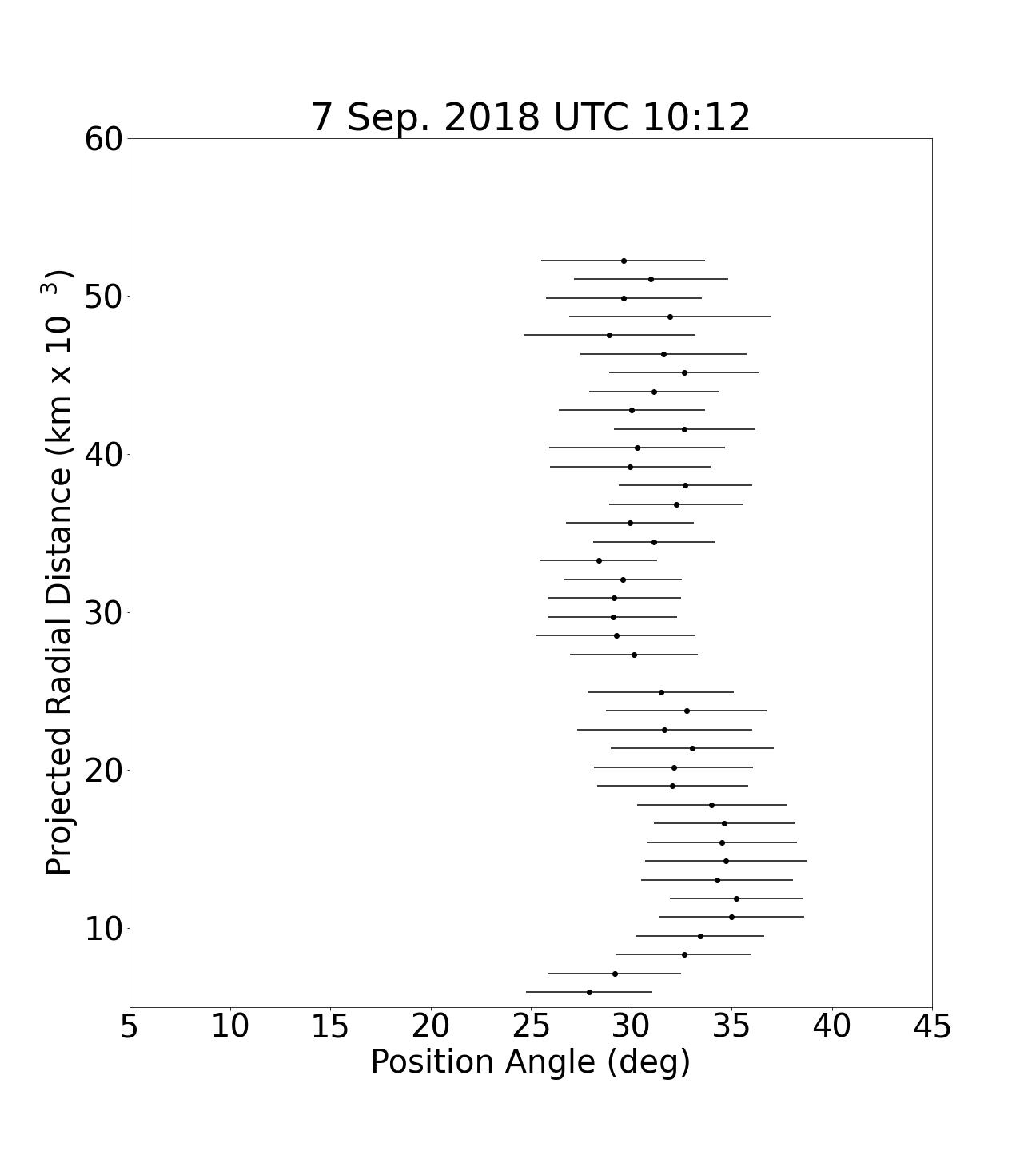}
    \includegraphics[width=0.9\textwidth]{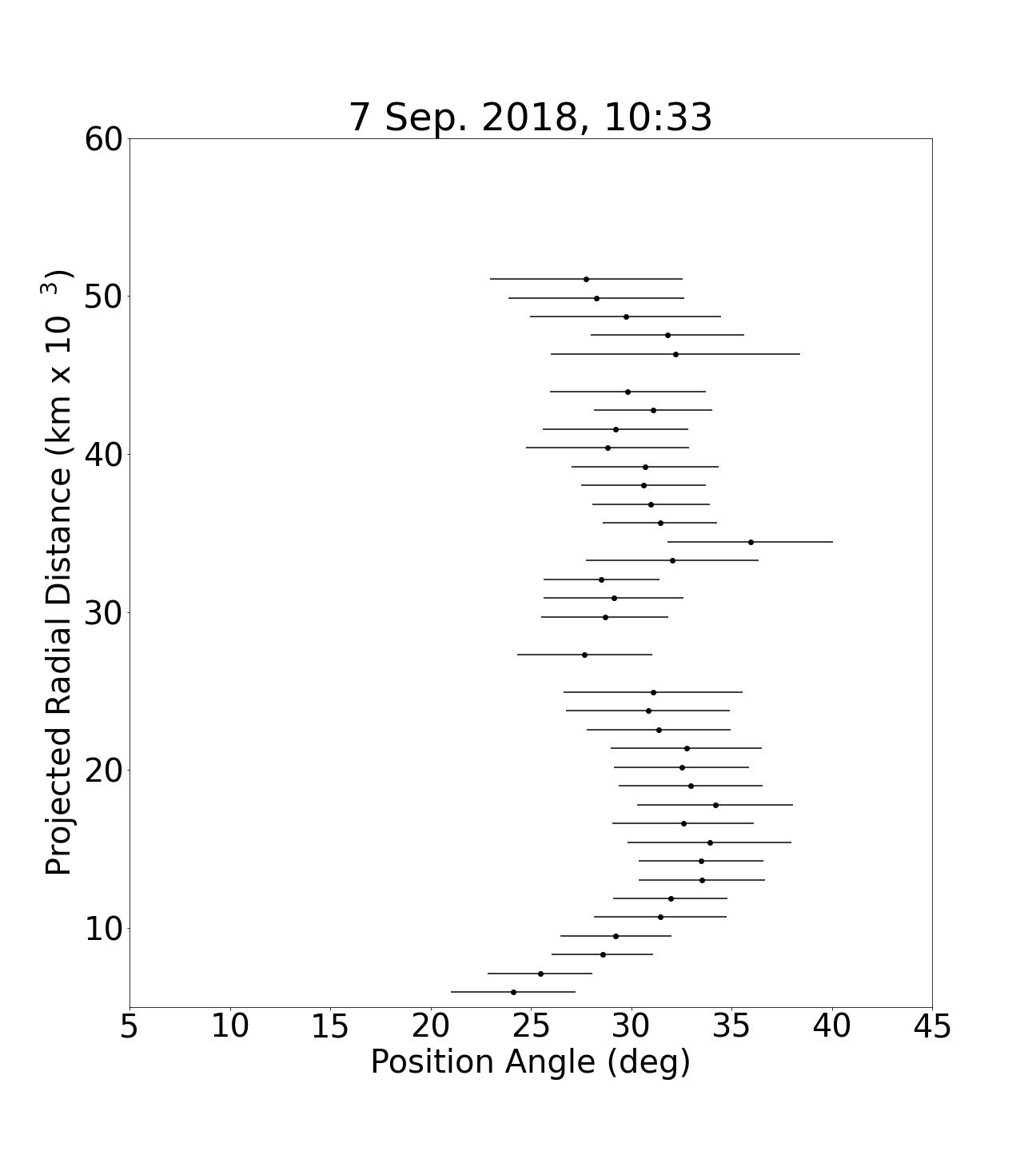}
    \includegraphics[width=0.9\textwidth]{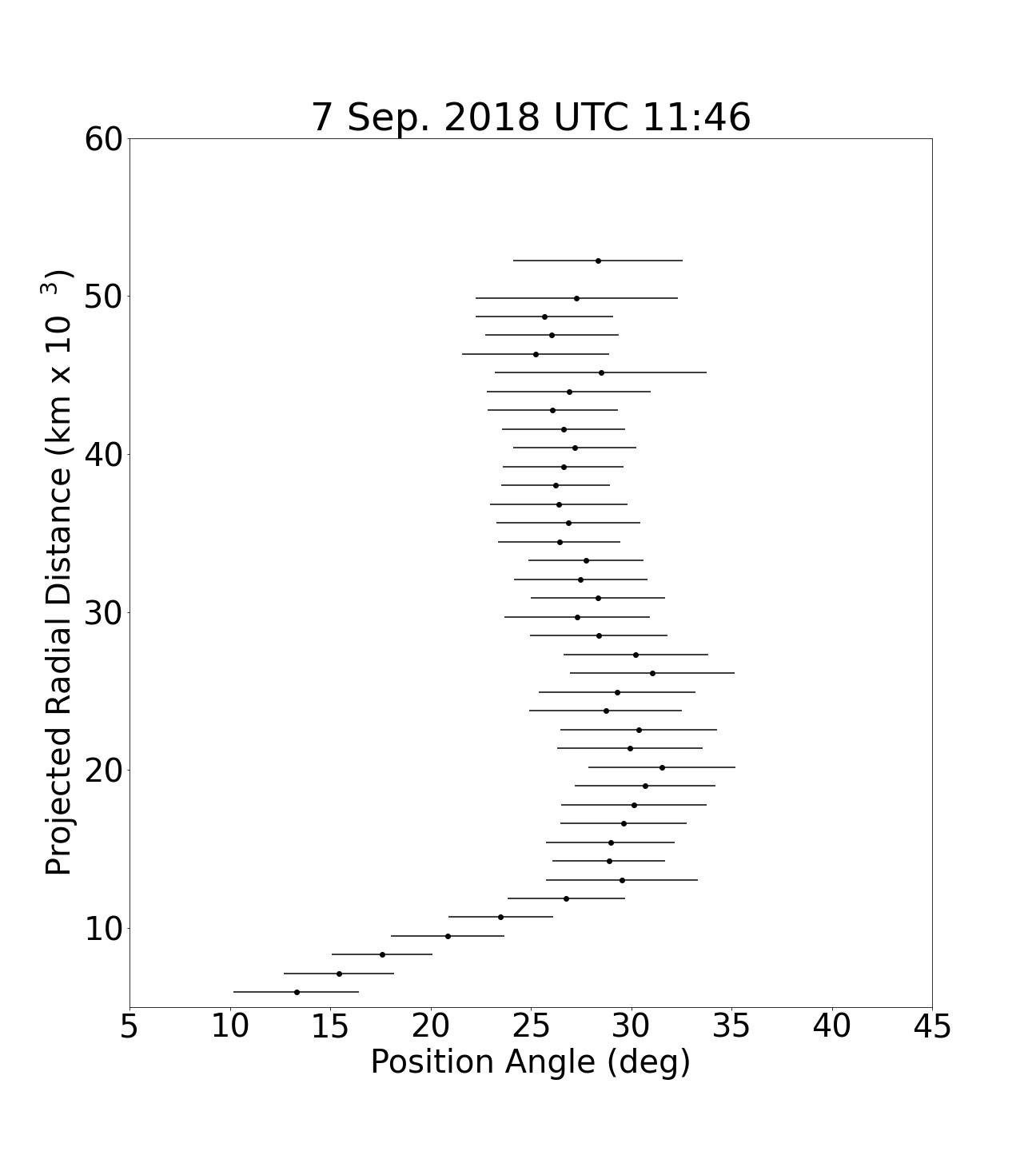}
\end{minipage}
\hspace{0.05\textwidth}
\begin{minipage}[t]{.45\textwidth}
\centering
    \includegraphics[width=0.9\textwidth]{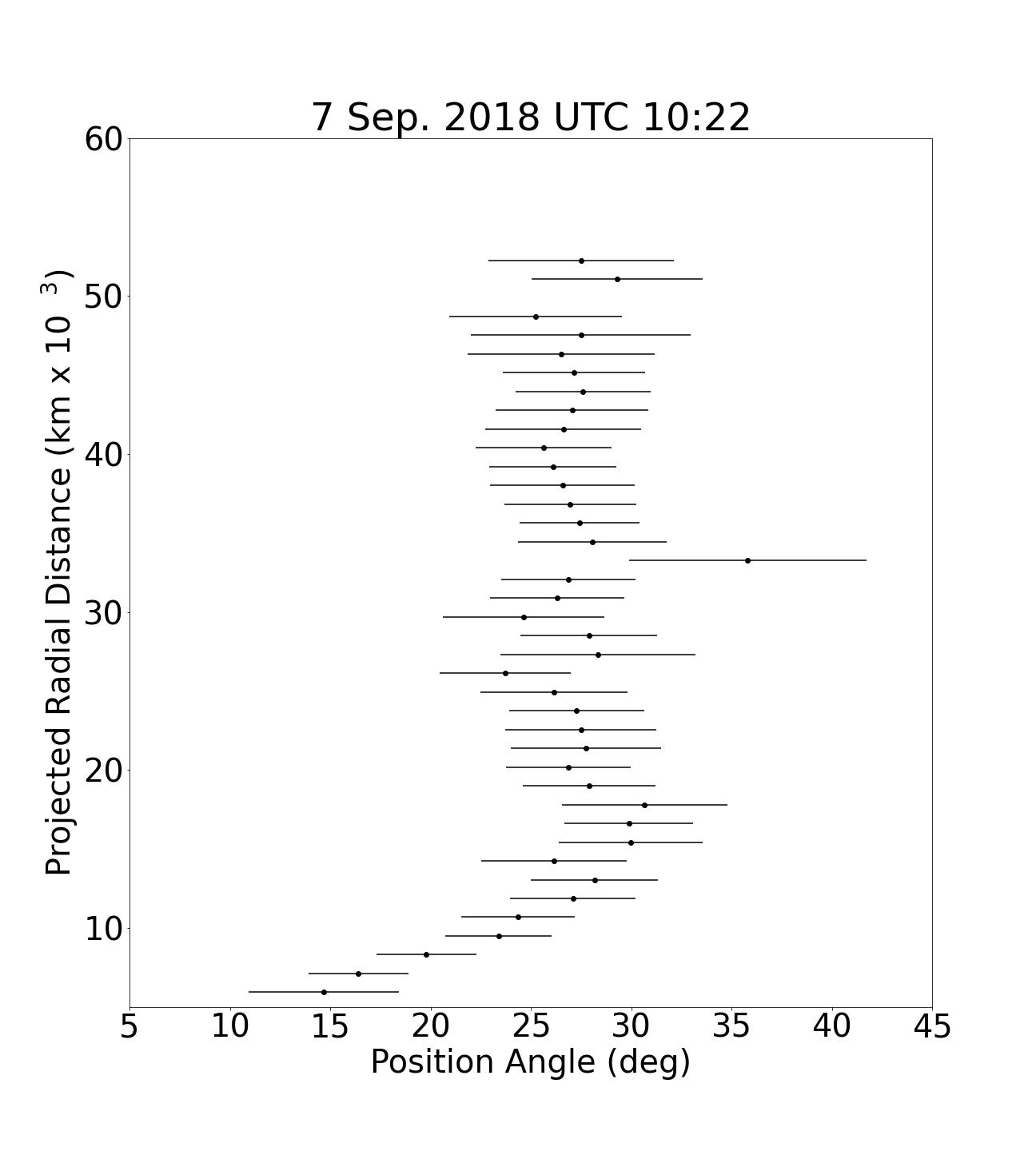}
    \includegraphics[width=0.9\textwidth]{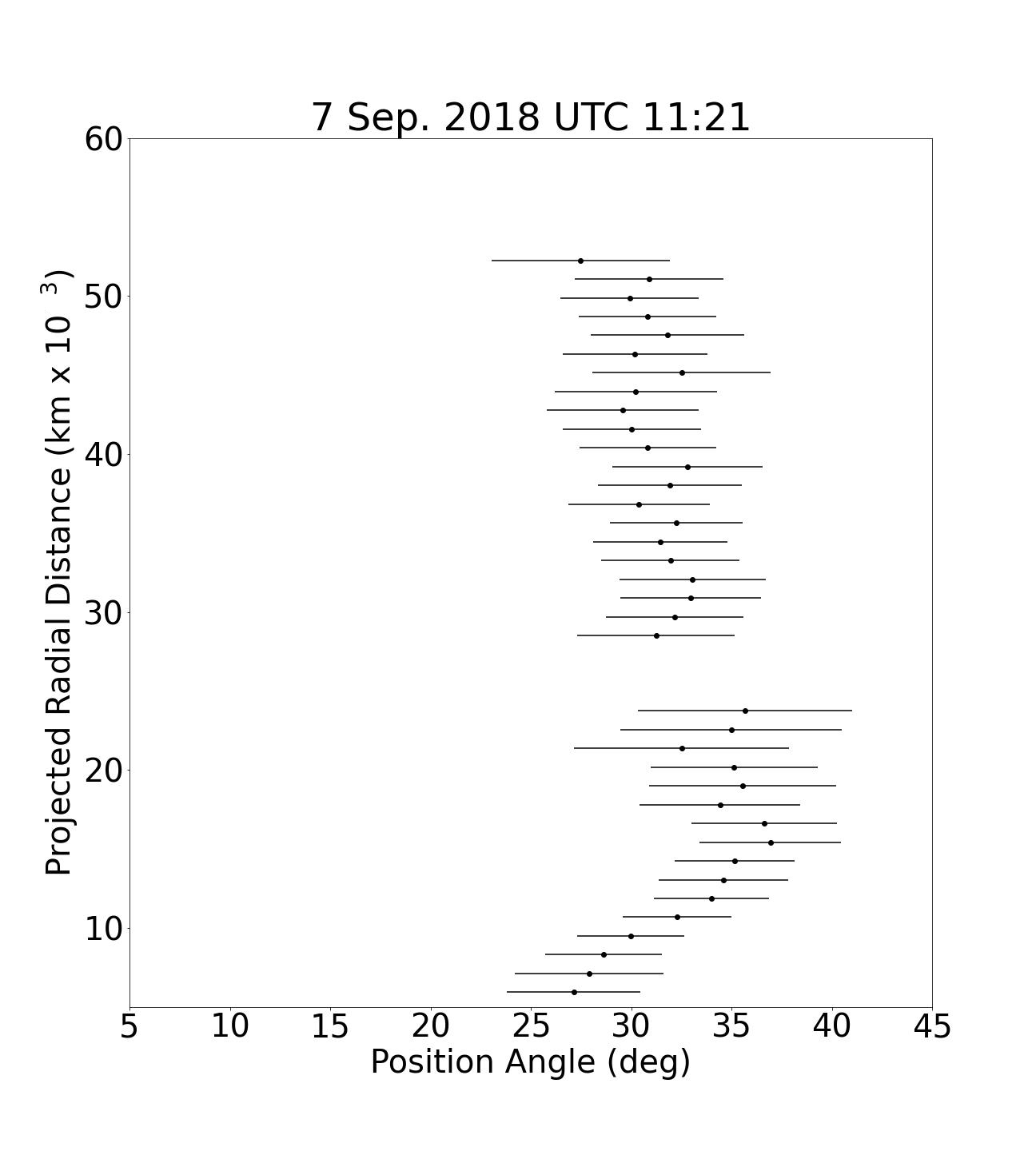}
    \vspace{2.5cm}
    \caption{Northern Jet morphology for 7 September 2018 at five different times spread out across the night; this night does not clearly exemplify the same corkscrew pattern seen in previous nights, nor does it show as drastic of outward movements.\label{morphology_night3}}
    \vspace{1cm}
\end{minipage}
\end{figure}

\begin{figure}[h]
\centering
\begin{minipage}[t]{.45\textwidth}
\centering
    \includegraphics[width=0.9\textwidth]{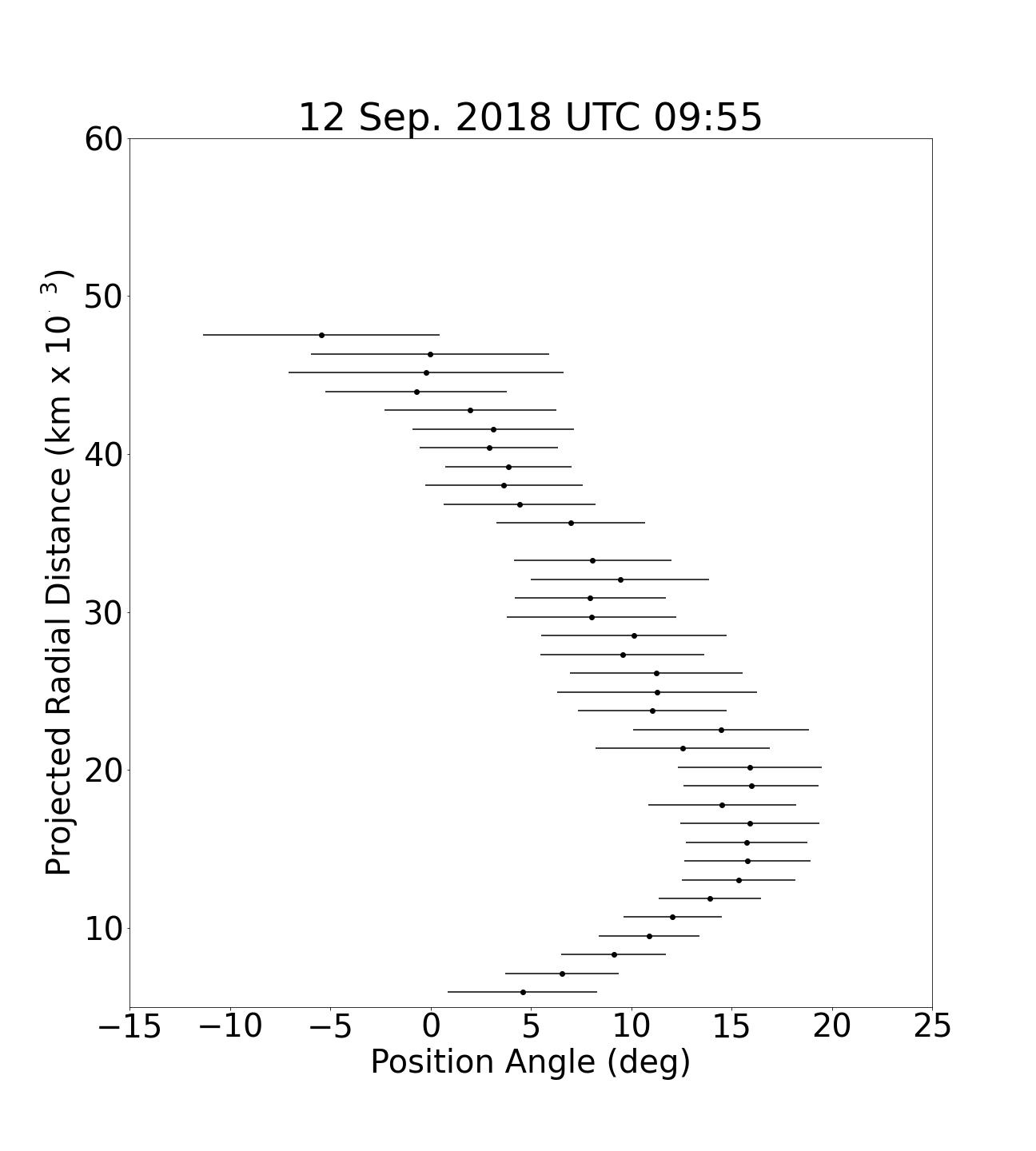}
    \includegraphics[width=0.9\textwidth]{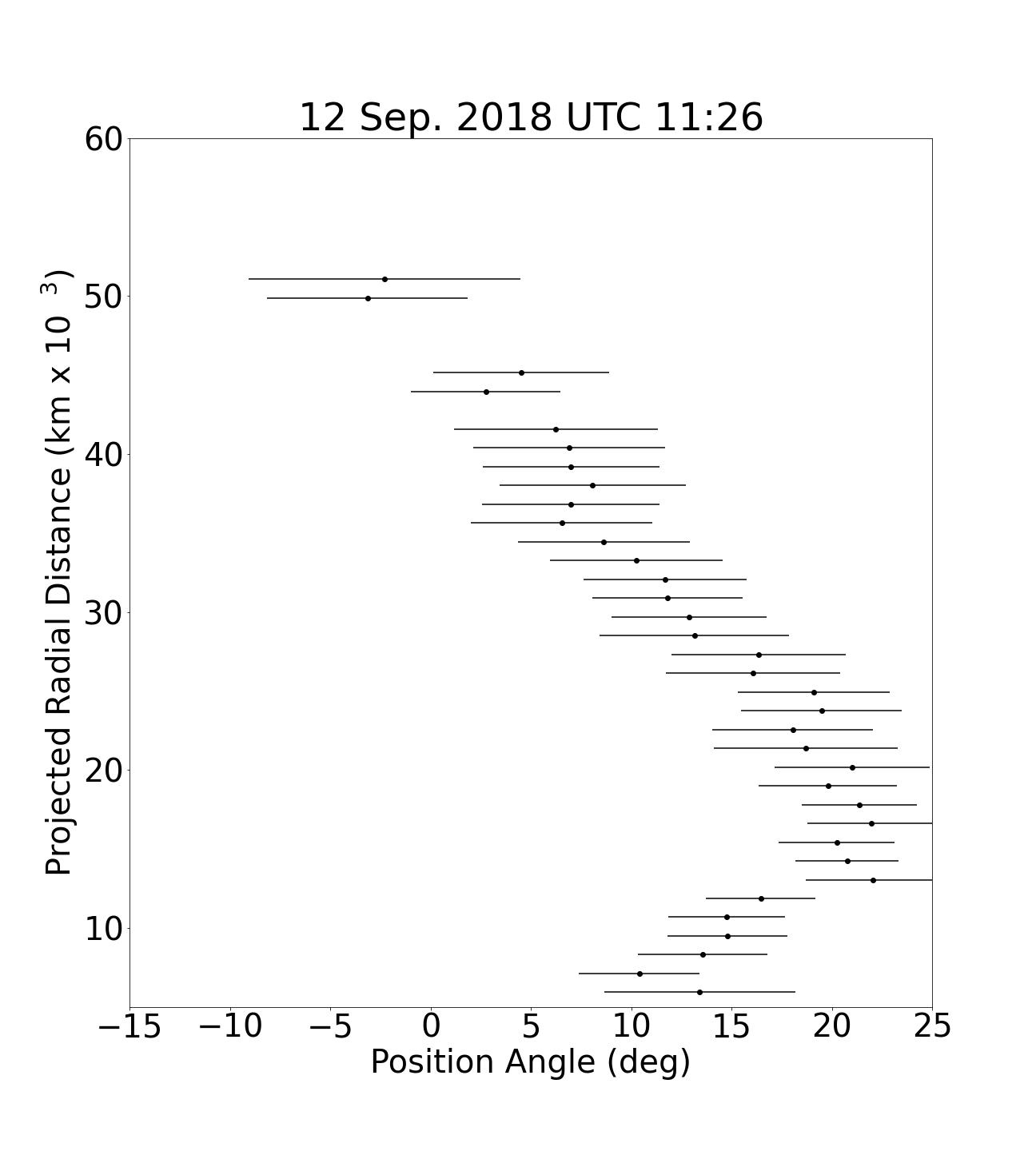}
    \end{minipage}
\hspace{0.05\textwidth}
\begin{minipage}[t]{.45\textwidth}
\centering
    \includegraphics[width=0.9\textwidth]{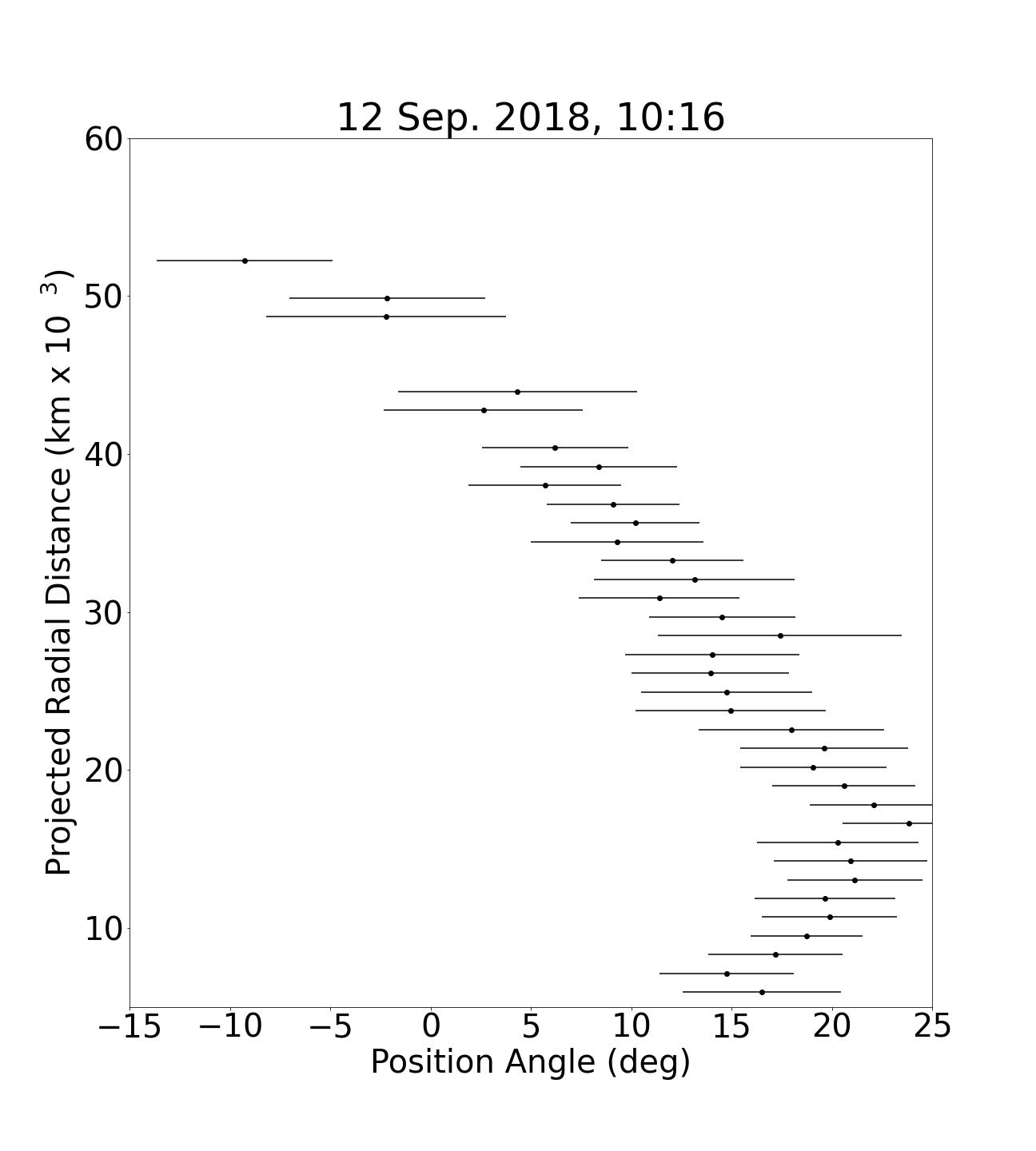}
    \includegraphics[width=0.9\textwidth]{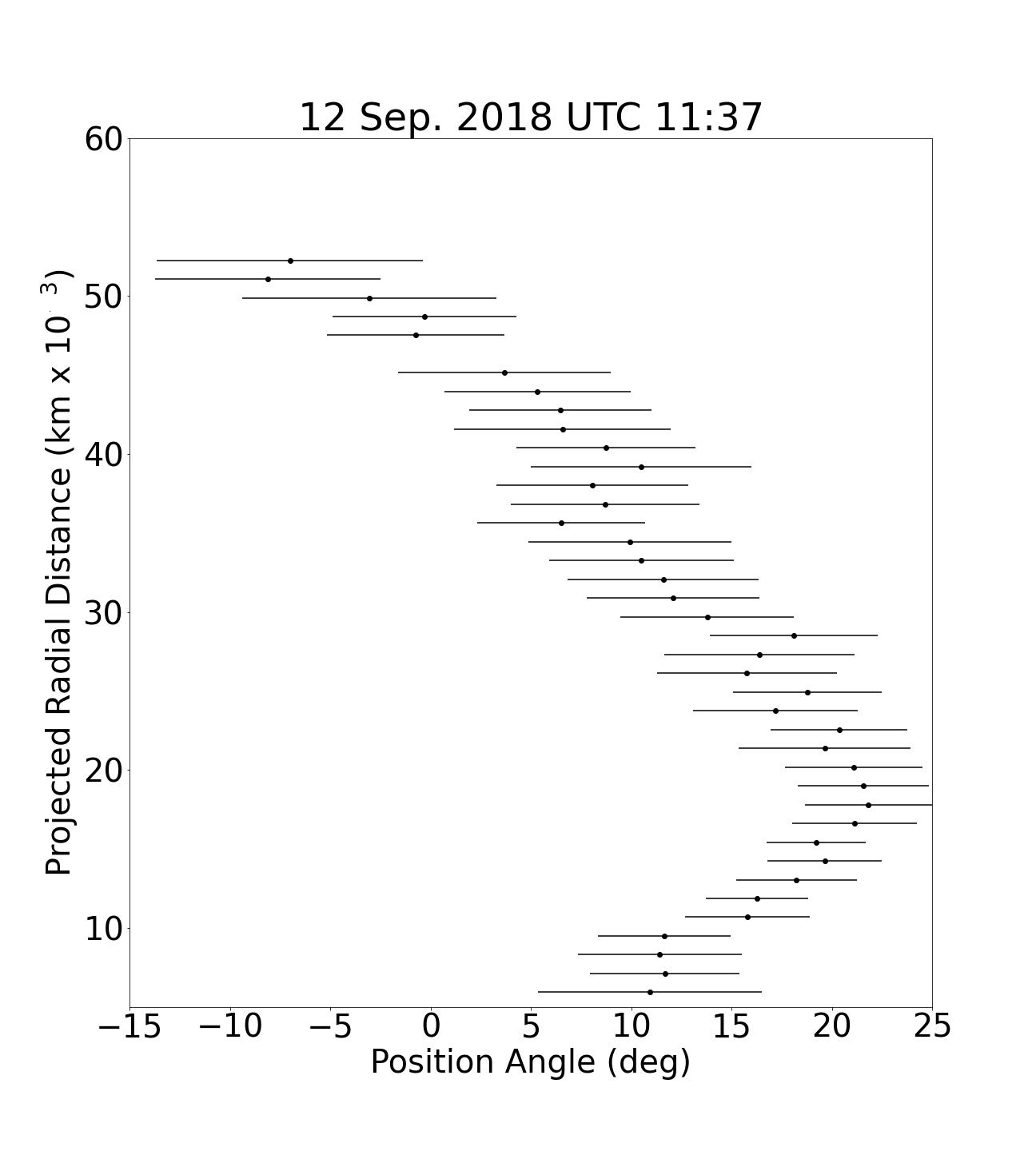}
\end{minipage}
    \caption{Northern Jet morphology for 12 September 2018 at four different times spread out across the night, showing a definite change in the curvature.\label{morphology_night4}}
\end{figure}

\begin{figure}[h]
\centering
\begin{minipage}[t]{.45\textwidth}
\centering
    \includegraphics[width=0.9\textwidth]{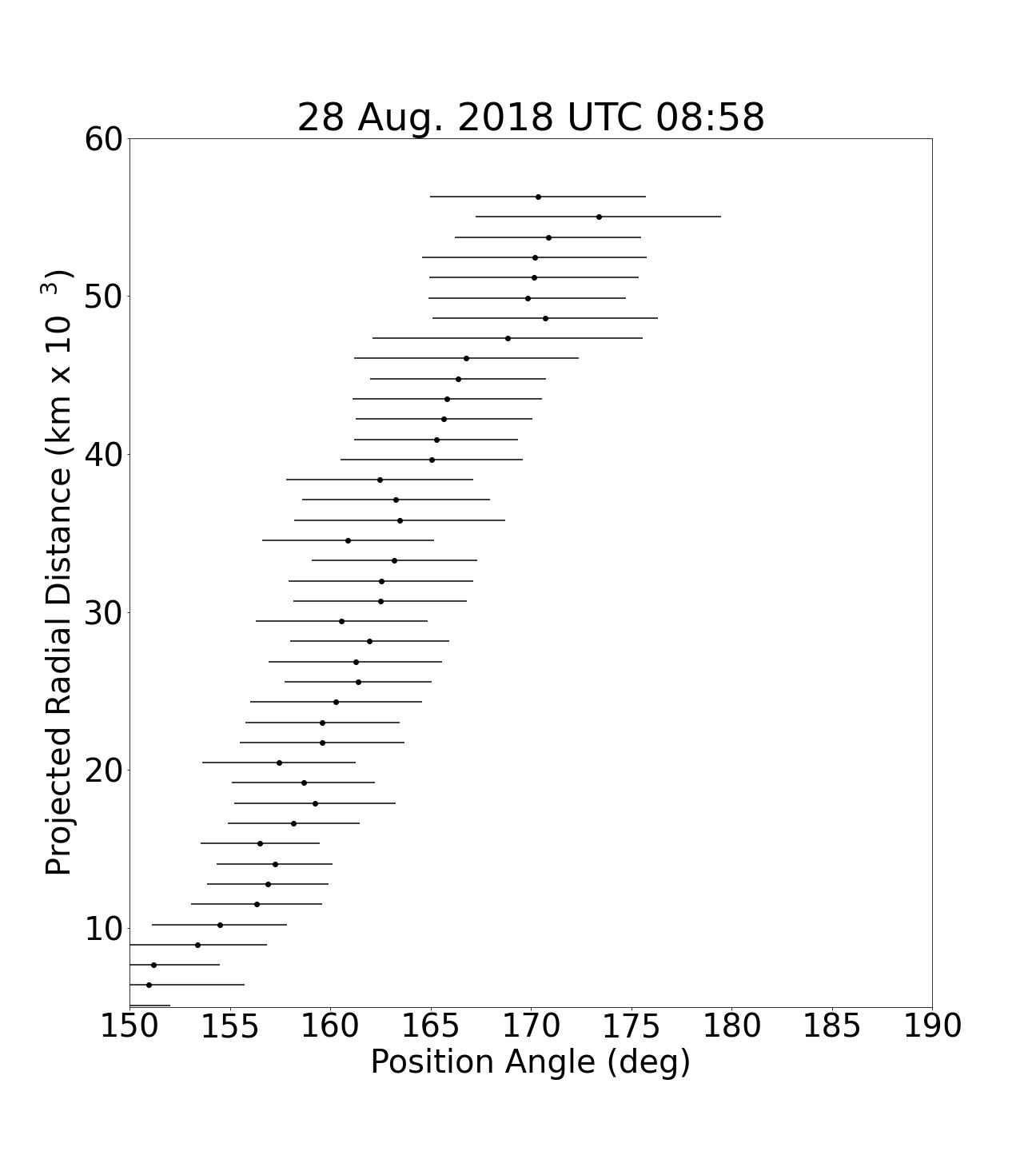}
    \includegraphics[width=0.9\textwidth]{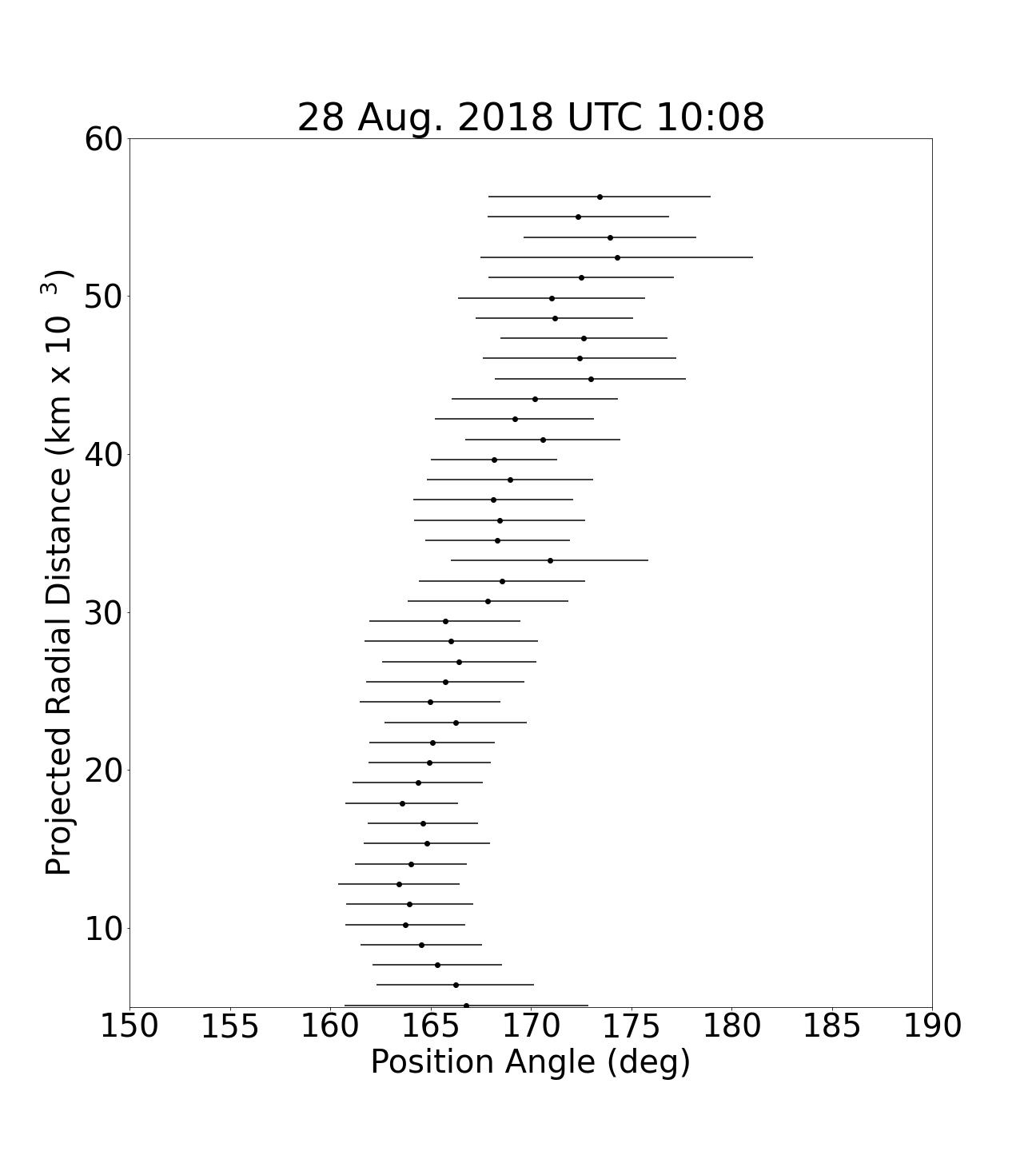}
    \includegraphics[width=0.9\textwidth]{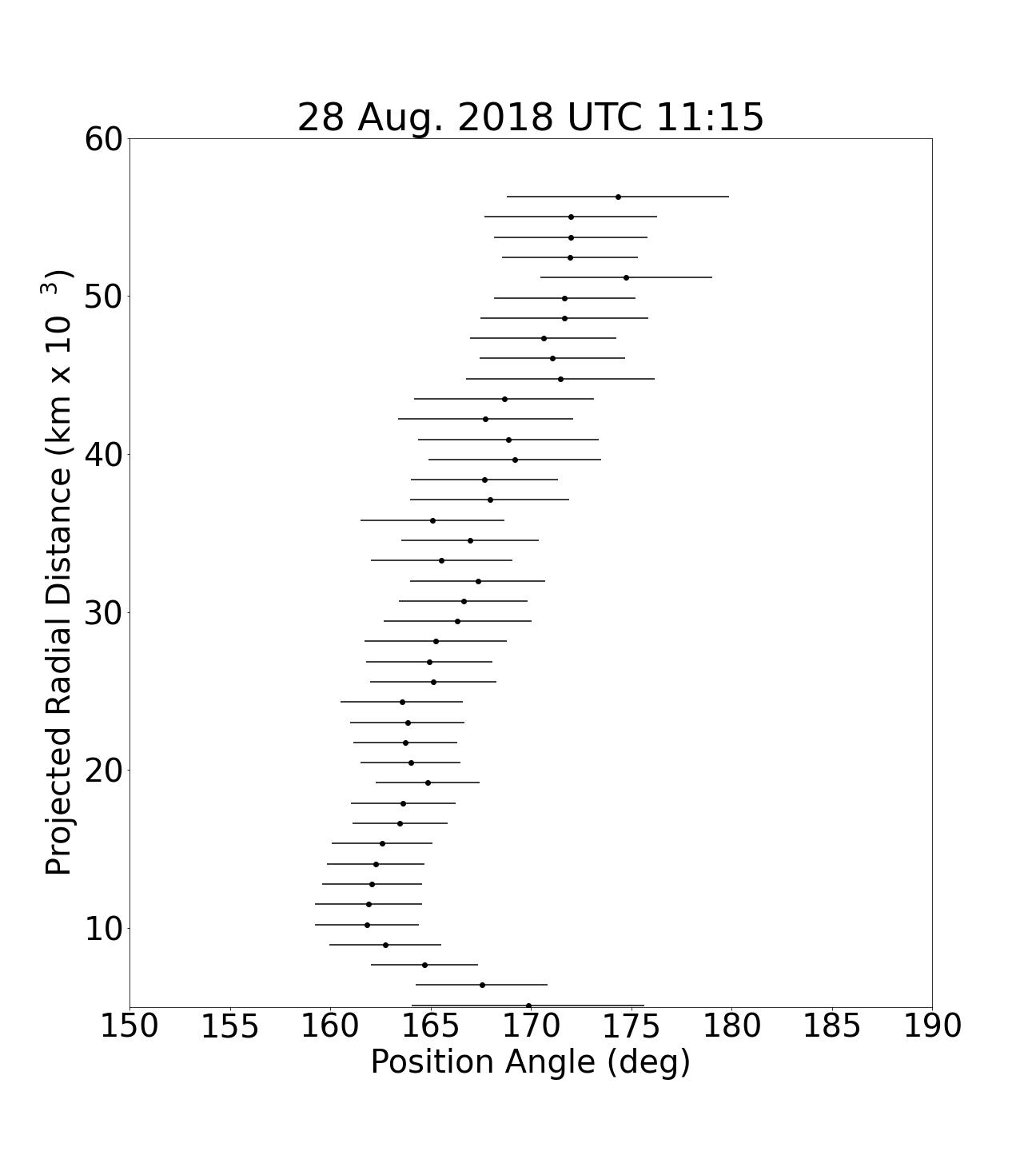}
\end{minipage}
\hspace{0.05\textwidth}
\begin{minipage}[t]{.45\textwidth}
\centering
    \includegraphics[width=0.9\textwidth]{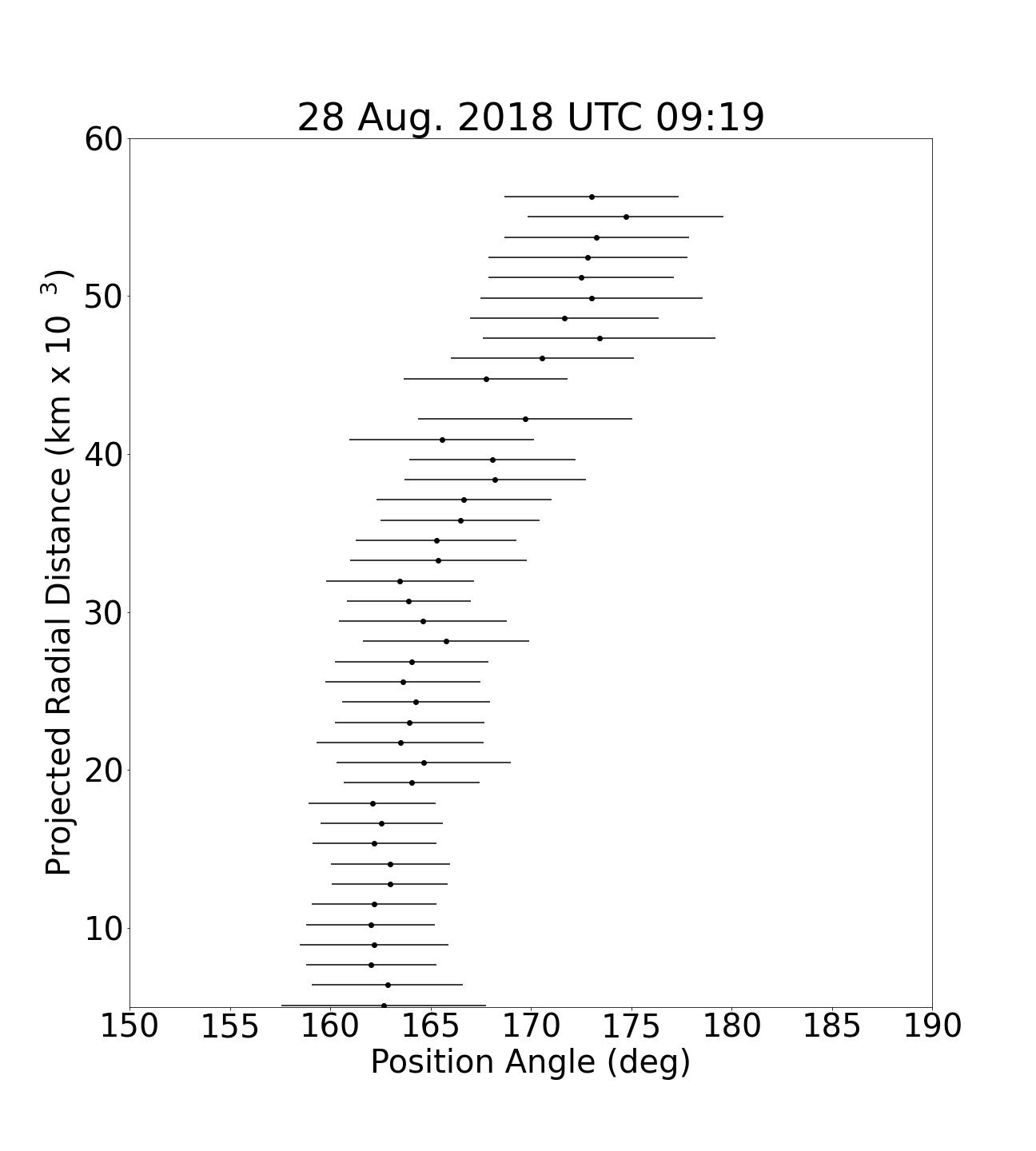}
    \includegraphics[width=0.9\textwidth]{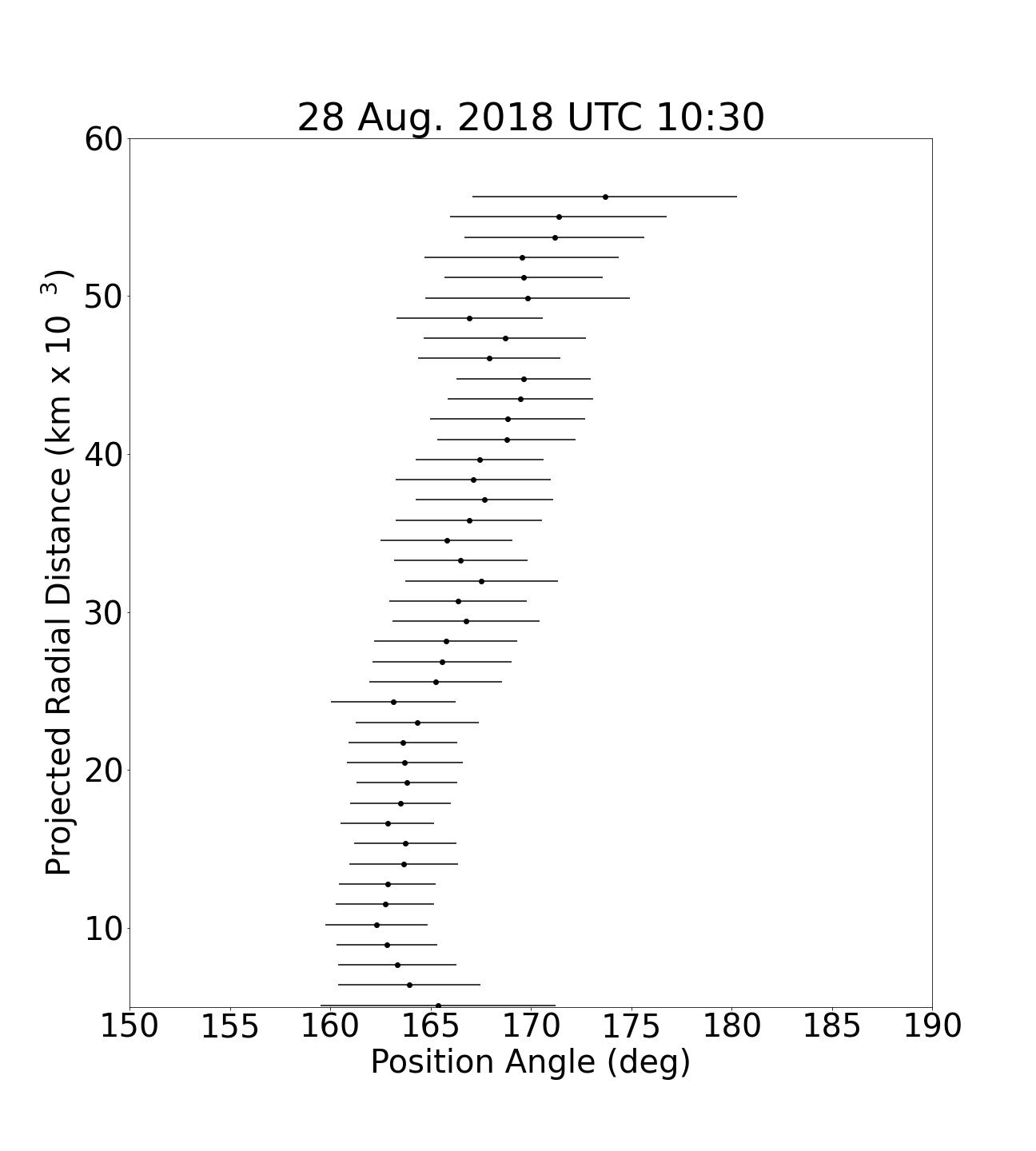}
    \vspace{2.5cm}
    \caption{Southern Jet morphology for 28 August 2018 at five different times spread out across the night, showing a curved jet most likely due to radiation pressure, with some near-nucleus curvature.\label{morphology_night1_Jet2}}
    \vspace{1cm}
\end{minipage}
\end{figure}

\begin{figure}[h]
\centering
\begin{minipage}[t]{.45\textwidth}
\centering
    \includegraphics[width=0.9\textwidth]{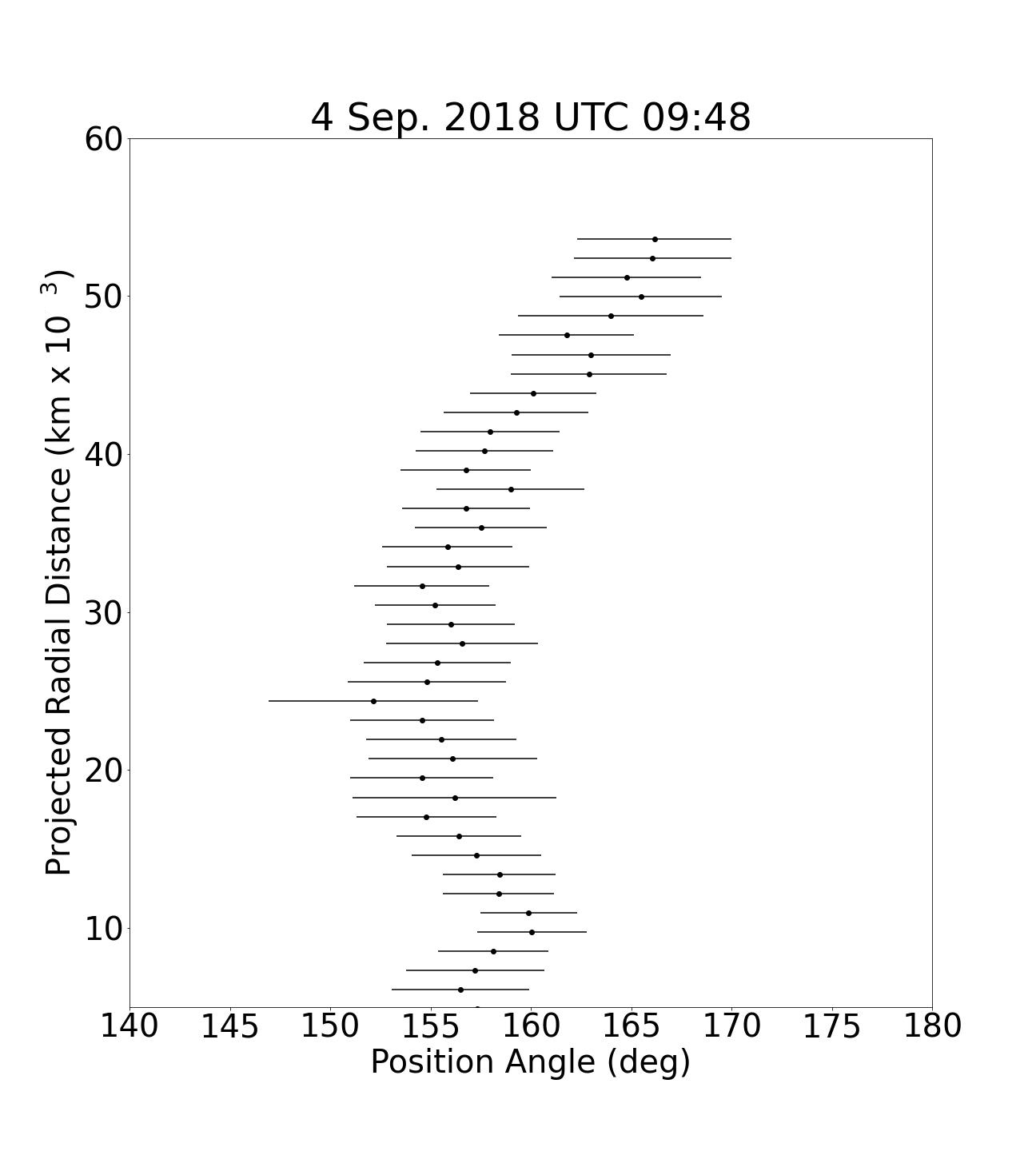}
    \includegraphics[width=0.9\textwidth]{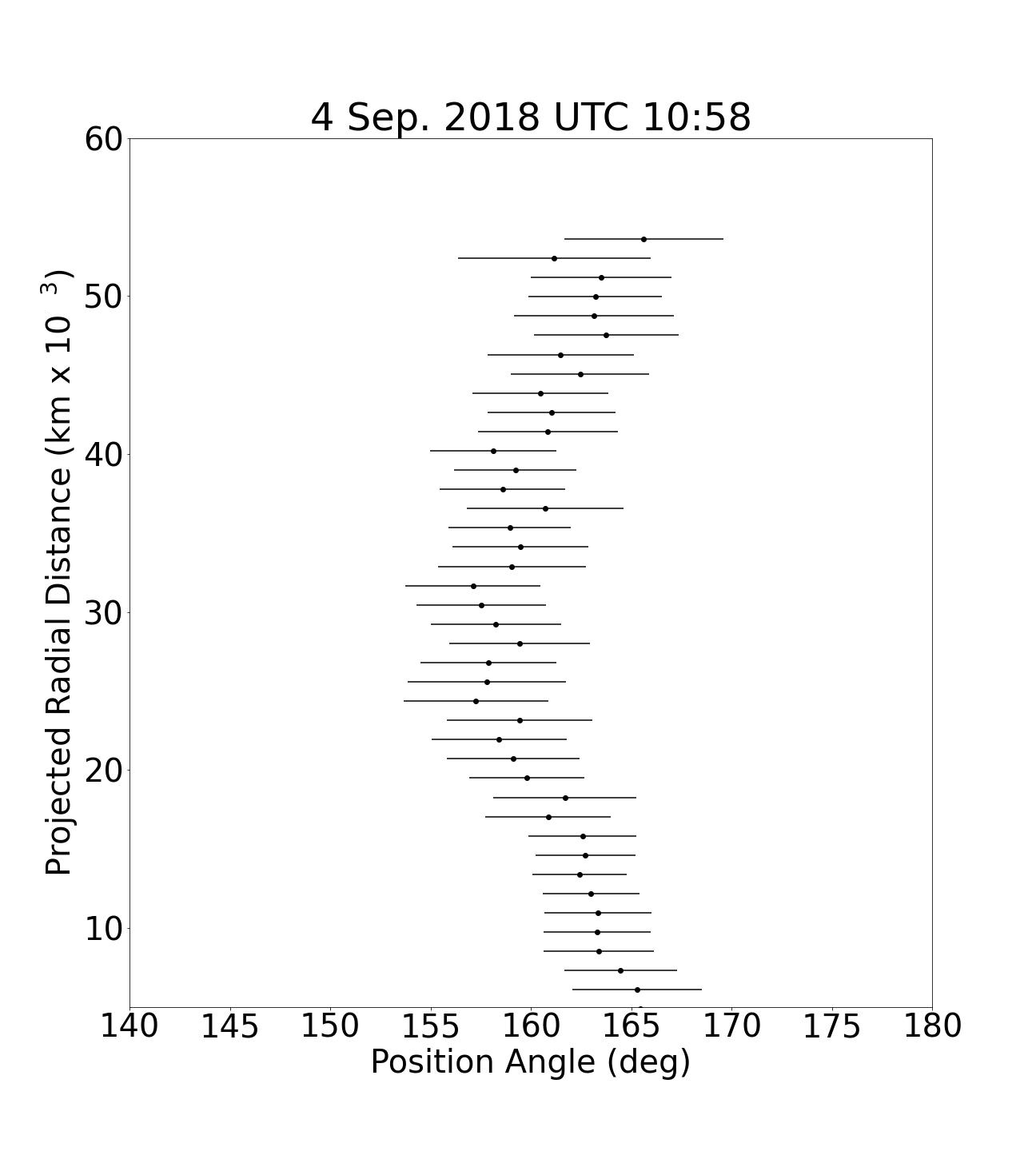}
    \includegraphics[width=0.9\textwidth]{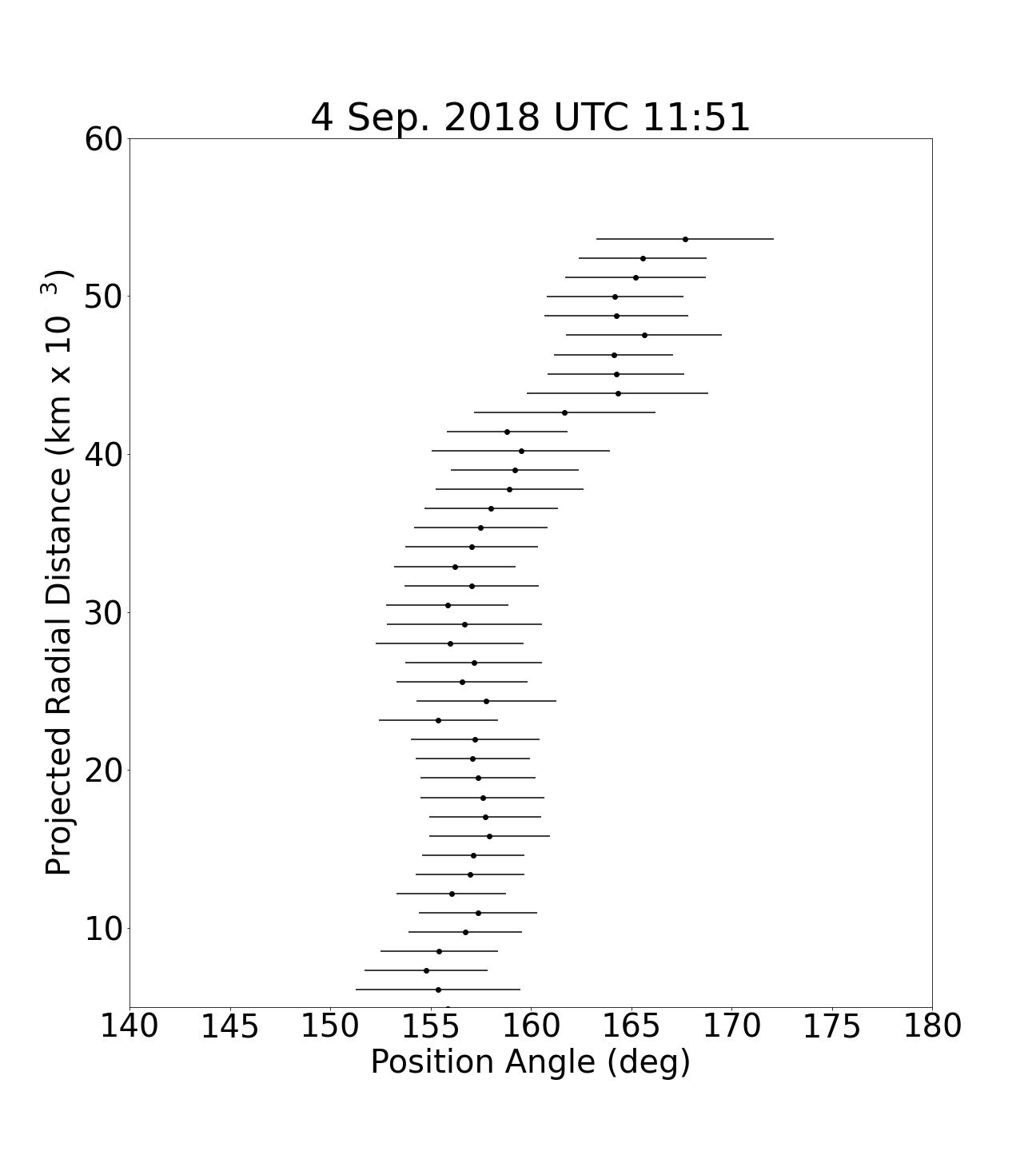}
\end{minipage}
\hspace{0.05\textwidth}
\begin{minipage}[t]{.45\textwidth}
\centering
    \includegraphics[width=0.9\textwidth]{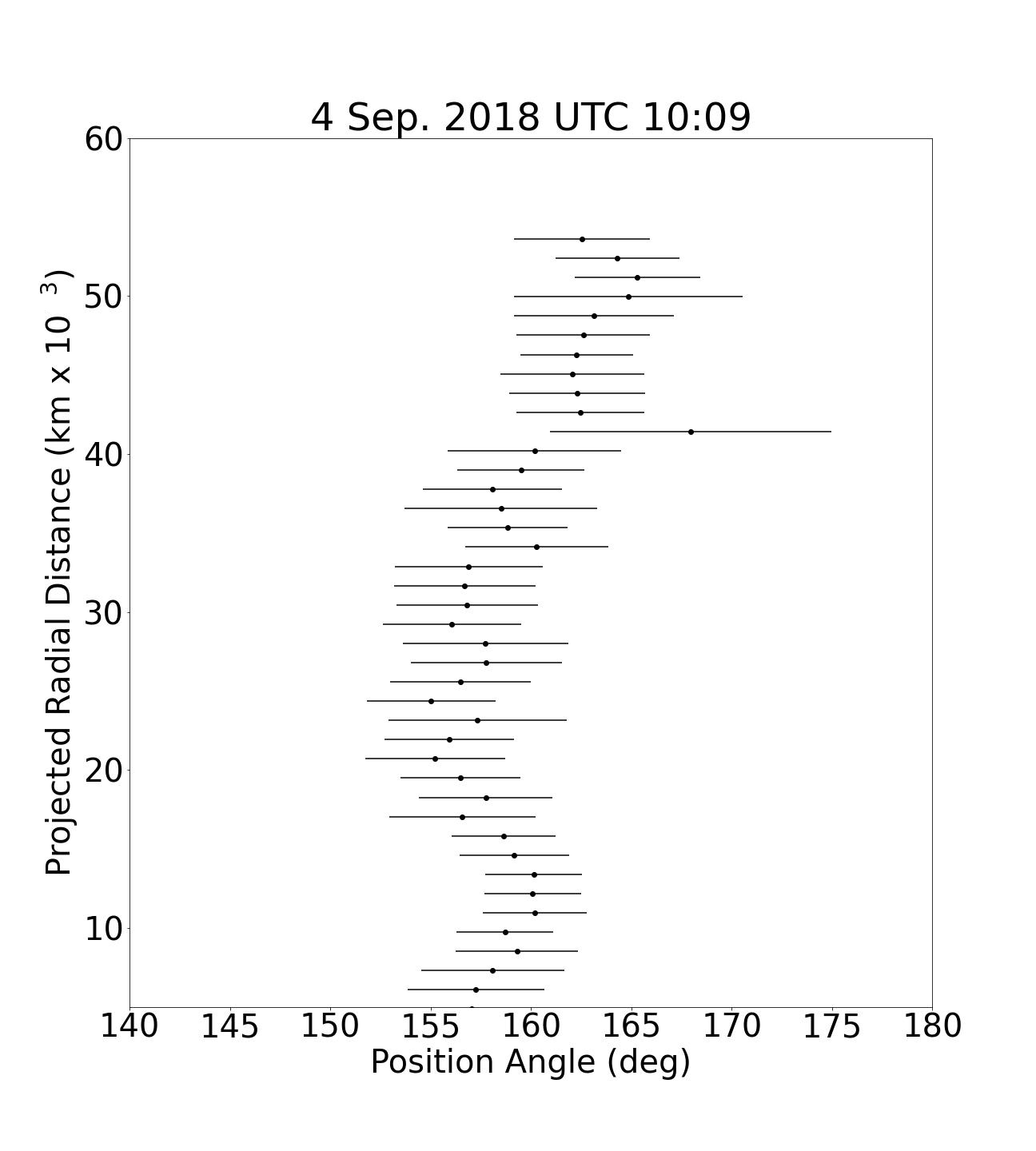}
    \includegraphics[width=0.9\textwidth]{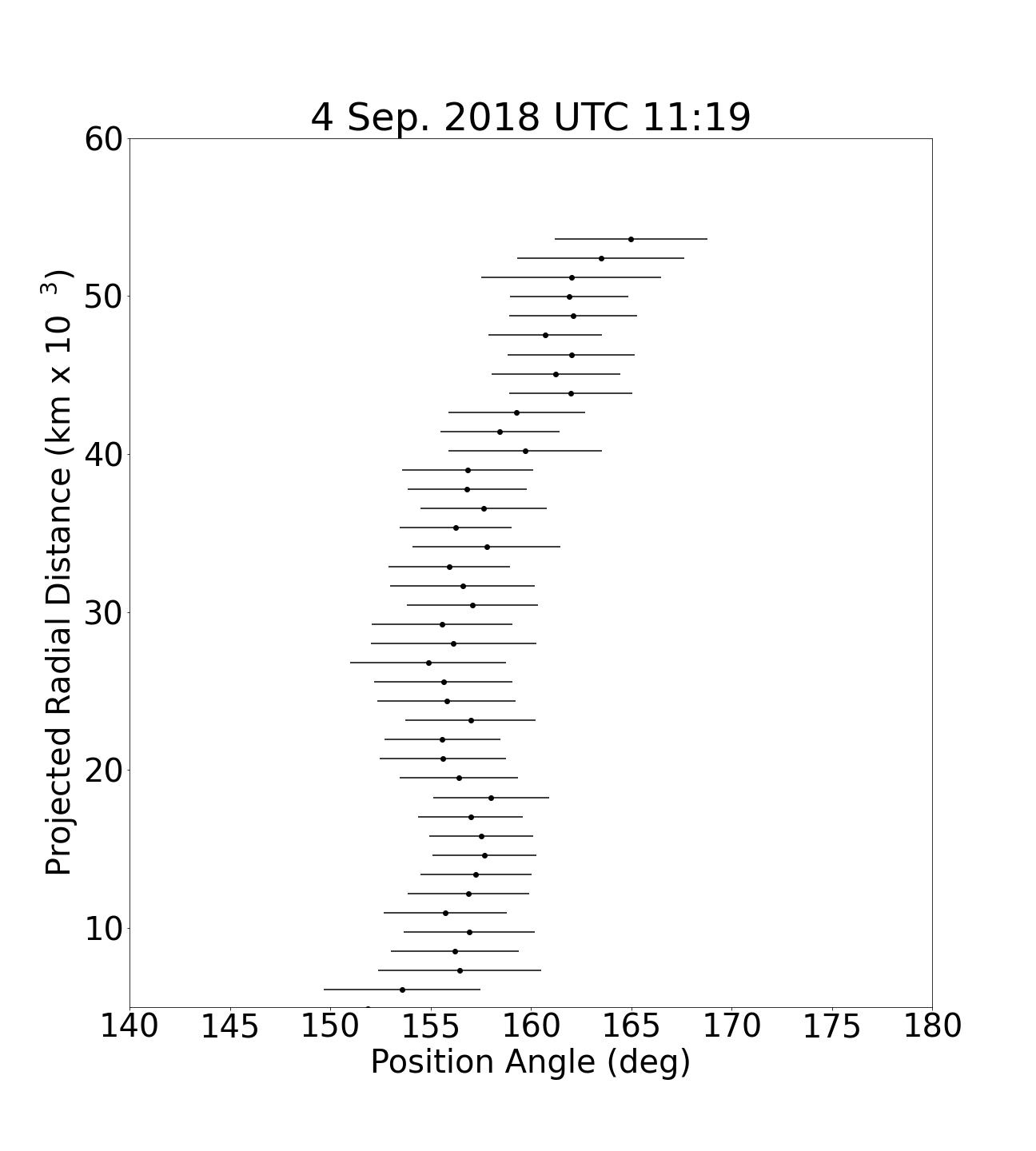}
    \vspace{2.5cm}
    \caption{Southern Jet morphology for 4 September 2018 at five different times spread out across the night, showing a potential corkscrew pattern more visible at less than 20,000 km.\label{morphology_night2_Jet2}}
    \vspace{1cm}
\end{minipage}
\end{figure}

\begin{figure}[h]
\centering
\begin{minipage}[t]{.45\textwidth}
\centering
    \includegraphics[width=0.9\textwidth]{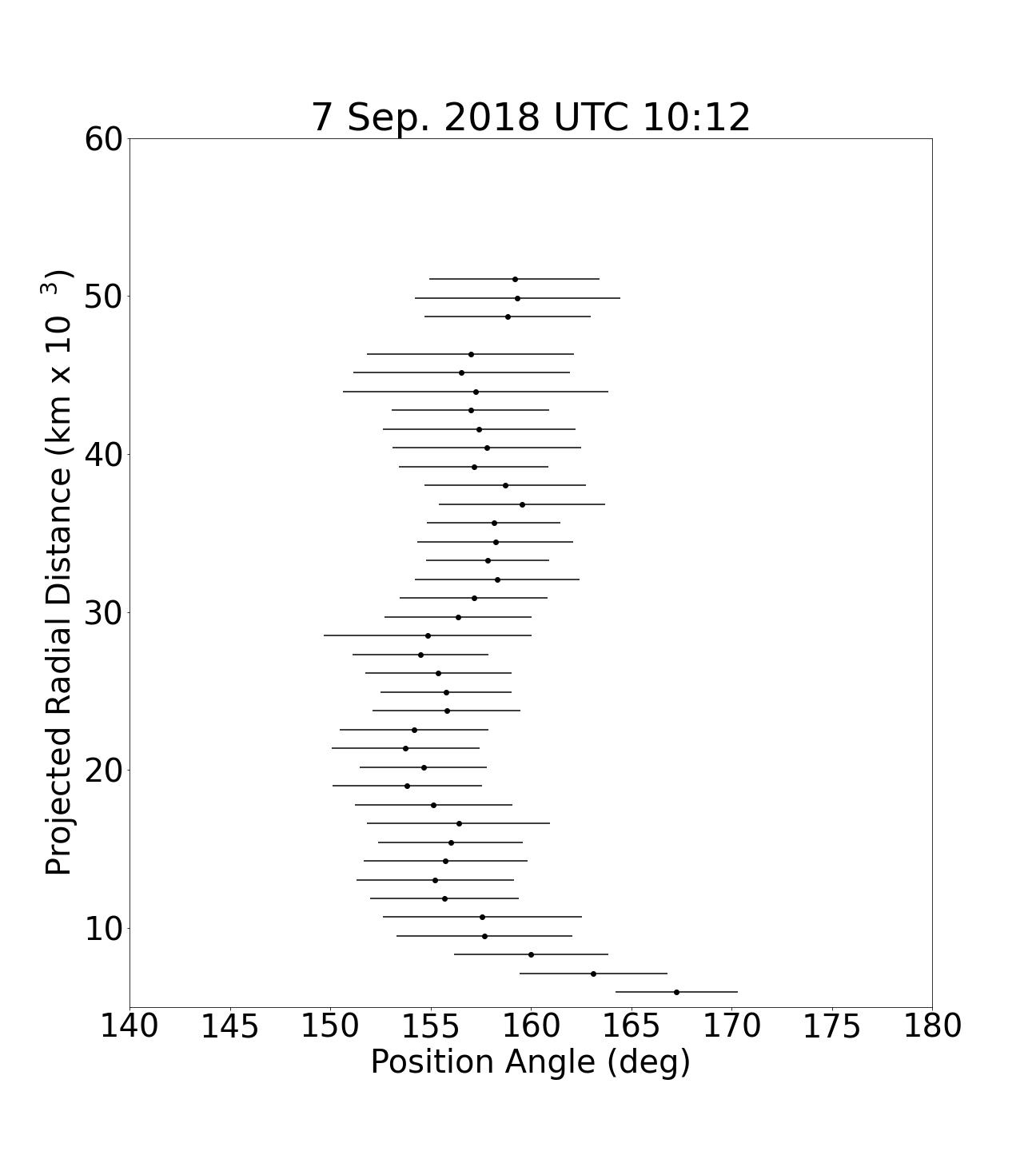}
    \includegraphics[width=0.9\textwidth]{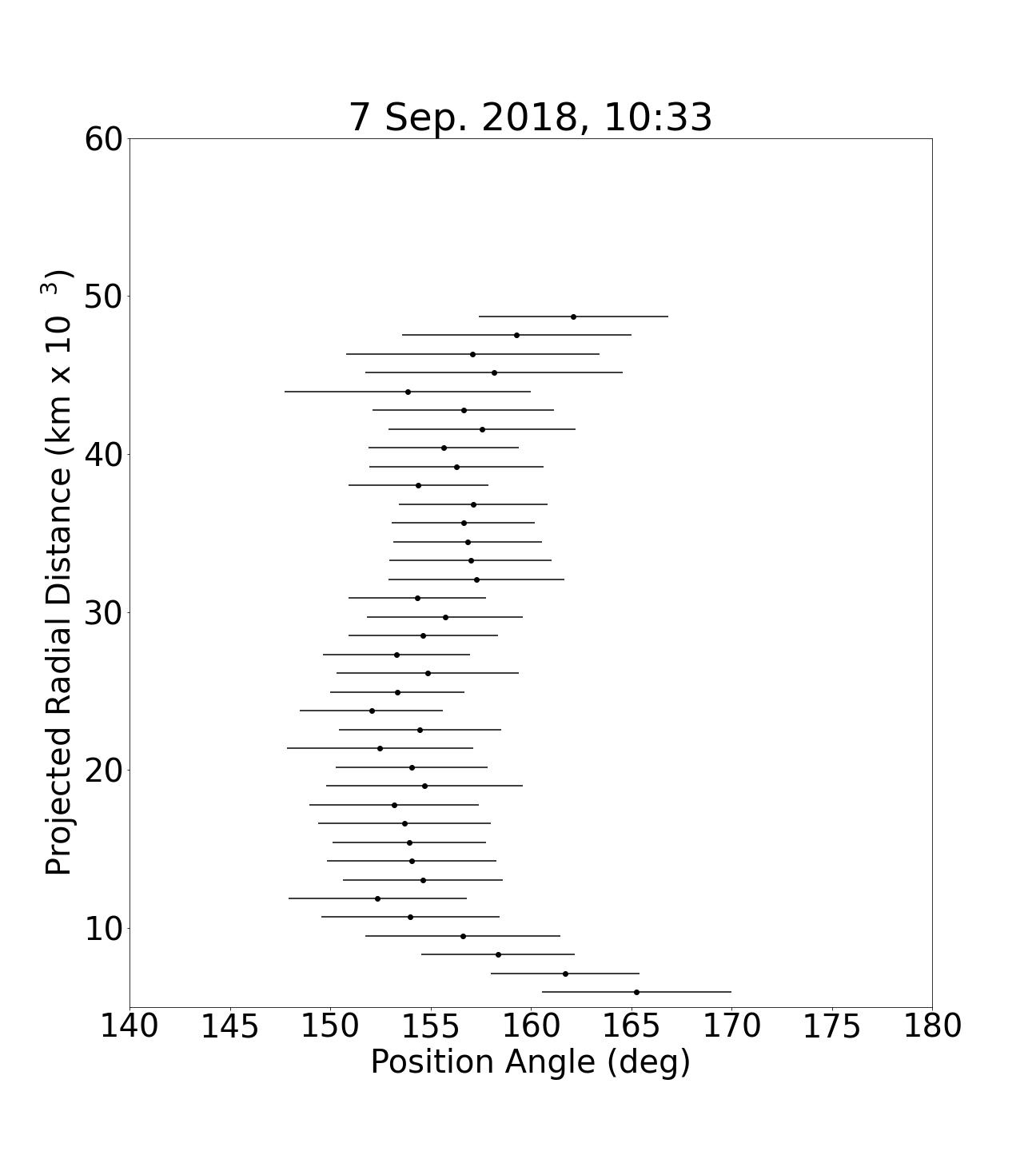}
    \includegraphics[width=0.9\textwidth]{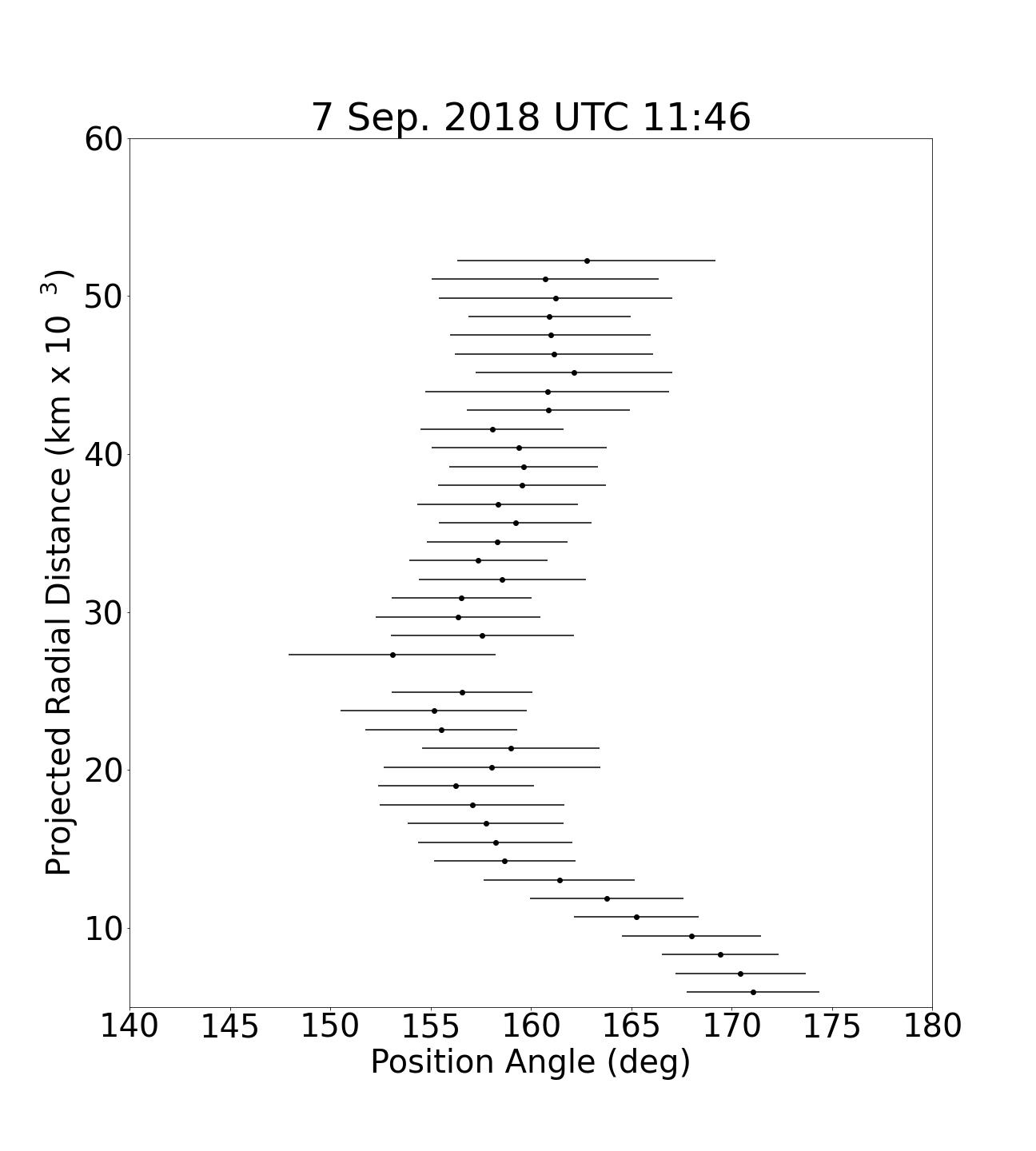}
\end{minipage}
\hspace{0.05\textwidth}
\begin{minipage}[t]{.45\textwidth}
\centering
    \includegraphics[width=0.9\textwidth]{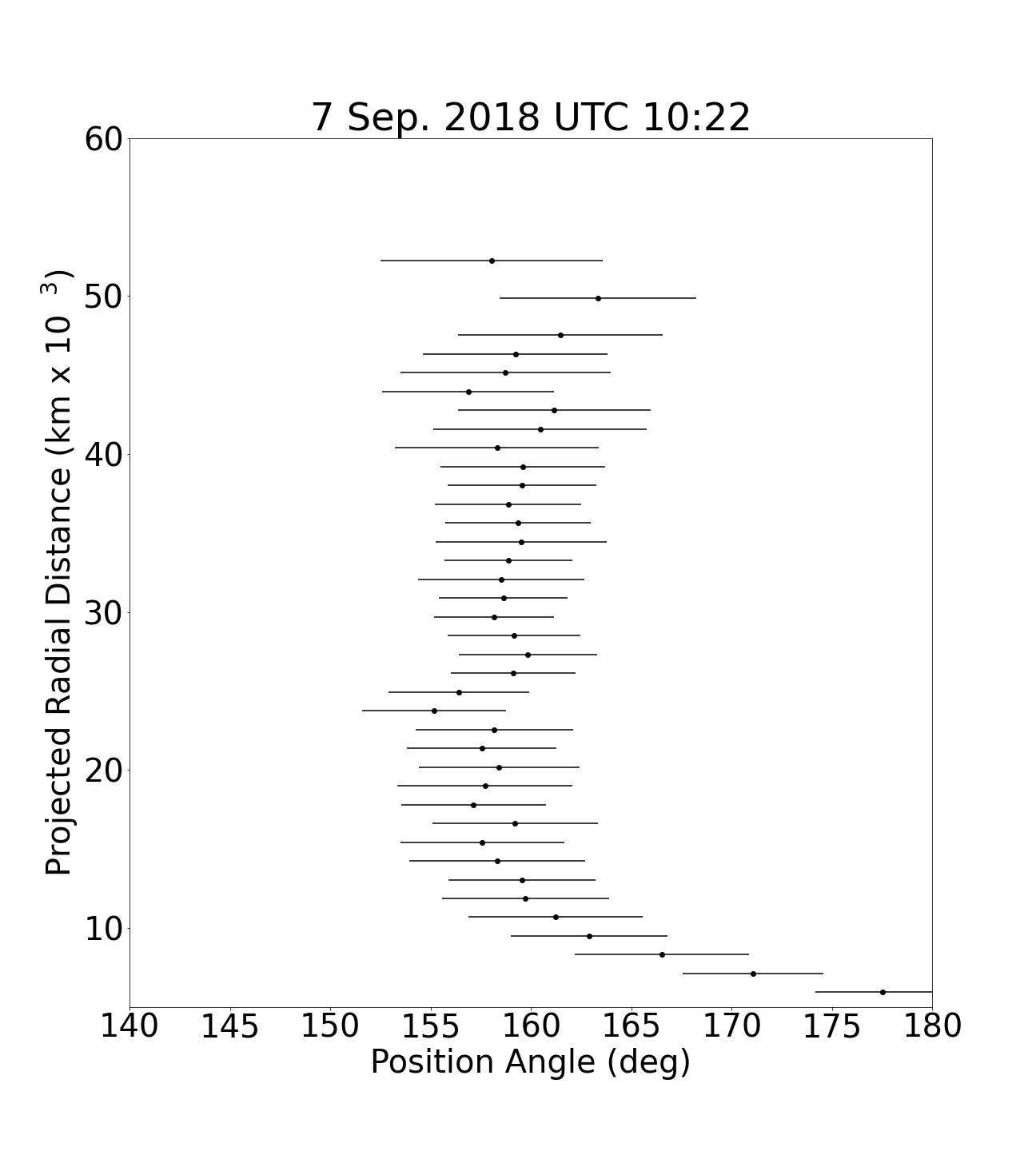}
    \includegraphics[width=0.9\textwidth]{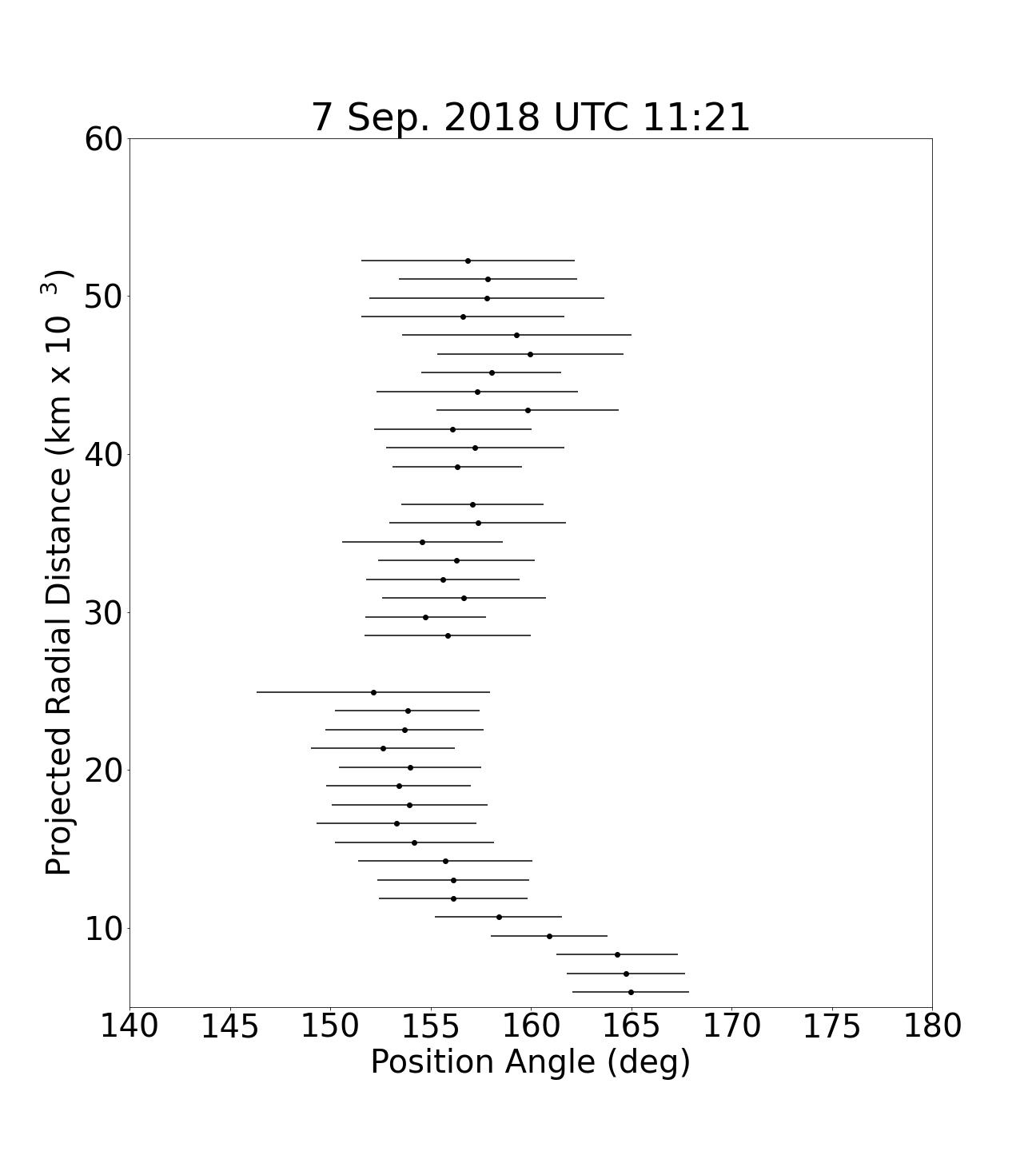}
    \vspace{2.5cm}
    \caption{Southern Jet morphology for 7 September 2018 at five different times spread out across the night, showing a strong curvature near the nucleus that is most prominent in the images taken at 11:21 and 11:46 UTC, potentially representing a corkscrew pattern.\label{morphology_night3_Jet2}}
    \vspace{1cm}
\end{minipage}
\end{figure}

\begin{figure}[h]
\centering
\begin{minipage}[t]{.45\textwidth}
\centering
    \includegraphics[width=0.9\textwidth]{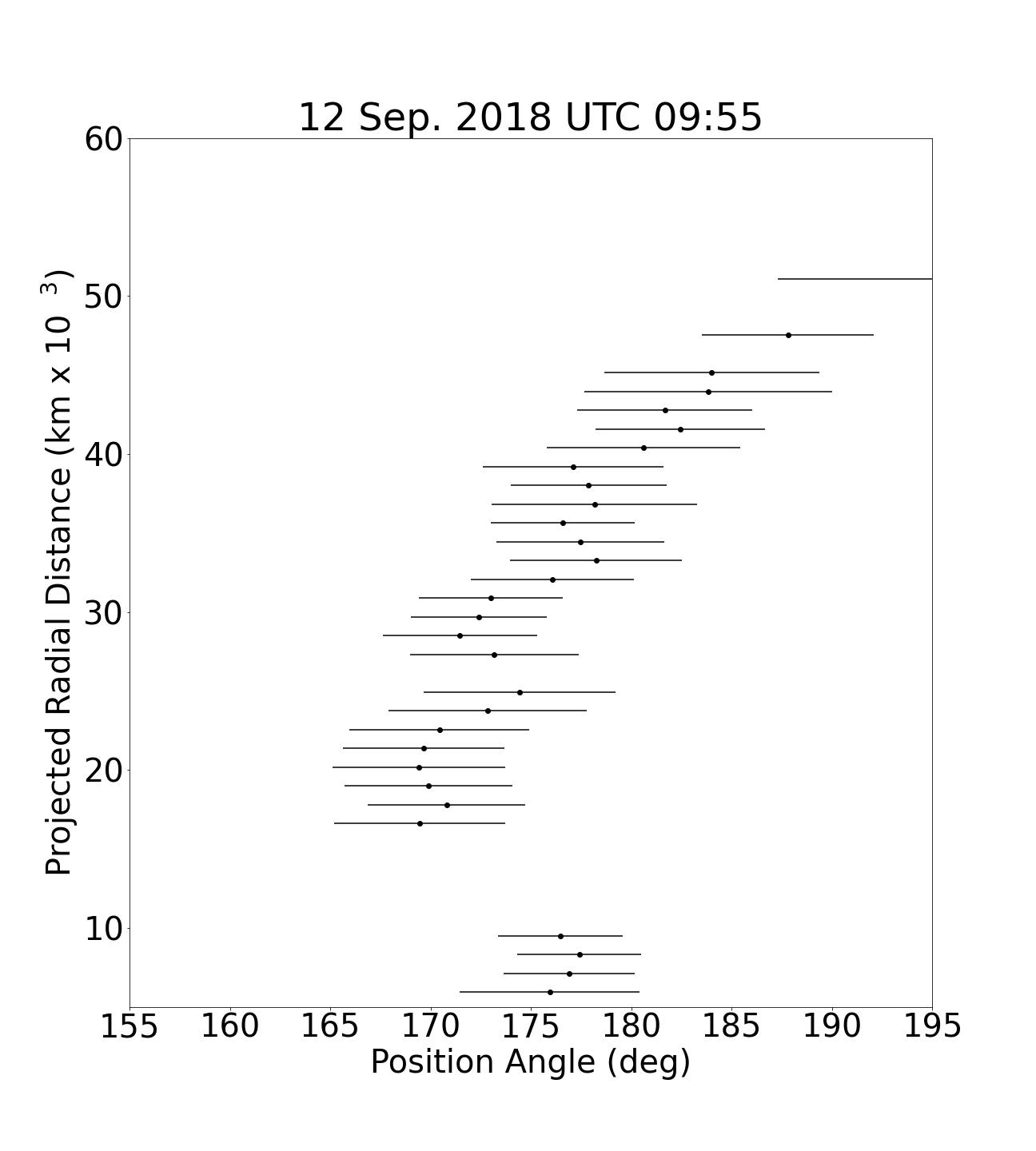}
    \includegraphics[width=0.9\textwidth]{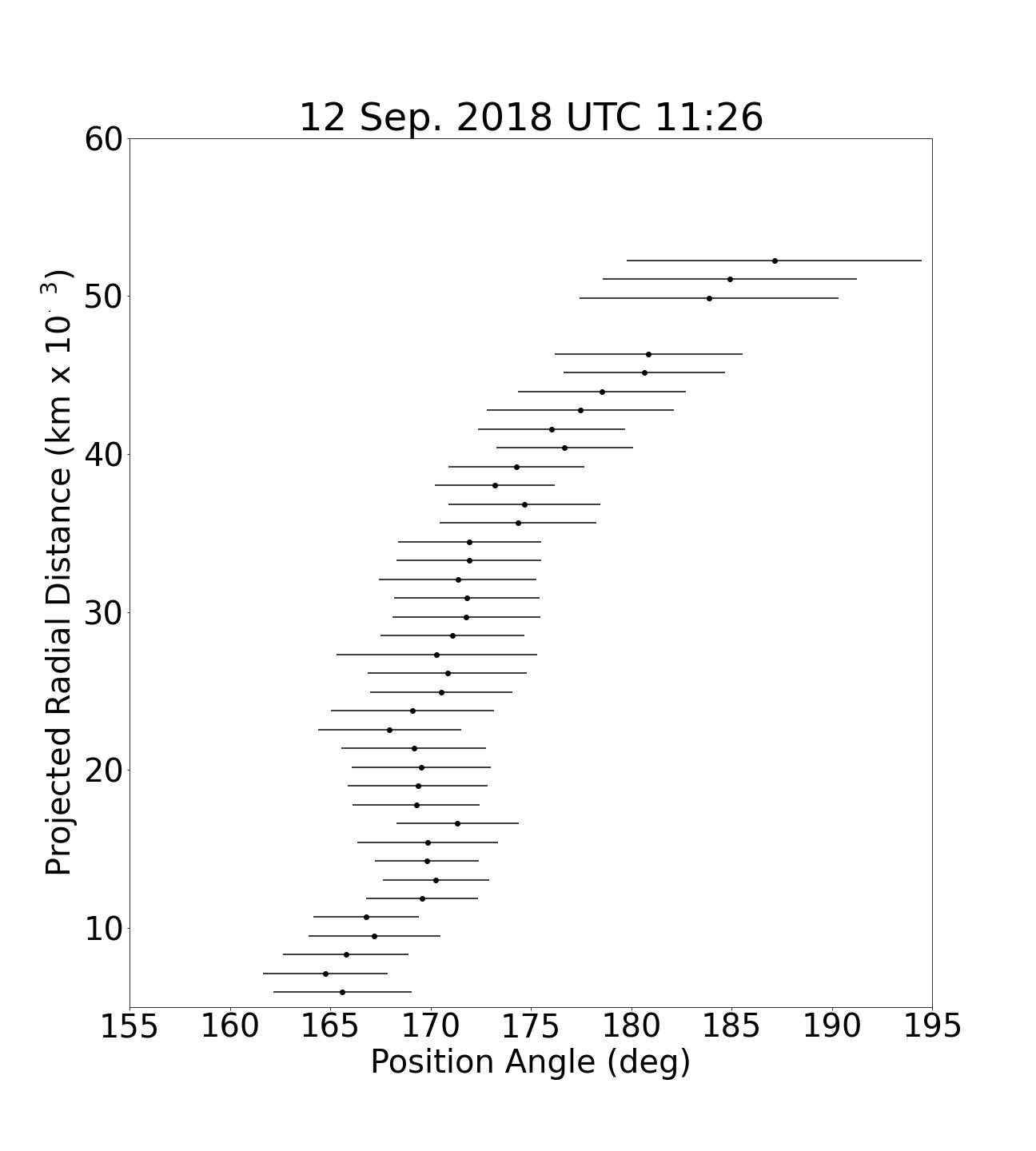}
    \end{minipage}
\hspace{0.05\textwidth}
\begin{minipage}[t]{.45\textwidth}
\centering
    \includegraphics[width=0.9\textwidth]{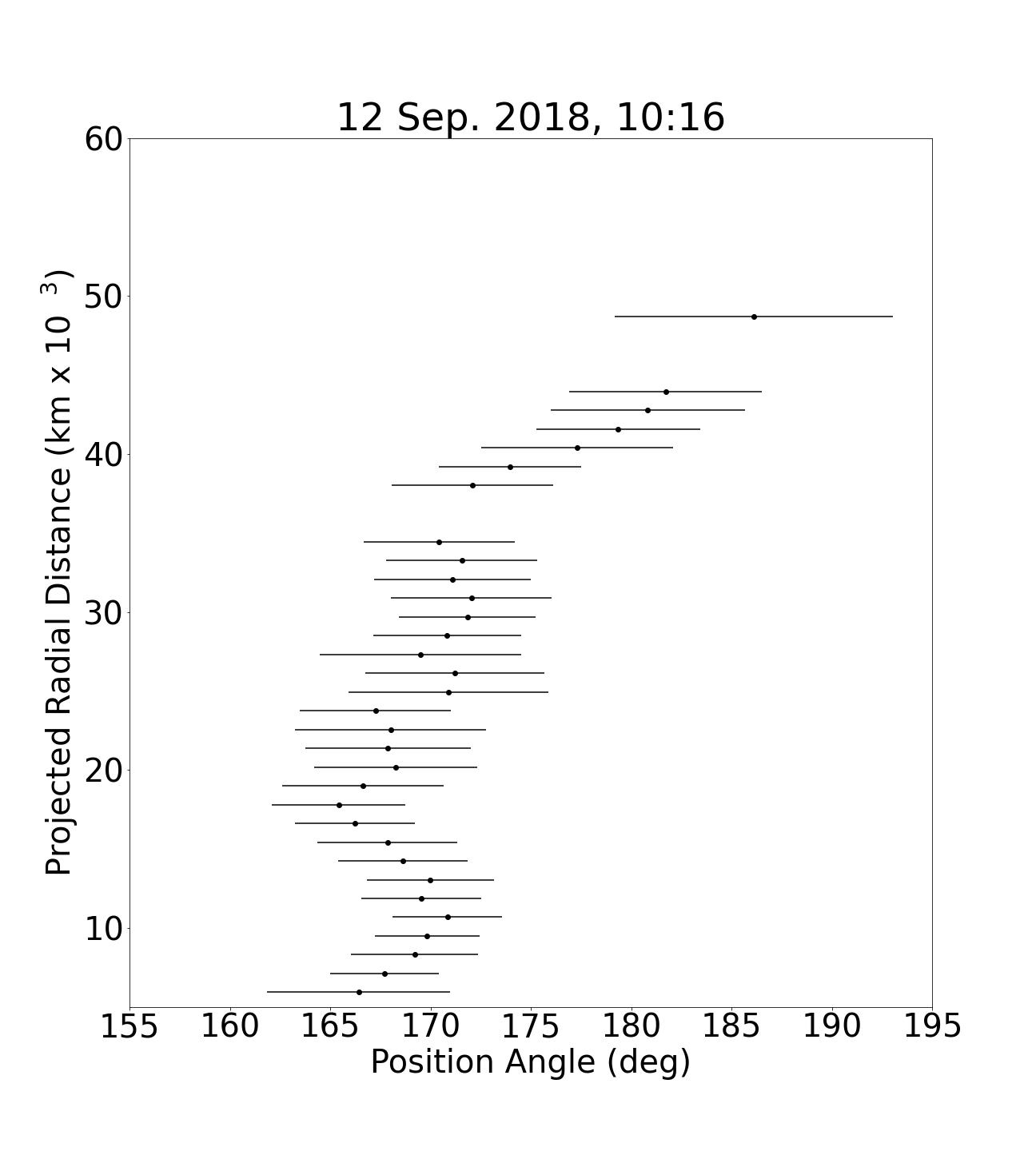}
    \includegraphics[width=0.9\textwidth]{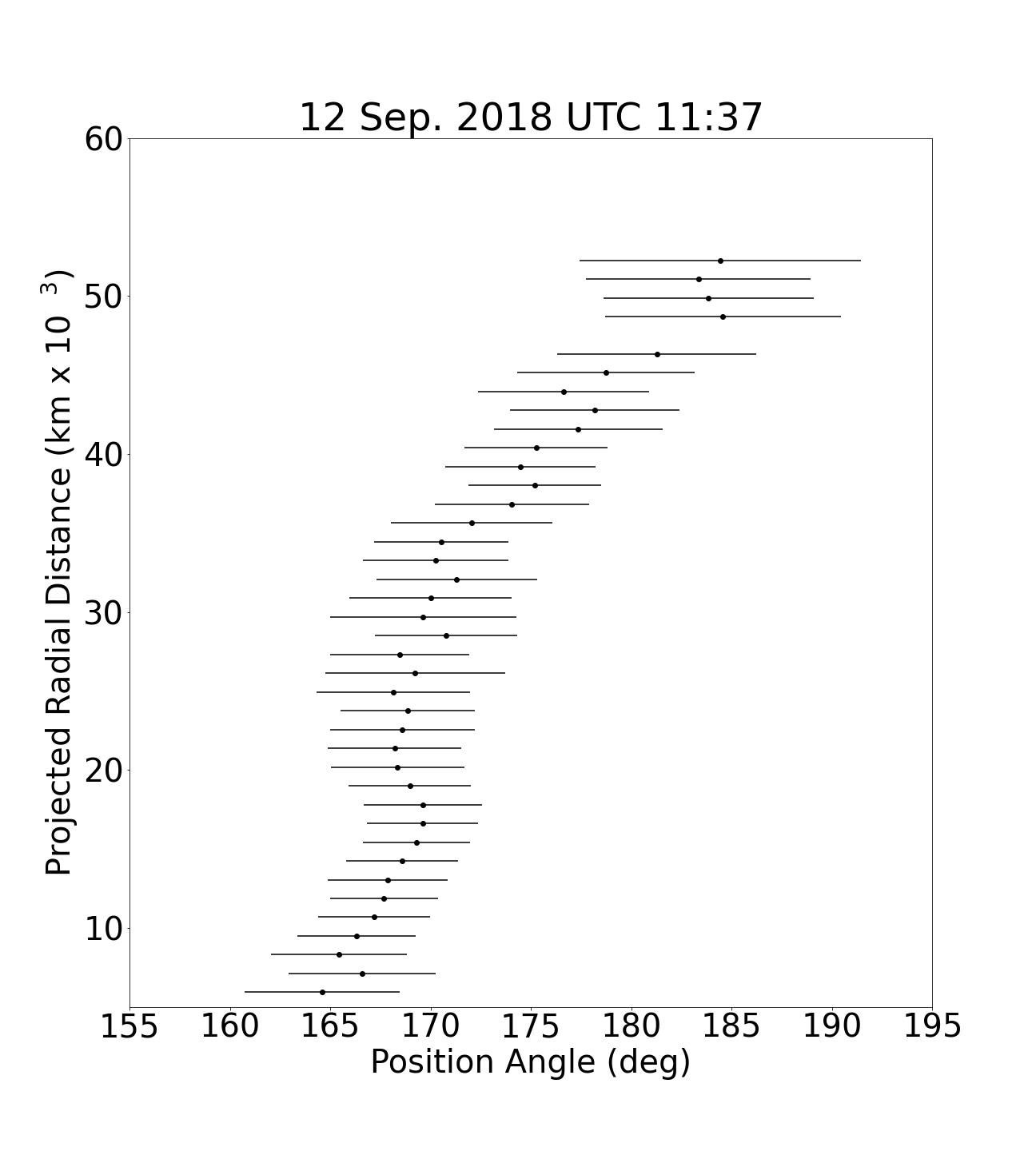}
\end{minipage}
    \caption{Southern Jet morphology for 12 September 2018 at four different times spread out across the night, showing a potential corkscrew pattern more visible at less than 20,000 km and a strong curvature at greater distances most likely due to solar radiation pressure.\label{morphology_night4_Jet2}}
\end{figure}

We analytically determine the azimuthal location of each jet feature as a function of projected distance for both the Northern and Southern Jets as explained in \S \ref{peak_distance}. Figures \ref{morphology_night1}--\ref{morphology_night4} and Figures \ref{morphology_night1_Jet2}--\ref{morphology_night4_Jet2} show the morphology of the Northern and Southern Jets, respectively, spaced over each night for 28 August 2018, 4 September 2018, 7 September 2018, and 12 September 2018. For the Northern Jet, there is a clear corkscrew pattern shown in several of these figures while the Southern Jet does not show as clear a pattern. We measure the behavior of the corkscrew pattern in the Northern Jet by determining the position angle of its central location, and its width. These results are summarized in Table \ref{jet1_position_angles}. The small width of the corkscrew pattern suggests that the jet originates at a location near the pole ranging from $\sim$4$^{\circ}$ to $\sim$6$^{\circ}$ away, indicating a cometocentric latitude of approximately $\sim$86$^{\circ}$ to $\sim$84$^{\circ}$. Additionally, considering the center of the corkscrew, we can determine the axis along which the comet must rotate for each night of observations.

\begin{table}[ht]
\centering
\caption{Measured position angle of the Northern Jet's corkscrew center and the width of the pattern.}
\label{jet1_position_angles}
\begin{tabular}{lccccc}
\textbf{ } & \textbf{Position} & \textbf{Width of}\\
\textbf{Date} & \textbf{Angle} (deg) & \textbf{Corkscrew} (deg)\\\hline
2018-08-28      & 20 $\pm$ 4 & 11 $\pm$ 4  \\
2018-09-04      & 29 $\pm$ 3 & 11 $\pm$ 3 \\
2018-09-07      & 26 $\pm$ 4 & 8 $\pm$ 4 \\
2018-09-12      & 15 $\pm$ 3 & 9  $\pm$ 3 \\
\end{tabular}
\end{table}

\subsection{Constraining the Pole Vector \label{Pole_vector_section}}
On a given night, the sky-plane projected rotational pole direction coincides with the measured position angle of the corkcrew's center, as shown in Table \ref{jet1_position_angles}. This, however, does not provide us with the non-projected information, which can be obtained by having a sufficient change in observing geometry of the comet. Utilizing similar methods as described in \citet{samarasinha_2002} and \citet{Farnham_2002}, we are able to constrain the pole vector using the Northern Jet.

We can determine the direction, in Right Ascension, $\alpha$, and Declination, $\delta$, of the pole vector of 21P/GZ as described below. First, we define the Earth's direction from the frame of reference of the comet as $\alpha_{\earth}$ and $\delta_{\earth}$ for the Right Ascension and Declination, respectively. Next, using a spherical to Cartesian transformation, we can describe the Cartesian coordinates of the 
pole vector of 21P/GZ, $x_p$, $y_p$, and $z_p$, as:
\begin{align}
    x_p &= \cos{\delta}\cos{\alpha} \\
    y_p &= \cos{\delta}\sin{\alpha} \\
    z_p &= \sin{\delta}.
\end{align}
The next step is to perform a transformation between the comet's frame of reference and the Earth's frame of reference. In this step, we are projecting the comet's frame of reference into the plane of the sky where the ``Northern" component \textendash\ labeled as $x_N$, $y_N$, and $z_N$ \textendash\ will be the component projected on the skyplane's north direction while the ``Eastern" component \textendash\ labeled as $x_E$, $y_E$, and $z_E$ \textendash\ will be the component projected onto the skyplane's eastern direction. Thus:
\begin{align}
    x_N &= - \sin{\delta_{\earth}}\cos{\alpha_{\earth}} \\
    y_N &= \sin{\delta_{\earth}}\sin{\alpha_{\earth}} \\
    z_N &= \cos{\delta_{\earth}},
\end{align}
and 
\begin{align}
    x_E &= \sin{\alpha_{\earth}} \\
    y_E &= - \cos{\alpha_{\earth}} \\
    z_E &= 0.
\end{align}
The final step is then to project the comet's pole vector onto the skyplane \textendash\ $Pole_N$ and $Pole_E$ for the northern and eastern components of the pole vector, respectively \textendash\ to be converted to a measurable position angle, and compared to our results. The pole vector projected onto the skyplane is given by: 
\begin{align}
    Pole_N &=x_p \cdot x_N + y_p \cdot y_N +z_p \cdot z_N \\
    Pole_E &=x_p \cdot x_E + y_p \cdot y_E +z_p \cdot z_E,
\end{align}
with the projected position angle of the pole position, $PA$, being given by:
\begin{equation}
    PA=atan2(Pole_N,Pole_E).   
\end{equation}

\begin{figure}
    \centering
    \hspace{-.45cm}
    \includegraphics[width=0.92\textwidth]{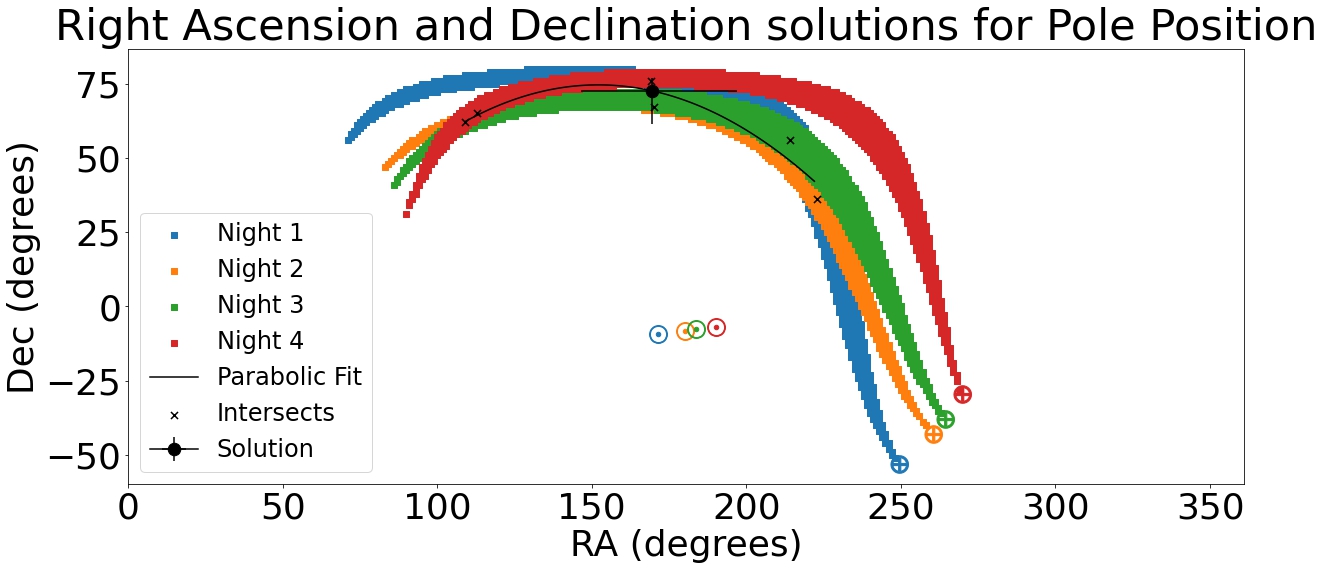}
    \includegraphics[width=0.9\textwidth]{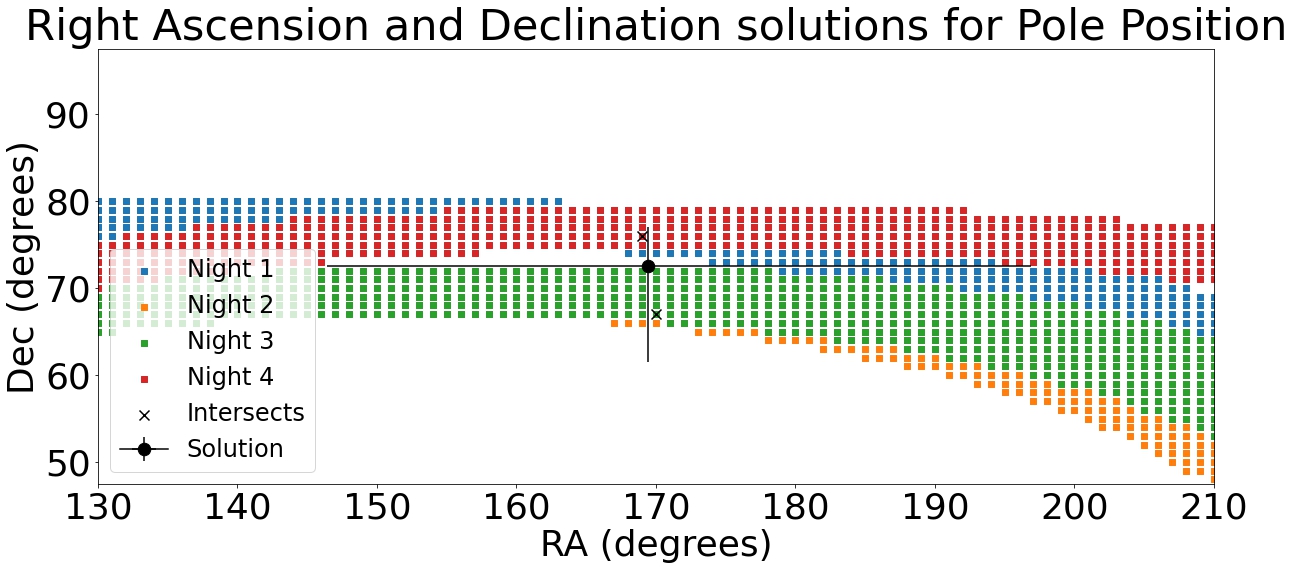}
    \caption{\textbf{Top:} Potential pole vector solutions for 28 August, and 4, 7 and 12 September 2018 as projected onto the celestial sphere. Intersections between pairs of nights are marked with $\times$, and the parabolic fit to the intersection points is shown in black. The $\odot$ and $\oplus$ indicate the Solar and Earth directions from the comet's frame of reference on the days corresponding to their color in the legend. The pole vector solution is marked at a Right Ascension of ${169^{+28}_{-23}}^{\circ}$ and a Declination of ${73^{+5}_{-11}}^{\circ}$. The error bars are computed by taking the standard deviations of the locations of the intersections on either side of the solution in both RA and DEC. \textbf{Bottom:} The same results zoomed in at the intersection point show that solutions for all four nights nearly overlap the pole vector solution and reveal the resolution of our grid search.}
    \label{pole_position}
\end{figure}

Where $atan2(Pole_N,Pole_E)$ is the angle between the North direction on the sky plane and the vector created by Pole$_N$ and Pole$_E.$ To determine the possible pole positions that correspond to our measurements in Table \ref{jet1_position_angles}, we conduct a grid search over all Right Ascensions and Declinations of potential pole vectors and reject all values which are not within the uncertainty of our measured projected pole position angles. Figure \ref{pole_position} shows the acceptable pole positions for all four nights of data.

To find the pole solution, all intersection points between pairs of nights are mapped, with six intersection points in total, as shown in Figure \ref{pole_position}. This produces a family of intersections for the pole position. The family of intersections is then fitted with a parabola, which closely matches the distribution of intersection points. The Right Ascension and Declination at the halfway point of the parabola's arc length are taken as the center of the family of solutions and the pole position of 21P/GZ. The position of 21P/GZ's pole vector is found to be at a Right Ascension of ${169^{+28}_{-23}}^{\circ}$ and a Declination of ${73^{+5}_{-11}}^{\circ}$. While the Right Ascension error is large, it is offset by the pole vector's high declination. The opposite pole vector would logically point to a Right Ascension of ${349^{+28}_{-23}}^{\circ}$ and a Declination of ${-73^{+11}_{-5}}^{\circ}$. Due to the lack of information surrounding the spin direction, we cannot conclude which pole corresponds to the conventionally defined positive pole. The pole vector of 169$^{\circ}$ and 73$^{\circ}$ in Right Ascension and Declination, respectively, is equivalent to an orbital longitude of 199$^{\circ}$ and obliquity of 10$^{\circ}$ in orbital coordinates, where orbital longitude is measured from the Sun-perihelion direction in a prograde sense in the orbital plane of comet 21P/GZ.

\subsection{Using our Determined Period to Calculate the $X$ parameter\label{x_parameter}}

When a comet approaches the sun, it releases gas that creates torques on its nucleus; these torques affect the rotation of the comet over time. \citet{x_param}
introduce the $X$ parameter, calculated by:

\begin{equation}
\label{equation_1}
    X=\frac{|\Delta P| R^2}{P^2 \zeta},
\end{equation}
where $\Delta P$ is the change in rotational period per orbit, $R$ is the radius of the nucleus, $P$ is the period of the comet, and $\zeta$ is the amount of water lost each orbit per unit surface area. The $X$ parameters of comets are expressed as the ratio between their $X$ parameter and that of comet 2P/Encke. Most comets have an $X/X_{Encke}$ within a factor of five \citep{x_parameter2}. We use our measured period, along with a period of 9.5 hours measured in the 1985 21P/GZ apparition by \citet{Leibowitz_1986} to determine the change in period per orbit. Table \ref{xparam_inputvalues} provides $\Delta P$ obtained from a combination of our measurements and those of \citet{Leibowitz_1986} along with $R$ and $\zeta$ values obtained from \citet{x_parameter2}. Using Equation \ref{equation_1} we calculate that $X_{GZ}/X_{Encke} \approx$ 3.4 $\pm$ 0.1 for a period of 7.39 $\pm$ 0.01 hours or $X_{GZ}/X_{Encke} \approx$ 1.9 $\pm$ 0.1 for a period of 10.66 $\pm$ 0.01 hours. This further supports the conclusion found in \citet{x_parameter2} that the $X$ parameter is a constant to within a factor of a few unity and suggests that 21P/GZ has not undergone abnormal torque effects in the recent past. 

\begin{table}[ht]
\centering
\caption{Values used to calculate $X_{GZ}/X_{Encke}$. $\Delta P$ is calculated from the change in period from \citet{Leibowitz_1986} and our determined period, which was measured 5 orbits later. The other values are obtained from \citet{x_parameter2}.}
\label{xparam_inputvalues}
\begin{tabular}{l|cccccc}
Comet & \textbf{$\Delta P$} (minutes) & \textbf{$R$} (km) & P (hours)  & \textbf{$\zeta/\zeta_{Encke}$}\\ \hline
21P/GZ (7.39 hour period) & 25.3 & 1.0 & 9.5 & 0.43 \\
21P/GZ (10.66 hour period) & 14.4 & 1.0 & 9.5 & 0.43 \\
2P/Encke & 4.0 & 2.4 & 11.0 & 1.0
\end{tabular}
\end{table}

\subsection{Unanswered Questions about Jet Morphology \label{Jet2_morphology}\label{npa_rotation}}

The morphology of the Southern Jet, as shown in Figures \ref{morphology_night1_Jet2} to \ref{morphology_night4_Jet2}, does not follow a clear and consistent pattern. However, it is possible to observe that at large cometocentric distances, the Southern Jet appears to be curving away from the sunward direction ($PA_{\odot}\approx 90^{\circ}$). This can be explained easily through solar radiation pressure. The relativistic momentum of a photon, $p$, is given by:
\begin{equation}
    p=\frac{E}{c},
    \label{momentum_equation}
\end{equation}
where $E$ is the energy of the photon and $c$ is the speed of light in a vacuum. The average energy that the CN molecules will then receive can be calculated using the fluorescence efficiency, $g$ as follows: 
\begin{equation}
    E= gt,
    \label{fluorescence_energy}
\end{equation}
where $t$ represents the amount of time a molecule spends in the coma. Combining Equations \ref{momentum_equation} and \ref{fluorescence_energy} provides:
\begin{equation}
    p=\frac{gt}{c}.
    \label{momentum_fluo}
\end{equation}
Knowing that $F=dp/dt$, we can calculate the force due to radiation pressure using Equation \ref{momentum_fluo} to obtain $F_{rad}=g/c$. Finally, using Newton's second law, $a=\frac{F}{m}$, we can solve for the acceleration due to radiation pressure, $a_{rad}$ in the direction opposite of the solar direction to obtain:
\begin{equation}
    a_{rad} = \frac{g}{cm_{CN}},
\end{equation}
where $m_{CN}$ is the mass of a CN molecule. Using the fluorescence efficiencies from \citet{Schleicher_fluorescence} for 21P/GZ on 28 August, 2018, we find the acceleration due to radiation is $\sim$ 0.25 cm s$^{-2}$. This decreases to 0.21 cm s$^{-2}$ by 12 September, 2018 due to a decreased fluorescence efficiency, $g$.

If we assume all the acceleration to be perpendicular to the motion of the jet, and we assume a measured outflow velocity, v=730 $\pm$ 30 m s$^{-1}$, at 50,000 km from the nucleus, the jet would have moved 5,900 $\pm$ 500 km in the anti-sunward direction, the equivalent of $6 \pm 1 ^{\circ}$. This is consistent with the curvature at higher distances as seen, for example, in Figure \ref{morphology_night1_Jet2}. 

\section{Conclusion}
We observed comet 21P/GZ over four nights from 28 August 2018 to 12 September 2018. We azimuthally enhanced our data to reveal the coma morphology of CN jet features. We used a newly developed method that measured the projected radial distance of the peak flux of a jet versus time, then using a Lomb-Scargle Periodogram, we obtained a rotational periodicity of either 7.39 $\pm$ 0.01 or 10.66 $\pm$ 0.01 hours. From the rotational periodicity of 21P/GZ, and previously obtained measurements dating from 1985 \citep{Leibowitz_1986}, we were able to measure the $X$ parameter, as defined in \citet{x_param}, and obtain $X_{GZ}/X_{Encke} \approx$ 3.4 $\pm$ 0.1 for a periodicity 7.39 $\pm$ 0.01 hours or $X_{GZ}/X_{Encke} \approx$ 1.9 $\pm$ 0.1 for a periodicity 10.66 $\pm$ 0.01 hours, which is consistent with contemporary literature.

Additionally, we were able to obtain a maximum projected CN gas outflow velocity for the Northern Jet of 730 $\pm$ 30 m s$^{-1}$ and for the Southern Jet of 740 $\pm$ 30 m s$^{-1}$.

We further analyzed both jets' morphologies to constrain rotational states. We determined that the Northern Jet possessed a corkscrew pattern, approximately 10$^{\circ}$ in width, suggesting a proximity to the pole position. From that analysis, we were able to constrain a pole position at a Right Ascension of ${169^{+28}_{-23}}^{\circ}$ and a Declination of ${73^{+5}_{-11}}^{\circ}$, with undetermined sense of rotation. From the analysis of the Southern Jet, we concluded that radiation pressures had a significant effect on the curvature at large cometocentric distances.

\section{Acknowledgements}
We would like to thank the invaluable time and effort given by Dr. Walt Harris in the mentoring necessary for the completion of this work. We would like to thank Dr. Beatrice Mueller for her contribution to the data analysis process. We would like to thank everyone who participated in the data gathering process, including all students from the Lunar and Planetary Laboratory course PTYS 495B/595B (Fall 2018). We would like to thank the Steward Observatory technical staff for the amount of time they dedicated to this project. Finally, we would like to thank the SSO grant NNX16A670G awarded to Walt Harris and the NESSF grant 80NSSC18K1241 awarded to Cassandra Lejoly (PI: Harris) for allowing this work to be completed.

\bibliography{references.bib}{}
\bibliographystyle{aasjournal}

\end{document}